\documentclass[trackchanges,preprint2]{aastex62}

\usepackage{amsmath}
\usepackage{breqn}
\usepackage{verbatim}
\usepackage{multirow}
\usepackage{booktabs}
\usepackage{mathrsfs}

\newcommand{\vk}{v_{\rm kick}}
\newcommand{\Mbh}{M_{\bullet}}
\newcommand{\ia}{i_a}
\newcommand{\ib}{i_b}
\newcommand{\ie}{i_e}

\graphicspath{{./}{Figures/}}

\submitjournal{ApJ}

\shorttitle{The Means to an END}
\shortauthors{Akiba \& Madigan}

\begin{document}

\title{Anisotropic Star Clusters around Recoiling Supermassive Black Holes}

\author[0000-0002-0647-718X]{Tatsuya Akiba}
\affiliation{JILA and Department of Astrophysical and Planetary Sciences, CU Boulder, Boulder, CO 80309, USA}
\email{tatsuya.akiba@colorado.edu}

\author[0000-0002-1119-5769]{Ann-Marie Madigan}
\affiliation{JILA and Department of Astrophysical and Planetary Sciences, CU Boulder, Boulder, CO 80309, USA}

\begin{abstract}
Gravitational wave recoil kicks from merging supermassive black hole binaries can have a profound effect on the surrounding stellar population. In this work, we study the dynamic and kinematic properties of nuclear star clusters following a recoil kick. We show that these post-kick structures present unique signatures that can provide key insight to observational searches for recoiling supermassive black holes. In \citet{Akiba2021}, we showed that an in-plane recoil kick turns a circular disk into a lopsided, eccentric disk such as the one we observe in the Andromeda nucleus. Building on this work, here we explore many recoil kick angles as well as initial stellar configurations.  
For a circular disk of stars, an in-plane kick causes strong apsidal alignment with a significant fraction of the disk becoming retrograde at large radii. If initial orbits are highly eccentric, an in-plane kick forms a bar-like structure made up of two anti-aligned lopsided disks. An out-of-plane kick causes clustering in the argument of periapsis, $\omega$, regardless of the initial eccentricity distribution. Initially isotropic configurations form anisotropies in the form of a torus of eccentric orbits oriented perpendicular to the recoil kick. Post-kick surface density and velocity maps are presented in each case to highlight the distinct, observable structures of these systems.
\end{abstract}


\section{Introduction} \label{sec:intro}

Binary supermassive black holes are a natural consequence of galaxy mergers \citep{Beg80}, and gravitational waves are emitted during their inspiral and eventual coalescence. In an asymmetric black hole binary (i.e., different masses and/or spins), anisotropic gravitational waves carry away linear momentum in addition to angular momentum and energy. Most radiation occurs near time of merger \citep{Gonzalez2007a, Lousto2008, Blanchet2005} and the remnant black hole receives a recoil kick \citep{Peres1962, Bekenstein1973, Wiseman92,Campanelli2007, Herrmann07}. Typical kick velocities of $\mathcal{O}(100)~ \rm{km \ s^{-1}}$ result in damped oscillations of the merger remnant within the galactic potential with its eventual return to the nucleus on $\mathcal{O}$($10^6$--$10^9$~yr) timescales \citep{Merritt2004, Madau2004, KomossaMerritt2008}.  In rare cases, the kick velocity may exceed the escape velocity of the galaxy leading to black hole ejection \citep{Gualandris2008}.
Detections of recoiling supermassive black holes would provide key constraints on black hole binary mass and spin evolution, on gravitational wave event rates, and the merger evolution of galaxies. 

Observational searches for recoiling black holes have thus far focused on active galaxies. 
If a recoiling black hole is accreting at the time of kick, it carries with it its accretion disk and broad line region. The recoiling remnant may then be observed as an offset active galactic nucleus (AGN) \citep{Madau2004,Blecha2008}. Numerous such candidates have been discovered either spectroscopically via Doppler-shifted optical broad lines and/or via kpc-scale spatial offsets \citep[e.g.,][]{KomossaZhouLu2008,Chiaberge2018,Kim2018,Hogg2021}. However, most black holes are not actively accreting as AGN \citep{Kormendy2013} which greatly limits candidate numbers. It is, furthermore, difficult to rule out alternate possibilities such as gaseous outflows, AGN jet activity,  and dual AGN \citep[e.g.,][]{Comerford2017}. 

In this paper, we explore the effect that the gravitational wave recoil kick has on the surrounding stellar population, as the basis for using the distinct structures to detect recoiling systems. 
\citet{Komossa2008} and \citet{Merritt2009} first explored this topic finding that a recoiling black hole carries with it a ``hypercompact stellar system'' (HCSS). This system may reveal itself via tidal disruption flares or accretion of gas from stellar winds onto the black hole.  
Furthermore, the properties of HCSSs contain information about  the merger histories of galaxies, binary black hole evolution, and the distribution of gravitational-wave kicks.

In \citet{Akiba2021} we explored the post-kick orbits of the stars within a HCSS. To make the problem analytically tractable, we focused on the simple case of the stars initially being distributed in a circular disk, inspired by the theory of circumbinary star formation in a self-gravitating accretion disk \citep{Pac78,Goo03}. We furthermore considered the effect of an in-plane recoil kick motivated by \citet{Bogdanovic2007}. We showed that the kick experienced by the merger remnant can directly result in the formation of a lopsided stellar disk of eccentric, aligned orbits, a configuration we call an {\it eccentric nuclear disk (END)}. Eccentric disks are fairly common in our local universe \citep{Lauer2005}. The double nucleus of M31 \citep{Lauer1993}, for instance, is well-explained by an eccentric disk \citep{Tremaine1995}. If the eccentric disk was formed via gravitational wave recoil, its dynamic and kinematic properties can illuminate M31's merger history.

To measure alignment of stellar orbits we use the eccentricity vector, 
\begin{equation}
    \vec{e} = \frac{\vec{v} \times \vec{j}}{G\Mbh} - \frac{\vec{r}}{r} \ ,
    \label{eqn:eccentricity_vector}
\end{equation}
\noindent where $\vec{r}$ is the radius vector, $\vec{v}$ is the velocity vector, $\vec{j}$ is the angular momentum vector, and $\Mbh$ is the mass of the central black hole. The eccentricity vector points from the apoapsis to the periapsis of an orbit and its magnitude is equal to the scalar eccentricity. In \citet{Akiba2021}, we showed that a circular orbit in the $x$-$y$ plane about a central object that is given an in-plane recoil kick in the $x$-direction will have a preferential direction of eccentricity vectors,

\begin{equation}
    \langle \vec{e} \rangle = \frac{3}{2} \frac{\vk}{v_*} \ \hat{y} \ ,
\label{eqn:mean_ecc_vec}
\end{equation}

\noindent where $\vk$ is the recoil kick speed and $v_*$ is the initial speed of the star. That is, a lopsided disk of stars will form around the recoiling black hole perpendicular to its direction of motion. We showed using $N$-body simulations that the eccentric nuclear disk maintains its apsidal-alignment over tens of thousands of orbits (the limit of our simulations) and that this leads to a greatly enhanced rate of tidal disruption events.

Here we extend our original study in \citet{Akiba2021} to more general initial stellar distributions and a range of kick directions. Our goal is to systematically explore these parameter spaces and present a catalog of asymmetric stellar distributions following a recoil kick imparted on the central object. We note that while we focus our attention on stars surrounding a recoiling supermassive black hole, the dynamics explored here are directly applicable to other astrophysical contexts such as the orbits of planetary material following a white dwarf natal kick or a supernova kick. 

We begin in Section~\ref{sec:circ_in} with stars once again in a circular disk experiencing an in-plane kick, and sequentially add complexity. In Section~\ref{sec:circ_out} we add an out-of-plane component to the kick, and in Section~\ref{sec:ecc} we give non-zero values of eccentricity to the stars. In Section~\ref{sec:inc}, we relax the constraint of the stars being in a disk and expand our exploration to spherical distributions. In each section we confirm analytic predictions with numerical simulations. We summarize our findings and conclude in Section~\ref{sec:conclusion}.

\section{Circular Disk with an In-plane Kick Revisited}
\label{sec:circ_in}

\subsection{Analytics: circular disk, in-plane kick}

In \citet{Akiba2021}, we examined the kick magnitude required to induce apsidal alignment in a disk with a very narrow radial range. In reality, we are more interested in the inverse problem. The recoil kick is determined by relativistic effects, which in theory can be calculated using the black hole orbits, masses and spins. The surrounding stellar disk, which spans a wide range of semi-major axes, reacts to this kick with certain radial ranges responding more strongly than others. In \citet{Akiba2021}, we  hypothesized that there are three distinct radial regimes of interest based on comparing the circular speed of stars, $v_*$, and the recoil kick speed, $\vk$. When $v_* \gg \vk$, the stellar orbits are negligibly affected and the $\sim$circular disk is retained. When $v_* \sim \vk$, the apsidal alignment is strong and an eccentric nuclear disk forms. Finally, when $v_* \ll \vk$, stars become unbound from the black hole. However, due to the narrow radial range investigated, we overlooked a key feature --- the presence of a \textit{predominantly retrograde disk at large radii}.

\begin{figure*}[t!]
\centering
\includegraphics[width=0.8\textwidth]{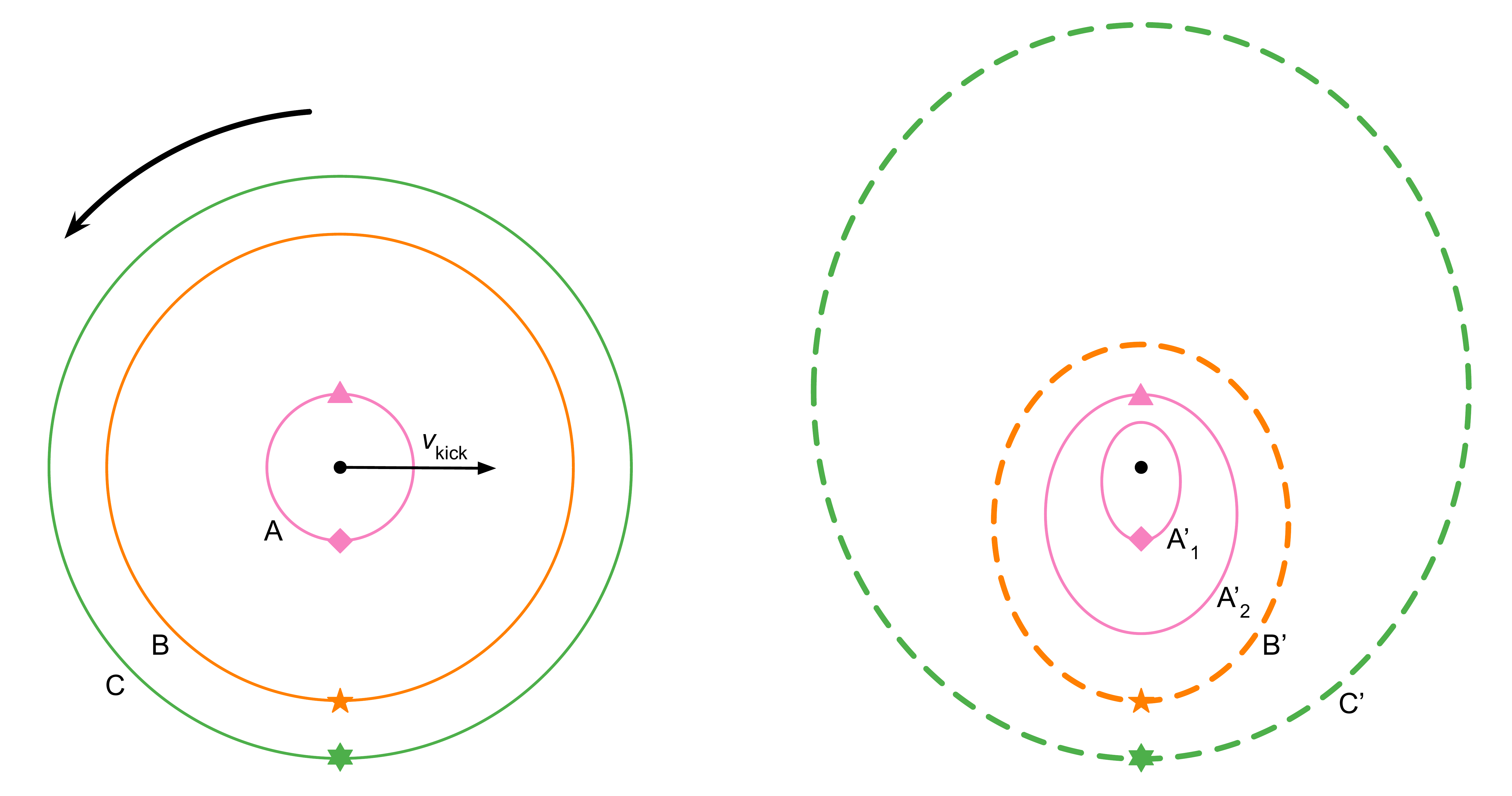}
\caption{\textbf{Circular disk with an in-plane recoil kick.} 
A schematic diagram illustrating four pre- and post-kick orbits. (\textit{Left:}) Initial circular orbits A, B, and C with markers indicating the stars' positions at the time of kick, and the arrow labeled $\vk$ is the black hole kick direction. The stars initially move on counter-clockwise (prograde) orbits. (\textit{Right:}) The post-kick eccentric orbits where dashed lines indicate where the stars move on clockwise (retrograde) orbits. Thus, we expect a retrograde population to emerge above a certain radius in the post-kick stellar distribution. \\}
\label{fig:kick_schematic}
\end{figure*}

A schematic diagram illustrating the emergence of a retrograde disk is shown in Figure \ref{fig:kick_schematic}. In the left panel, three different initial circular orbits A, B, and C are shown in solid lines with several possible star locations indicated by markers. The recoil kick is shown by the arrow labeled $\vk$. Orbit A is the inner region investigated in \citet{Akiba2021}: the top (pink triangle) star speeds up in the reference frame of the black hole moving onto a larger semi-major axis orbit ($\rm{A}'_2$) while the bottom (pink diamond) star slows down and goes into a smaller semi-major axis orbit ($\rm{A}'_1$), as shown in the right panel, with both orbits remaining prograde. The triangle star will remain bound for kick velocities $\vk < (\sqrt{2}-1) v_*$, which correspond to radii \[ r < \frac{(\sqrt{2}-1)^2 G\Mbh}{\vk^2} \ . \]
Apsidal alignment will be strongest where $\langle \vec{e} \rangle \sim 1$. Using Equation \ref{eqn:mean_ecc_vec}, we define this point to be the \textit{characteristic radius} of the eccentric nuclear disk,

\begin{equation}
r_c \equiv \frac{4G\Mbh}{9 \vk^2} \ .
\label{eqn:ecc_rad}
\end{equation}

\noindent The diamond star will have a prograde post-kick orbit if $\vk < v_*$, at radii \[ r < \frac{G\Mbh}{\vk^2} \ . \] Moving to larger radii, the stellar velocity can reverse and become retrograde in the reference frame of the black hole, illustrated by orbits B and C in the schematic. For orbit B, the (orange five-pointed) star moves onto a retrograde orbit with smaller semi-major axis ($\rm{B}'$). This happens when $v_* < \vk < 2 v_*$, or at radii \[\frac{G\Mbh}{\vk^2} < r < \frac{4G\Mbh}{\vk^2} \ . \] Even further out at orbit C, the (green six-pointed) star goes into a larger semi-major axis orbit that is retrograde ($\rm{C}'$). This occurs when $2 v_* < \vk < (\sqrt{2} + 1) v_*$, corresponding to radii \[\frac{4G\Mbh}{\vk^2} < r < \frac{(\sqrt{2} + 1)^2G\Mbh}{\vk^2} \ . \] Beyond $r_{\rm{out}} \approx 13 \ r_c$, no star remains bound to the recoiling black hole in this set-up.

\subsection{Numerical Set-Up} \label{sec:sim_setup}

To explore the post-kick radial structures further, we use the open-source, $N$-body simulation package \texttt{REBOUND} \citep{Rein2012}. We use code units of $G = 1$, $\Mbh = 1$, and characteristic radius $r_c = 1$ such that the period of a circular orbit at this radius is $P(r_c) = 2 \pi$. For astrophysical context, we will translate these units to $\Mbh = 4 \times 10^{6} \ M_{\odot}$ and $r_c = 0.765$ pc (given by Equation \ref{eqn:ecc_rad}) where we choose $\vk = 100$ km/s.

To study the instantaneous post-kick stellar distribution, we initialize $N = 5 \times 10^4$ stars in an axi-symmetric, thin disk spanning four orders of magnitude in semi-major axis space, $a = 0.005$--$50$, with surface density profile $\Sigma \propto a^{-1}$. Inclination is Rayleigh distributed with scale parameter $\sigma_i = 3^\circ$. Longitude of periapsis, $\varpi$, and mean anomaly, $\mathcal{M}$, are uniformly distributed in $[0, 2 \pi)$. Since the characteristic radius is $r_c = 0.765$ pc, the initial semi-major axis range translates to $a = 0.00382$--$38.2$ pc. The outer edge is beyond the radius of influence for a black hole of this mass, but we note the sensitive dependence of the characteristic radius on the kick velocity, $r_c \propto \vk^{-2}$. If the kick velocity doubles, $\vk = 200$ km/s, the characteristic radius drops by a factor of four, $r_c = 0.19$ pc, and so on. As we are investigating the distribution of stars instantaneously following the kick, the stars are massless in our simulations. Surface density and velocity maps are generated by evolving stars on their post-kick Keplerian orbits.

\subsection{Numerical Results: circular disk, in-plane kick}

\subsubsection{Radial Regions and Apsidal Alignment}

\label{sec:circ_in_rad}

\begin{figure*}[t!]
\centering
\includegraphics[width=0.85\linewidth]{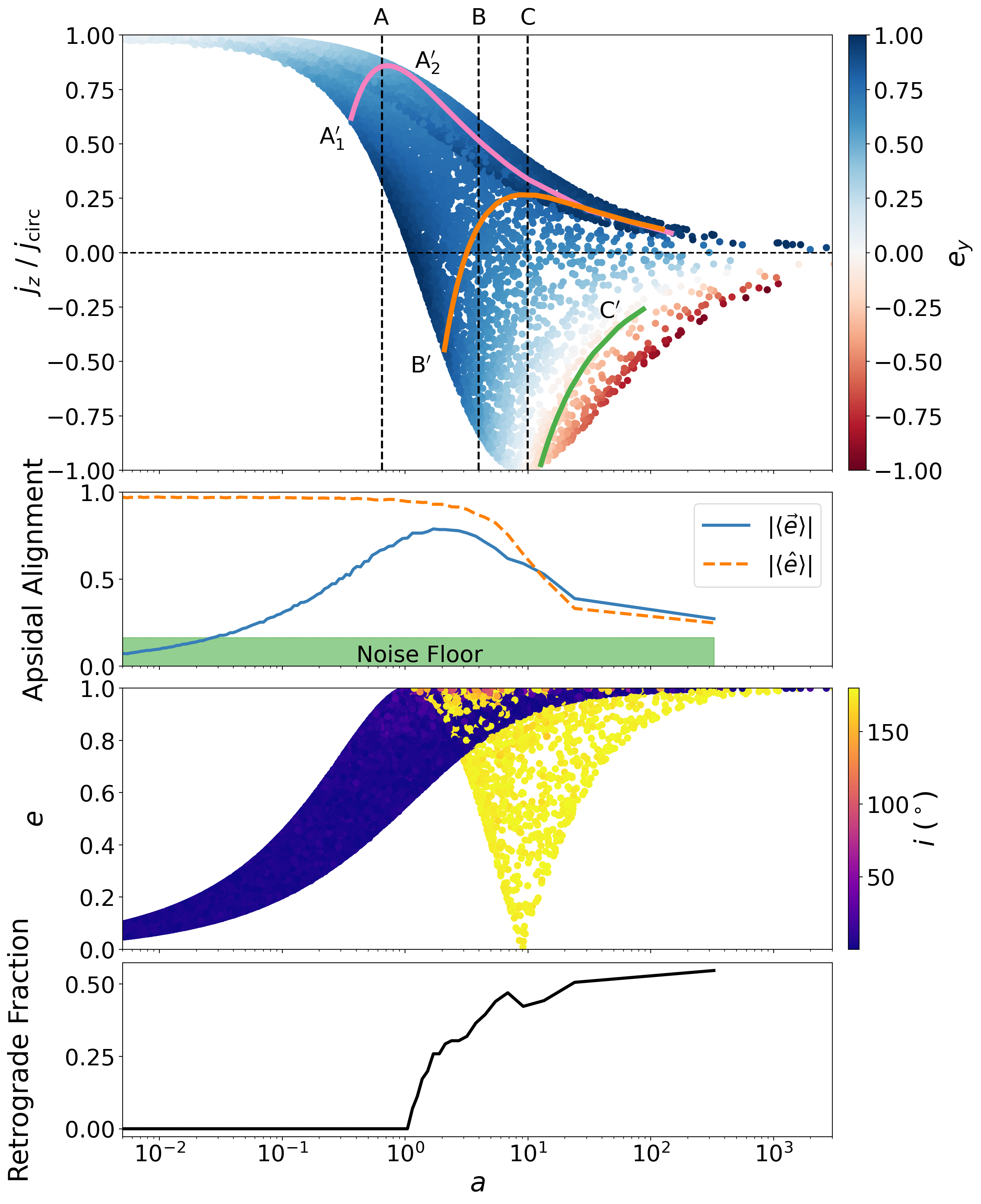}
\caption{\textbf{Circular disk with an in-plane recoil kick.} 
 (\textit{Top:}) The post-kick distribution of the $z$-component of angular momentum (normalized by the circular angular momentum at a given radius) as a function of post-kick semi-major axis scaled by the characteristic radius, with the color bar showing the $y$-component of the eccentricity vector. We show radii corresponding to the circular orbits illustrated in the schematic Figure \ref{fig:kick_schematic}, A, B, and C (black dashed lines), and the corresponding post-kick distribution emanating from those radii (pink, orange, and green solid lines). (\textit{Second from top:}) Two measures of post-kick apsidal alignment: the magnitude of the mean eccentricity vector (blue solid line), $| \langle \vec{e} \rangle|$, and the mean unit eccentricity vector (orange dashed line), $| \langle \hat{e} \rangle|$, binned by semi-major axis. The shaded noise floor shows the maximum expected deviation from 0 (at a $3 \sigma$ level) for an isotropic distribution. (\textit{Second from bottom:}) The eccentricity profile as a function of semi-major axis color-coded by orbital inclination. (\textit{Bottom:}) The retrograde fraction of the disk as a function of semi-major axis. The post-kick stellar orbits exhibit a rich structure of apsidal alignment, eccentricity, and retrograde fraction each with a sensitive dependence on semi-major axis.}
\label{fig:sma_ang_mom_mean_ecc_circ_in}
\end{figure*}

In Figure \ref{fig:sma_ang_mom_mean_ecc_circ_in} in the top panel, we show the post-kick distribution of the $z$-component of angular momentum (normalized by the circular angular momentum at a given radius), $j_z/j_{\rm{circ}}$, as a function of semi-major axis, with the color bar showing the $y$-component of the eccentricity vector. Prograde orbits are above ($j_z > 0$) and retrograde orbits are below ($j_z < 0$) the dashed black line. Circular orbits are farther from the dashed line with $|j_z/j_{\rm{circ}}| \sim 1$ and orbits become highly eccentric as $|j_z/j_{\rm{circ}}| \rightarrow 0$. Overlaid on top are the three radii schematically illustrated in Figure \ref{fig:kick_schematic}. All stars that start at an inner orbit A remain prograde, but some stars become eccentric with a smaller semi-major axis ($\rm{A}_1'$) whereas others take on a larger semi-major axis ($\rm{A}_2'$) depending upon their initial position with respect to the kick vector. Stars that start at B can become retrograde, once again depending on their initial position, but stars that flip orientation move to smaller semi-major axis orbits ($\rm{B}'$). The stars that remain on prograde orbits move to high eccentricity orbits. Stars beginning at C that remain bound all flip to retrograde orientation and take on a larger semi-major axis orbit ($\rm{C}'$). Based on these results, we identify five kinematically distinct post-kick radial regions instantaneously following the kick:

\begin{enumerate}

    \item At $a < 0.1$, $v_* \gg \vk$ and the circular orbits are mostly unaffected.
    \item At $0.1 < a < 1$, we have a prograde eccentric disk in the $+y$-direction with eccentricities generally increasing with increasing semi-major axis.
    \item At $1 < a < 10$, we have an eccentric disk in the same direction ($+y$) and an emerging retrograde population. The prograde stars become more eccentric whereas retrograde stars become more circular with increasing semi-major axis.
    \item At $10 < a \lesssim 3 \times 10^3$, we get a disk where all of the orbits are eccentric and the prograde orbits are aligned in the same direction as before ($+y$), but the retrograde orbits are aligned in the opposite direction ($-y$). Both prograde and retrograde orbits become more eccentric with increasing semi-major axis.
    \item Finally, at $a \gtrsim 3 \times 10^3$, we do not observe many stars bound to the black hole. For most reasonable choices of astrophysical scale, this region would be outside of the black hole's radius of influence.

\end{enumerate}

In the second from the top panel of Figure \ref{fig:sma_ang_mom_mean_ecc_circ_in}, we show two measures of apsidal alignment binned by semi-major axis. The mean eccentricity vector (solid blue line) is given by

\[ | \langle \vec{e} \rangle | = \left| \frac{\sum_i^N \vec{e}_i}{N} \right| \ , \]

\noindent where $N$ is the number of bound particles and $\vec{e}_i$ is the eccentricity vector of the $i$-th particle. The mean \textit{unit} eccentricity vector (dashed orange line) is similarly defined by replacing $\vec{e}$ with $\hat{e}$, the unit eccentricity vector. A perfectly aligned eccentric nuclear disk would have $| \langle \hat{e} \rangle | = 1$ and $| \langle \vec{e} \rangle | = \langle e \rangle$, the mean scalar eccentricity, whereas a random distribution of eccentricity vectors gives both $| \langle \hat{e} \rangle |$ and $| \langle \vec{e} \rangle | \sim 0$. The mean unit eccentricity vector was used in \citet{Akiba2021} and measures the alignment of eccentricity vectors independent of how eccentric the orbits are. For an eccentric disk to be stable, however, both the magnitude and alignment of eccentricity vectors are important for orbits to effectively gravitationally torque each other \citep{Madigan2018}. The difference between $| \langle \vec{e} \rangle |$ and $| \langle \hat{e} \rangle |$ is clearest at small semi-major axes. While the eccentricity vectors are extremely aligned with $| \langle \hat{e} \rangle | \sim 1$, the orbits remain $\sim$circular at small semi-major axis, so apsidal alignment is unlikely to be maintained long-term. $| \langle \vec{e} \rangle |$ captures this effect and gives us the expected peak in apsidal alignment near $r_c$. 
The noise floor shows the maximum expected $|\langle \vec{e} \rangle|$ or $|\langle \hat{e} \rangle|$ deviations (within 3$\sigma$) if the stellar distribution were isotropic. $\sigma = 1/\sqrt{N}$ for $|\langle \hat{e} \rangle|$ and $\sigma = \sqrt{\langle e^2 \rangle/N}$ for $|\langle \vec{e} \rangle|$. Since $1/\sqrt{N} > \sqrt{\langle e^2 \rangle/N}$ for bound stars, we plot $1/\sqrt{N}$ as the ``noise floor''. The eccentricity vectors become most aligned when $0.1 < a < 10$ as expected.

In the second from the bottom panel of Figure \ref{fig:sma_ang_mom_mean_ecc_circ_in}, we show the eccentricity profile as a function of semi-major axis with the inclination distribution shown with the color bar. The prograde population has a predominantly positive eccentricity gradient, $de/da > 0$, whereas the retrograde population hits a minimum eccentricity at $a \sim 10$ with eccentricities increasing both below and above. Finally, in the bottom panel of Figure \ref{fig:sma_ang_mom_mean_ecc_circ_in}, we show the retrograde fraction as a function of semi-major axis. We see that the retrograde population emerges for semi-major axis larger than $\sim r_c$, and the fraction is $\gtrsim 40 \%$ for $a > 10$. This significant retrograde fraction produces distinct density and velocity maps offering a unique opportunity to identify recoiling black holes using their surrounding star clusters. 

\subsubsection{Post-Kick Density and Velocity Profile}

\begin{figure}[t!]
\centering
\includegraphics[width=\linewidth]{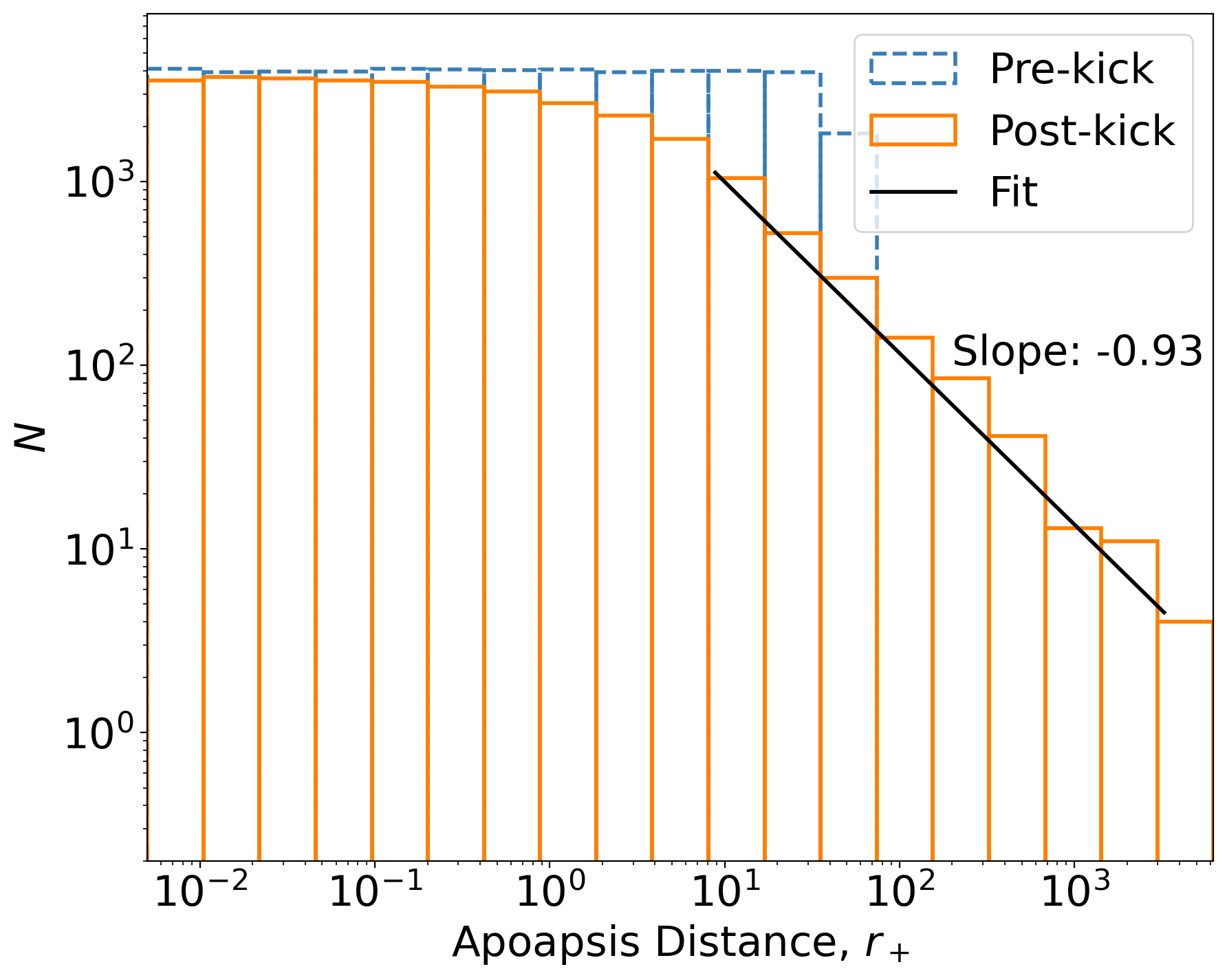}
\caption{\textbf{Circular disk with an in-plane recoil kick.} 
Pre- and post-kick number density as a function of apoapsis distance, $r_+$. The initial pre-kick number density is flat, but the post-kick number density extends to higher apoapsis distances with a turn around from a flat region to a power-law region around $r_+ \approx 4$. At $r_+ \gtrsim 4$, the number density is roughly distributed as $N \propto r^{-0.93}$ according to our linear fit.}
\label{fig:sigma_mean_r_circ_in}
\end{figure}

The stellar density profile changes as a result of the recoil kick. Since we start with a circular disk of stars, the initial radius of a stellar orbit is well-defined. Post-kick, however, many of the stellar orbits obtain significant eccentricities. In this case, stars change their distance from the central object dramatically over the course of their orbit. Since stars on eccentric orbits spend most of their time near apoapsis, we assume that characterizing the number density as a function of the apoapsis distance, $r_+$, will be the most accurate in an orbit-averaged sense. In Figure \ref{fig:sigma_mean_r_circ_in}, we show how the number density as a function of apoapsis distance changes following the kick. Before the kick, the number density is flat with apoapsis distance, consistent with $\Sigma \propto a^{-1}$. Post-kick, the stellar distribution extends to larger apoapsis distances and there appears a turnover from a flat region to a power-law sloped region beyond an apoapsis distance $r_+ \approx 4$. Beyond this radius, the approximate number density profile is $N \propto r^{-0.93}$, which corresponds to a cuspy surface density profile $\Sigma \propto r^{-1.93}$. In other words, the number density remains relatively unchanged at small radii and becomes significantly steeper at larger radii.

\begin{figure*}[t!]
\centering
\includegraphics[width=\linewidth]{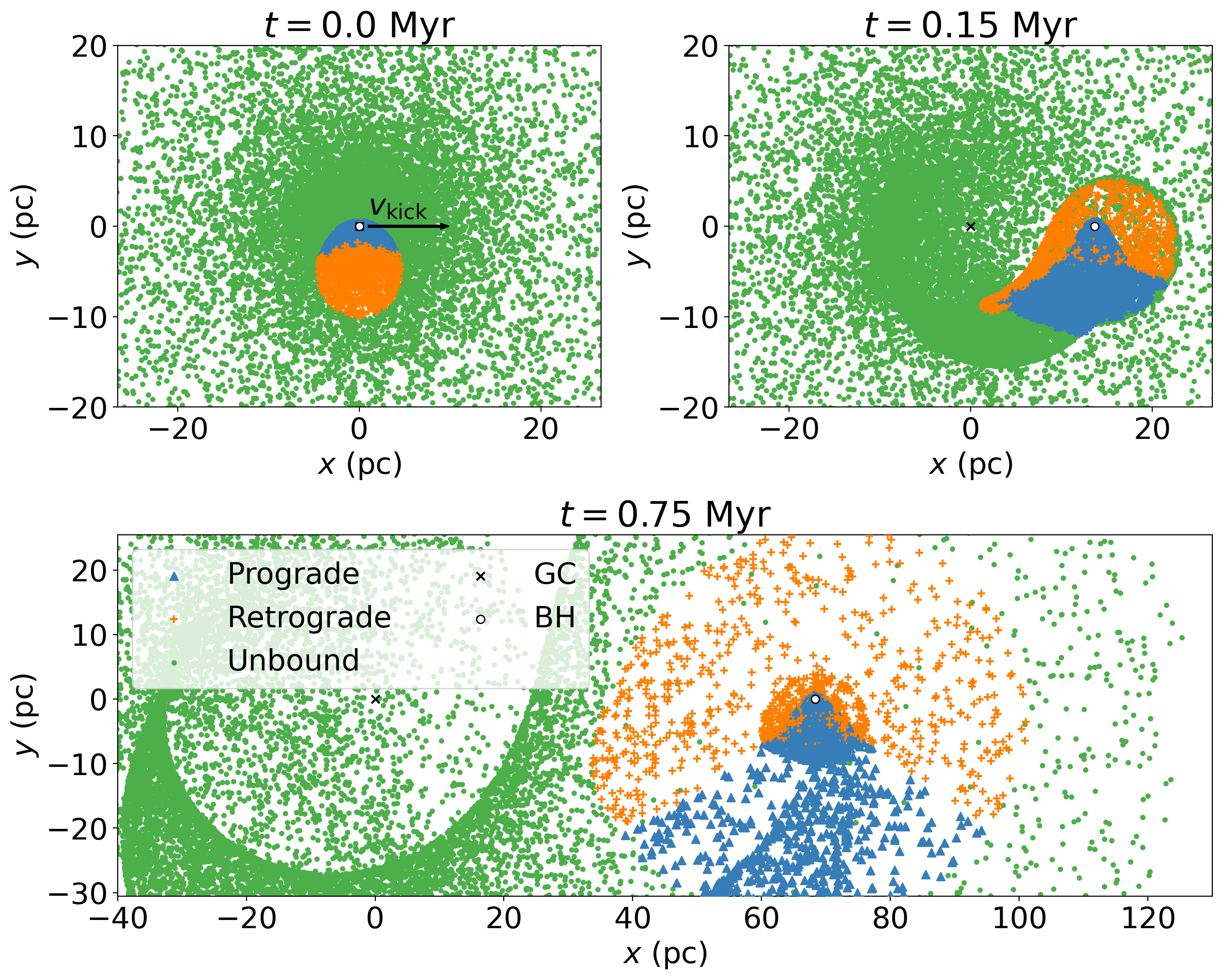}
\caption{\textbf{Circular disk with an in-plane recoil kick.} 
The face-on, post-kick distribution of stars at (\textit{top left:}) $t = 0$, (\textit{top right:}) $t = 0.15$ Myr, and (\textit{bottom:}) $t = 0.75$ Myr. Bound stars are separated into prograde (blue triangle markers) and retrograde (orange ``+'' markers). The unbound stars are marked with green dots. The ``x'' marker shows the center of the galaxy and the ``o'' marker shows the recoiling black hole. The arrow labeled $\vk$ indicates the recoil kick direction. All positions are shown in the galactic center frame. Length scale and time units assume a black hole mass of $M_\bullet = 4 \times 10^6 \ M_{\odot}$ and a kick of $\vk = 100$ km/s. The prograde, bound orbits form an eccentric nuclear disk in the $+y$-direction. The unbound stars form an arc of over-density behind the recoiling black hole.}
\label{fig:pos_stars_circ_in}
\end{figure*}

\begin{figure*}[t!]
\centering
\includegraphics[width=0.9\linewidth]{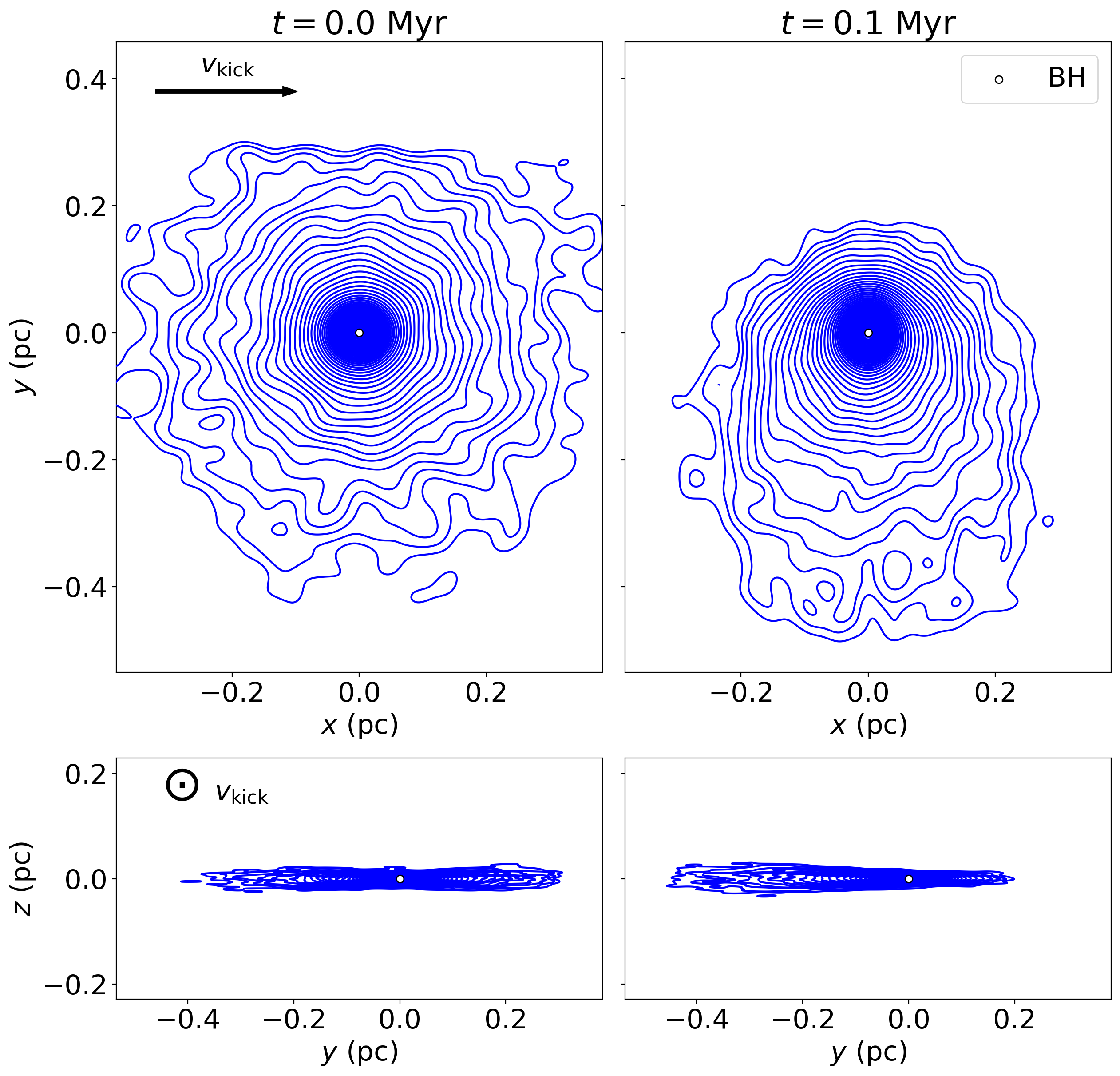}
\caption{\textbf{Circular disk with an in-plane recoil kick.} The post-kick surface density distribution of bound stars in the (\textit{top:}) face-on and (\textit{bottom:}) edge-on orientations at (\textit{left:}) $t = 0$ and (\textit{right:}) $t = 0.1$ Myr, zoomed-in to sub-parsec scales. On this scale, all stars are on prograde orbits. The recoil kick direction is marked in each orientation and labeled as $\vk$. The ``o'' marker shows the recoiling black hole, and density contours are computed in the reference frame of the recoiling black hole. Length scale and time units assume a black hole mass of $M_\bullet = 4 \times 10^6 \ M_{\odot}$ and a kick of $\vk = 100$ km/s. The post-kick density contours reveal an eccentric nuclear disk.}
\label{fig:sigma_zoom_circ_in}
\end{figure*}

\begin{figure*}[t!]
\centering
\includegraphics[width=\linewidth]{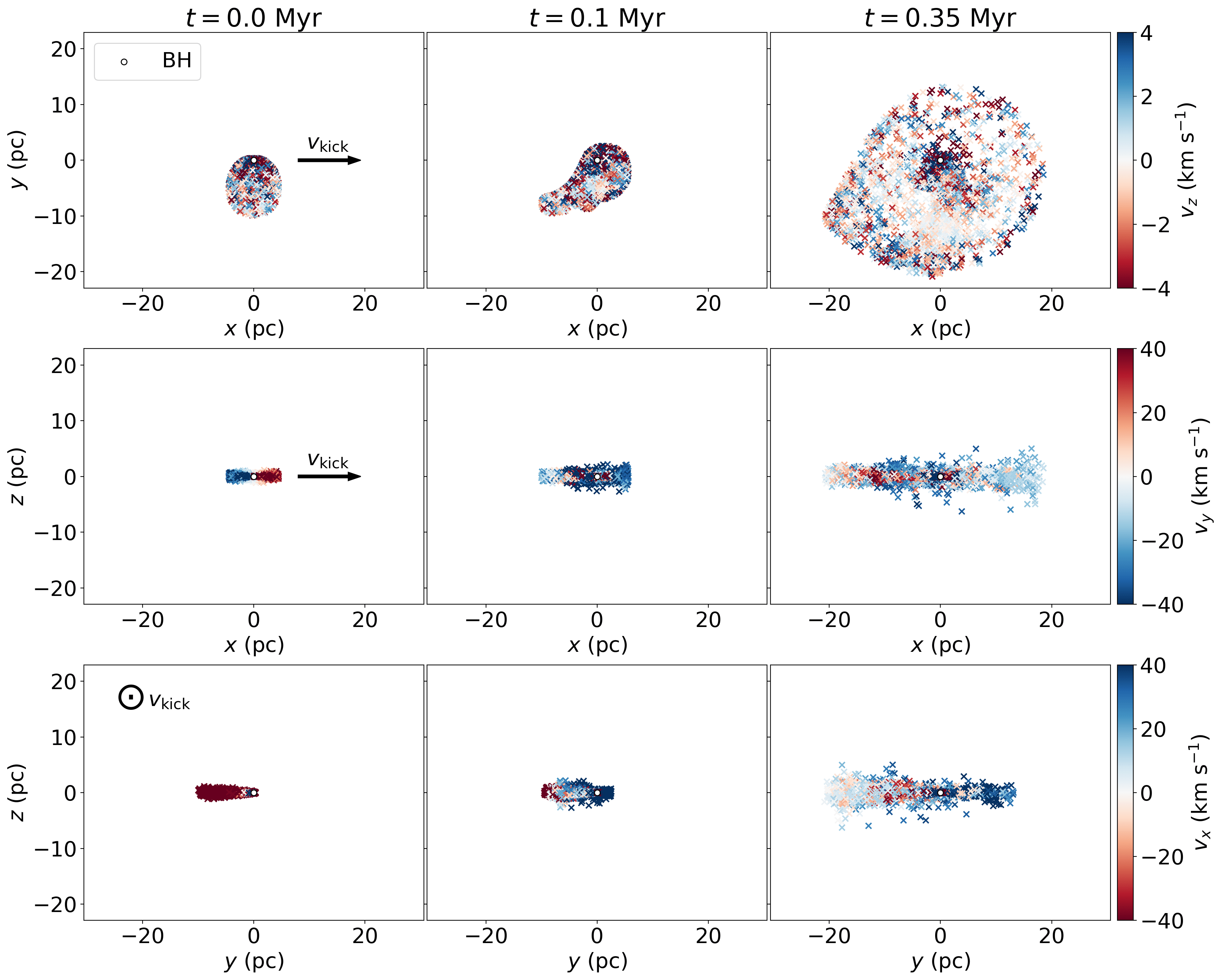}
\caption{
\textbf{Circular disk with an in-plane recoil kick.}
The line-of-sight velocity maps in the frame co-moving with the recoiling black hole (``o'' marker) in the (\textit{top:}) $x$-$y$ plane, (\textit{center:}) $x$-$z$ plane, and (\textit{bottom:}) $y$-$z$ plane at (\textit{left:}) $t=0$, (\textit{center:}) $t=0.1$ Myr, and (\textit{right:}) $t=0.35$ Myr. In each orientation, the color bar is adjusted such that approaching velocities are blue and receding velocities are red. The recoil kick direction is marked and labeled as $\vk$ for each row. Length scale and time units assume a black hole mass of $M_\bullet = 4 \times 10^6 \ M_{\odot}$ and a kick of $\vk = 100$ km/s. The edge-on orientations clearly show the presence of a retrograde disk at large radii: there exists a mix of blue-shifted and red-shifted stars on either side of the black hole.}
\label{fig:velocity_maps_circ_in}
\end{figure*}

While this radial number density profile is informative, the post-kick stellar distribution is non-axisymmetric and so we need to explore the full two-dimensional profile. We plot the post-kick distribution of stars in the face-on orientation in Figure \ref{fig:pos_stars_circ_in} at three different times following the kick. We split the population into bound and unbound stars (green dots), and the bound stars are further separated into ones on prograde (blue triangles) and retrograde (orange ``+'' markers) orbits. The kick is in the $+x$-direction as indicated by the arrow, the galactic center is marked by the ``x'' symbol and the recoiling black hole is shown with the ``o'' symbol. Positions are computed in the galactic center frame. To obtain physical units, we assume a black hole mass of $M_\bullet = 4 \times 10^6 \ M_{\odot}$ recoiling at $\vk = 100$ km/s. We see the asymmetry immediately emerge at $t=0$: stars are more likely to remain bound to the black hole if initially located toward the $-y$-direction, and one can clearly see the predominantly retrograde population at large radii. As we move forward in time, the prograde population reveals its apsidal alignment in the $+y$-direction, consistent with Figure \ref{fig:sma_ang_mom_mean_ecc_circ_in}. The unbound stars also exhibit structure: at $t=0.15$ Myr, they form a large spiral which leaves a dense arc of unbound stars behind the recoiling black hole. We note that the stellar orbits do not interact with each other and therefore do not precess. We also do not take into account an external potential for the galaxy. 

We investigate the density structure at smaller radii using a kernel density estimation. The resulting surface density contours are shown in Figure \ref{fig:sigma_zoom_circ_in} at two different times following the kick for both the face-on (top) and edge-on (bottom) orientations. The ``o'' marker shows the recoiling black hole, and all contours are computed with respect to the black hole frame of reference. Notice that we have zoomed-in to sub-parsec scales, so the structure we see here concerns the bound, prograde population of stars. At $t=0$, the structure is symmetric since the stars have not yet evolved on their new orbits. At $t=0.1$ Myr, we can clearly see that the stars have arranged themselves into a lopsided disk.

In Figure \ref{fig:velocity_maps_circ_in}, we plot the line-of-sight velocity of the stars in the face-on ($x$-$y$) orientation as well as both edge-on ($x$-$z$ and $y$-$z$) orientations at three different times following the kick. In each orientation, the color bar is adjusted such that approaching velocities are blue and receding velocities are red. All velocities are calculated in the black hole's reference frame. In the face-on orientation, no clear structure can be seen. In the $x$-$z$ plane, the initial symmetry breaks down over time with different radial ranges exhibiting different velocity signatures. While the disk is initially entirely prograde with the left side blue-shifted and the right side red-shifted, the retrograde population emerges over time, and at $t=0.35$ Myr, the right side is blue-shifted and the left side is red-shifted at large radii. Closer in, the prograde population dominates the velocity signal: the left side is blue-shifted and the right side is red-shifted. The $y$-$z$ plane line-of-sight velocities clearly show the emergence of the retrograde orbits from the beginning: the stars that remain bound at large radii are primarily located toward the $-y$-direction and are all red-shifted since they are on retrograde orbits.

\section{Introducing Out-of-plane Kicks}
\label{sec:circ_out}

\subsection{Analytics: circular disk, arbitrary kick direction}
\label{sec:anal_out}

Now we expand our toy model in \citet{Akiba2021} by considering a recoil kick with an out-of-plane component. For simplicity, we once again assume that the circular orbit is initially in the $x$-$y$ plane, and without loss of generality, we assume that the kick has some positive $x$-component as well as some $z$-component. Say the magnitude of the kick is $\vk$ and it makes an angle $\alpha$ with the $z$-axis, where $0 \leq \alpha \leq \pi$. The initial circular orbit has position and velocity given by

\[ \vec{r} = r \cos(\theta) \ \hat{x} + r \sin(\theta) \ \hat{y} \equiv x \ \hat{x} + y \ \hat{y} \ , \]
\[ \vec{v} = - v_* \sin(\theta) \ \hat{x} + v_* \cos(\theta) \ \hat{y} \equiv v_x \ \hat{x} + v_y \ \hat{y} \ . \]

\noindent Post-kick, the velocity vector in the reference frame of the black hole becomes

\begin{equation}
\vec{v}' = (v_x - \vk \sin(\alpha)) \ \hat{x} + v_y \ \hat{y} - \vk \cos(\alpha) \ \hat{z} \ .
\label{eqn:post_kick_vel}
\end{equation}

\noindent Using vector identities and subtracting the initial eccentricity vector yields the post-kick eccentricity vector components

\[ \begin{aligned}
G\Mbh e_x &= \frac{1}{2} r \sin(2 \theta) v_* \vk \sin(\alpha) \\ &+ r \cos(\theta) \vk^2 \cos^2(\alpha) \ ,
\end{aligned} \]

\[ \begin{aligned}
G\Mbh e_y &= r \sin(\theta) \vk^2 \\ &+ r (1 + \sin^2(\theta)) v_* \vk \sin(\alpha) \ ,
\end{aligned} \]

\[ \begin{aligned}
G\Mbh e_z &= -r \cos(\theta) \vk^2 \sin(\alpha) \cos(\alpha) \ .
\end{aligned} \]

\noindent Additionally, the post-kick angular momentum vector components are
    
\[ \begin{aligned}
& j_x = -r \sin(\theta) \vk \cos(\alpha) \ , \\
& j_y = r \cos(\theta) \vk \cos(\alpha) \ , \\
& j_z = r v_* + r \sin(\theta) \vk \sin(\alpha) \ .
\end{aligned} \]

\subsubsection{Limit: $\alpha = \pi/2$, an in-plane kick}

When the kick is completely in the plane of the disk, $\alpha = \pi/2$, the eccentricity and angular momentum vector components reduce to
    
\[ \begin{aligned}
& G\Mbh e_x = \frac{1}{2} r \sin(2 \theta) v_* \vk \sin(\alpha) \ , \\
& G\Mbh e_y = r \sin(\theta) \vk^2 + r (1 + \sin^2(\theta)) v_* \vk \ , \\
& G\Mbh e_z = 0 \ , \\
& j_x = 0 \ , \\
& j_y = 0 \ , \\
& j_z = r v_* + r \sin(\theta) \vk \ , 
\end{aligned} \]

\noindent so the orbit remains in the $x$-$y$ plane. If we average over $\theta: 0 \rightarrow 2 \pi$ for the eccentricity vector components, we recover Equation \ref{eqn:mean_ecc_vec}, the analytic result from \citet{Akiba2021}.

\subsubsection{Limit: $\alpha = 0$ (or $\pi$), an out-of-plane kick}

The other limiting case is if the kick is completely out-of-plane, $\alpha = 0$ (or $\pi$). In this limit,
    
\[ \begin{aligned} 
    & G\Mbh e_x = r \cos(\theta) \vk^2 \ , \\
    & G\Mbh e_y = r \sin(\theta) \vk^2 \ , \\
    & G\Mbh e_z = 0 \  \\
\implies & G\Mbh \vec{e} = \vk^2 \ \vec{r} \ \rm{, \ and}
\end{aligned} \]

\[ \begin{aligned}
    & j_x = \mp r \sin(\theta) \vk \ , \\
    & j_y = \pm r \cos(\theta) \vk \ , \\
    & j_z = r v_* \\
\implies & \vec{j} = \vec{j}_{\rm{init}} \pm \frac{r \vk}{v_*} \ \vec{v} \ ,
\end{aligned} \]

\noindent so the circular orbit develops an eccentricity in the direction of the star's initial position, and the orbital angular momentum inclines in the direction (opposite that) of the star's initial velocity vector. The result is that all of the orbits collectively roll over their major axes and the stars are at periapsis of these new elliptical orbits. In addition, we see that the post-kick eccentricity is simply given by $e = (\vk/v_*)^{2}$. This means that stars which initially orbit at radii smaller than \[ r_{\rm{out}} = \frac{G\Mbh}{\vk^2} \] will remain bound where as stars beyond this radius are all ejected. The inclination of the orbit monotonically increases with radius as the ratio $\vk/v_*$ increases, and at $r_{\rm{out}}$ the angular momentum vector is simply given by \[ \vec{j} = r v_* (\hat{z} + \hat{v}) \ , \] which means that the maximum possible inclination is 45$^\circ$.

\newpage

\subsection{Numerical Results: circular disk, out-of-plane kick}
\label{ss:circ-out}

\subsubsection{Distribution of Orbital Elements and Tilt Vector Alignment}

\begin{figure}[t!]
\centering
\includegraphics[width=\linewidth]{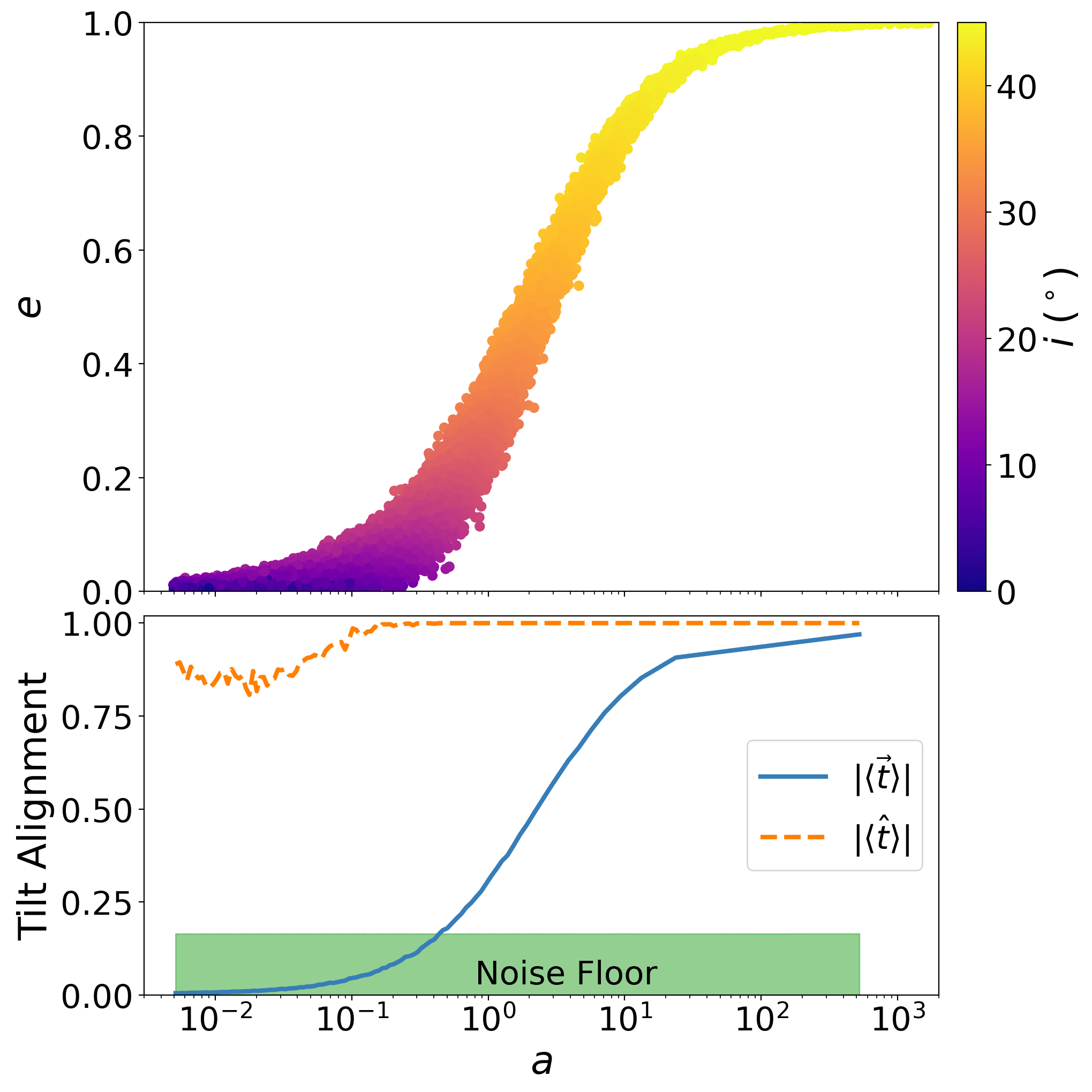}
\caption{\textbf{Circular disk with an out-of-plane recoil kick.} (\textit{Top:}) Distribution of the post-kick eccentricity as a function of semi-major axis with the inclination on the color bar. (\textit{Bottom:}) Two measures of $\omega$-clustering: the magnitude of the mean tilt vector (blue solid line), $|\langle \vec{t} \rangle|$, and the mean \textit{unit} tilt vector (orange dashed line), $|\langle \hat{t} \rangle|$, where the tilt vector is defined in Equation \ref{eqn:tilt_vec}, binned by semi-major axis. The shaded noise floor region shows the maximum expected deviation from 0 (within $3 \sigma$) if the distribution were isotropic. Both eccentricity and inclination monotonically increase with semi-major axis, and tilt vector alignment is statistically significant at most semi-major axes.}
\label{fig:ecc_sma_circ_out}
\end{figure}

\begin{figure}[t!]
\centering
\includegraphics[width=\linewidth]{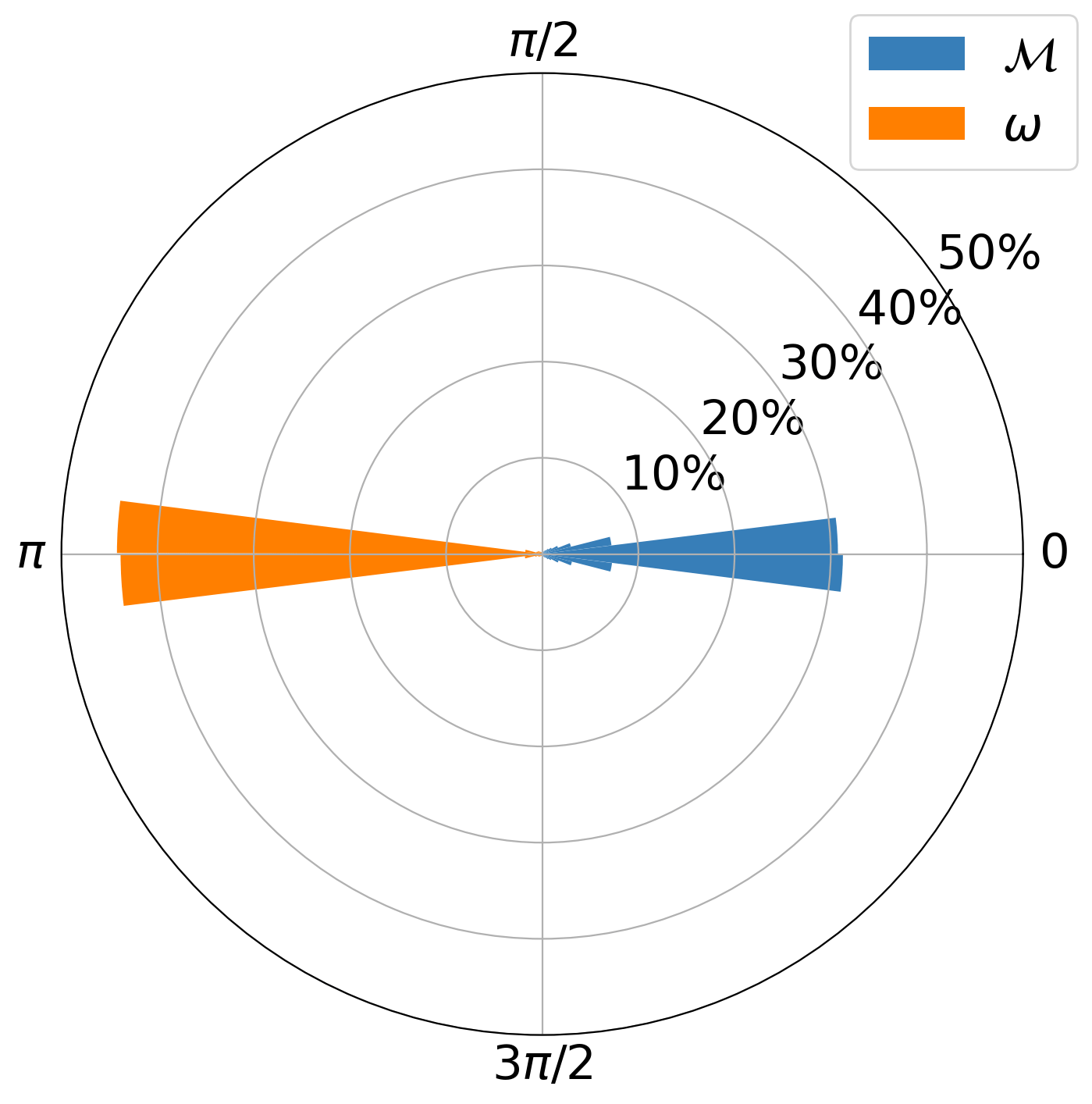}
\caption{\textbf{Circular disk with an out-of-plane recoil kick.} A histogram of the post-kick mean anomaly, $\mathcal{M}$, and argument of periapsis, $\omega$. Mean anomaly is strongly clustered around $0$ since every star begins its new orbit at periapsis, and $\omega$ is strongly clustered around $\pi$ since the new periapsis corresponds to the orbit's descending node for all orbits.}
\label{fig:omega_anom_hist_circ_out}
\end{figure}

\begin{figure*}[t!]
\centering
\includegraphics[width=\linewidth]{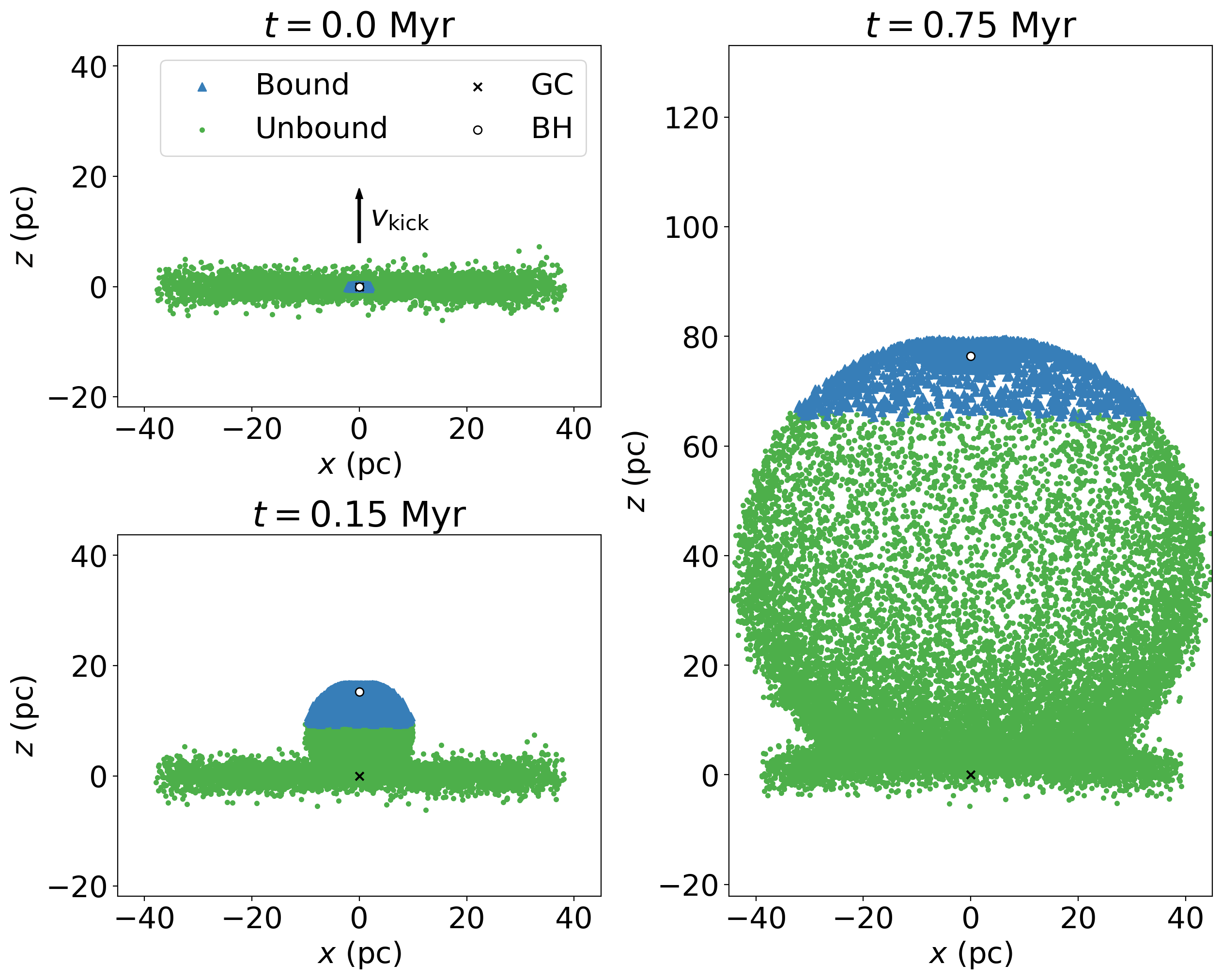}
\caption{\textbf{Circular disk with an out-of-plane recoil kick.} The edge-on, post-kick distribution of stars at (\textit{top left:}) $t = 0$, (\textit{bottom left:}) $t = 0.15$ Myr, and (\textit{right:}) $t = 0.75$ Myr. Bound and unbound stars are shown with blue triangle markers and green dots, respectively. The ``x'' marker shows the center of the galaxy and the ``o'' marker shows the recoiling black hole. The recoil is applied in the direction indicated by the arrow labeled $\vk$. All positions are computed in the galactic center frame of reference. Length scale and time units assume a black hole mass of $M_\bullet = 4 \times 10^6 \ M_{\odot}$ and a kick of $\vk = 100$ km/s. The post-kick distribution is a large ball of trailing stars with the bound stars forming a distinct dome-shape behind the recoiling black hole.}
\label{fig:pos_stars_circ_out_z}
\end{figure*}

\begin{figure*}[t!]
\centering
\includegraphics[width=\linewidth]{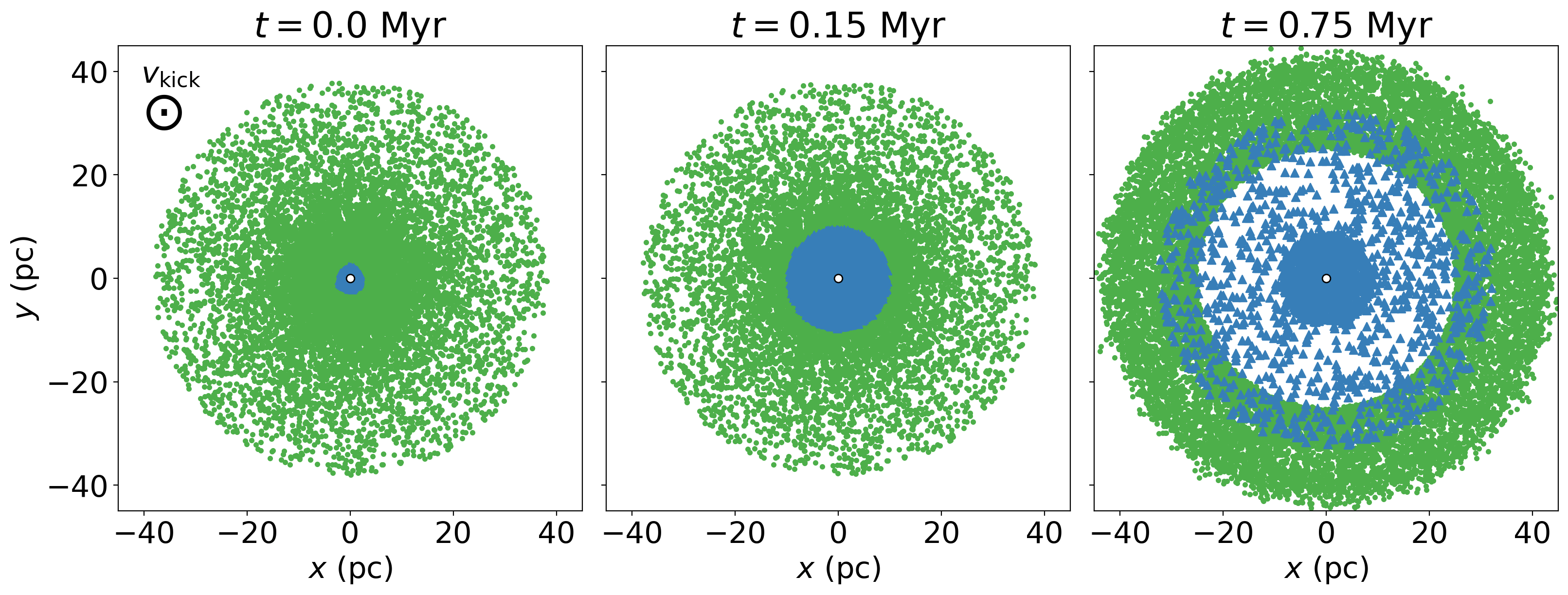}
\caption{\textbf{Circular disk with an out-of-plane recoil kick.} The face-on, post-kick distribution of stars at (\textit{left:}) $t = 0$, (\textit{center:}) $t = 0.15$ Myr, and (\textit{right:}) $t = 0.75$ Myr. Bound and unbound stars are shown with blue triangle markers and green dots, respectively. The ``x'' marker shows the center of the galaxy and the ``o'' marker shows the recoiling black hole. The recoil kick direction is out of the plane, marked and labeled as $\vk$. Length scale and time units assume a black hole mass of $M_\bullet = 4 \times 10^6 \ M_{\odot}$ and a kick of $\vk = 100$ km/s. The distribution is always axi-symmetric, but an under-density seems to emerge between the central region and the surrounding ring which both radially expand over time.}
\label{fig:pos_stars_circ_out_y}
\end{figure*}

\begin{figure*}[t!]
\centering
\includegraphics[width=\linewidth]{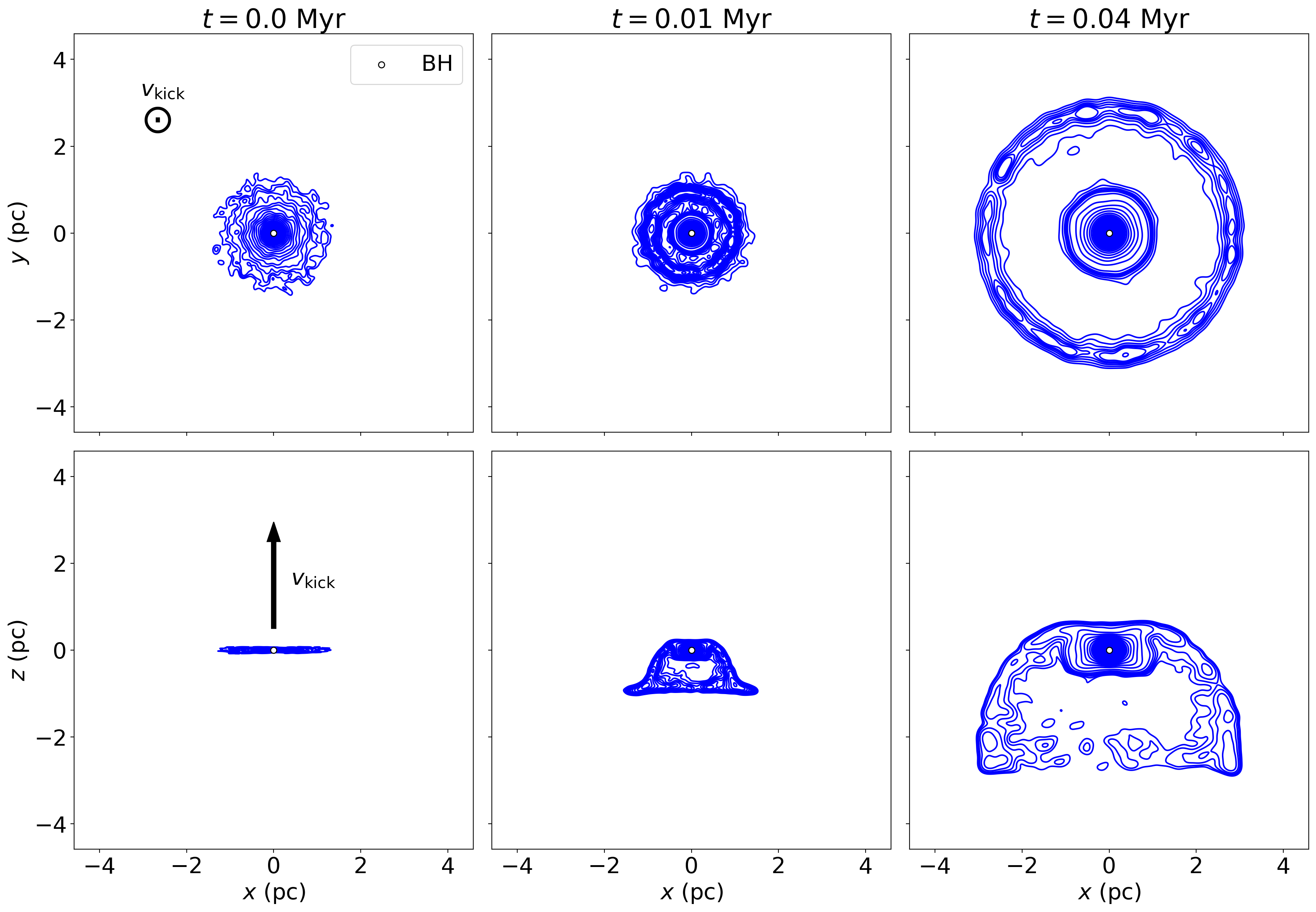}
\caption{\textbf{Circular disk with an out-of-plane recoil kick.} The post-kick surface density distribution of bound stars in the (\textit{top:}) face-on  and (\textit{bottom:}) edge-on orientations at (\textit{left:}) $t = 0$, (\textit{center:}) $t = 0.01$ Myr, and (\textit{right:}) $t = 0.04$ Myr. The ``o'' marker shows the recoiling black hole, and these density contours are calculated in the frame co-moving with the black hole. In each orientation, the recoil kick direction is marked and labeled as $\vk$. Length scale and time units assume a black hole mass of $M_\bullet = 4 \times 10^6 \ M_{\odot}$ and a kick of $\vk = 100$ km/s. The face-on orientation reveals the evolution of the central region and the surrounding ring over-density. The edge-on orientation shows the development of the characteristic dome-shape.}
\label{fig:sigma_zoom_circ_out}
\end{figure*}

\begin{figure*}[t!]
\centering
\includegraphics[width=\linewidth]{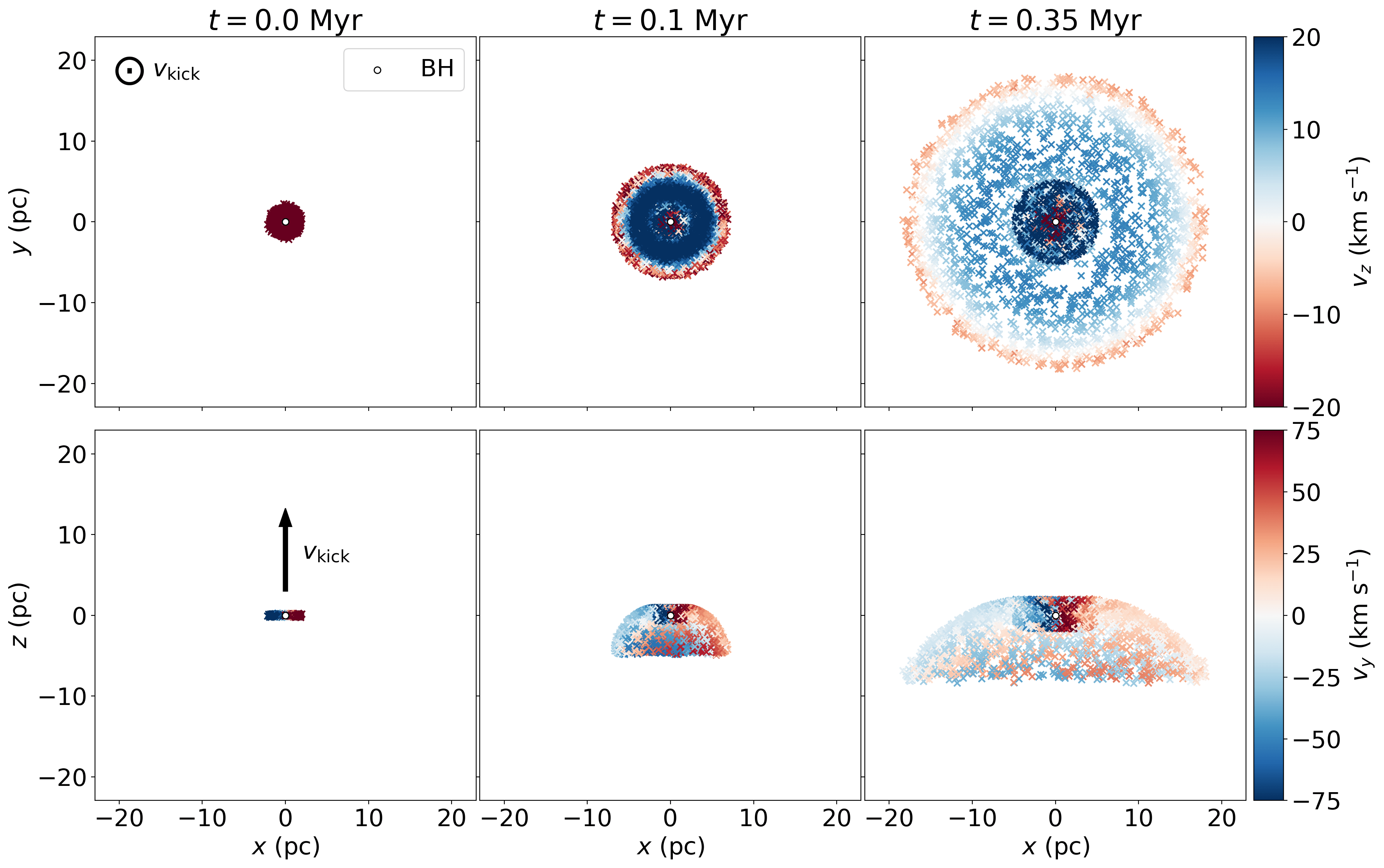}
\caption{\textbf{Circular disk with an out-of-plane recoil kick.} The line-of-sight velocity maps in the frame co-moving with the recoiling black hole (``o'' marker) in the (\textit{top:}) face-on and (\textit{bottom:}) edge-on  orientations at (\textit{left:}) $t=0$, (\textit{center:}) $t=0.1$ Myr, and (\textit{right:}) $t=0.35$ Myr. In each orientation, the color bar is adjusted such that approaching velocities are blue and receding velocities are red. The recoil kick direction is marked and labeled as $\vk$ for each row. Length scale and time units assume a black hole mass of $M_\bullet = 4 \times 10^6 \ M_{\odot}$ and a kick of $\vk = 100$ km/s. The strong correlation in anomalies and arguments of periapsis causes clear, circular patterns in line-of-sight velocity for the face-on orientation. The edge-on orientation shows that the dome-shaped structure maintains coherent rotation.}
\label{fig:velocity_maps_circ_out}
\end{figure*}

We numerically study the case of a circular disk with the central black hole receiving a purely out-of-plane kick. The set-up is identical to what is described in Section \ref{sec:sim_setup} with the only difference being that the kick is perpendicular to the disk plane. In the top panel of Figure \ref{fig:ecc_sma_circ_out}, we show the post-kick eccentricity distribution as a function of semi-major axis with the inclination color-coded. Both eccentricity and inclination monotonically increase with semi-major axis with the inclination plateauing at $\sim 45^\circ$ and eccentricity $e \sim 1$, as analytically predicted. In the bottom panel of Figure \ref{fig:ecc_sma_circ_out}, we introduce a new parameter --- a vector that can be used to measure clustering in the argument of periapsis, $\omega$, similar to how the eccentricity vector, $\vec{e}$, is a vector quantity that lets us measure clustering in the longitude of periapsis, $\varpi$. Note that $\varpi = \Omega + \omega$ where $\Omega$ is the longitude of ascending node. $\Omega$ is the angle between the reference direction ($+x$) and the ascending node, the point at which an orbit crosses the reference ($x$-$y$) plane and the star is moving upward (from $-z$ to $+z$). $\omega$, the argument of periapsis, is the angle between the ascending node and the orbit's periapsis. Thus, in a disk, $\varpi$ measures the angle between the reference direction and the periapsis. The new quantity which we call the \textit{tilt vector}, $\vec{t}$, has Cartesian components

\begin{equation}
\begin{aligned}
& t_1 = \cos(\Omega) e_x + \sin(\Omega) e_y \ , \\
& t_2 = -\sin(\Omega) e_x + \cos(\Omega) e_y \ , \\
& t_3 = e_z \ ,
\end{aligned}
\label{eqn:tilt_vec}
\end{equation}

\noindent where $e_x$, $e_y$, and $e_z$ are the $x$, $y$, and $z$-components of the eccentricity vector, respectively. This is a rotation of the eccentricity vector by an angle $-\Omega$, which allows us to calculate the position of the eccentricity vector from the ascending node of the orbit rather than from the reference direction. The ratio between $t_1$ and $t_3$ indicates the degree to which the orbit is tilting over their major axis versus minor axis. $t_1$ dominates when an orbit rolls over its major axis whereas $t_3$ dominates when an orbit pitches over its minor axis. It is important to note that the tilt vector is sensitive to our choice of reference plane and direction. Equation \ref{eqn:tilt_vec}, for instance, gives the tilt vector components when we choose the $x$-$y$ plane to be our reference plane with $+x$ being our reference direction.

The tilt vector is closely related to the inclination angles $i_a$ and $i_b$ as defined in \citet{Madigan2016} which measure the inclination components that roll and pitch, respectively (see Appendix~\ref{sec:appA}). Using the tilt vector, we can use the same measure of vector alignment we use for eccentricity vectors. In the bottom panel of Figure \ref{fig:ecc_sma_circ_out}, we show the mean tilt vector magnitude (blue solid line),

\[ | \langle \vec{t} \rangle | \equiv \left| \frac{\sum_i^N \vec{t}_i}{N} \right| \ , \] 

\noindent as well as the similarly-defined mean \textit{unit} tilt vector magnitude (orange dashed line) binned by semi-major axis. We notice that there is significant $\vec{t}$ and $\hat{t}$-alignment at most semi-major axis ranges, since all of the orbits tilt coherently over their major axes. Every star begins its new orbit at the descending node, which also corresponds to its new periapsis. This creates a highly non-uniform post-kick distribution of the mean anomaly, $\mathcal{M}$, in addition to the argument of periapsis, $\omega$, as shown in Figure \ref{fig:omega_anom_hist_circ_out}. Mean anomaly is clustered near 0 since stars are initially located at their periapsis, and $\omega$ is clustered near $\pi$ because the ascending node of the new orbit corresponds to the apoapsis. 

\subsubsection{{Post-Kick Density and Velocity Profile}}

The post-kick stellar distribution exhibits unique density and velocity profiles for the purely out-of-plane kick case as well. We show the edge-on distribution of the stars at three different times in Figure \ref{fig:pos_stars_circ_out_z}. The bound stars are denoted with blue triangles and the unbound stars are marked by green dots, with the black hole recoiling in the $+z$-direction as indicated by the arrow. The galactic center is shown with the ``x'' marker and the recoiling black hole is shown with the ``o'' marker. All positions are calculated in the galactic center frame. Length and time units assume a black hole mass of $M_\bullet = 4 \times 10^6 \ M_{\odot}$ and a kick of $\vk = 100$ km/s. At $t=0$, the distribution is an axi-symmetric disk where only stars within $r_{\rm{out}} \sim 2$ pc remain bound to the recoiling black hole. At subsequent times, we see a large ball of trailing stars: the bound stars form a ``dome'' shape behind the black hole and the unbound stars make up the rest of the spherical distribution following the black hole. For the same three times following the kick, we show the face-on distribution of stars in Figure \ref{fig:pos_stars_circ_out_y}. Since the kick is purely out-of-plane, the distribution remains axi-symmetric about the $z$-axis throughout, but the bound population expands significantly in radius, and at $t=0.75$ Myr, there appears an interesting density structure where there is a central over-density as well as a dense ring of stars surrounding it, with an under-density in between.

We further explore the density structure closer-in, at $\sim$parsec scales, using a kernel density estimation. We present both face-on (top) and edge-on orientations (bottom) of these zoomed-in surface density maps in Figure \ref{fig:sigma_zoom_circ_out} at three different times following the kick. The ``o'' marker shows the recoiling black hole, and the density contours are constructed in the frame co-moving with the black hole. In the face-on orientation, we see that the central over-density and the surrounding ring forms immediately following the kick with both components expanding in radius over time. In the edge-on orientation, the appearance is a dome-like shape: the stars that are most tightly bound to the black hole forms a puffy disk at small radii, and the stars at slightly larger radii form an arc that trails the recoiling black hole. The puffy disk and arc in the edge-on orientation correspond to the central over-density and the surrounding ring in the face-on orientation, respectively.

The alignment of the tilt vector (bottom panel of Figure \ref{fig:ecc_sma_circ_out}) accompanied by the strong mean anomaly correlation (Figure \ref{fig:omega_anom_hist_circ_out}) produces coherent structures in the line-of-sight velocity of stars. We show velocity maps in the reference frame co-moving with the black hole in both the face-on and edge-on orientations over time in Figure \ref{fig:velocity_maps_circ_out}. In each orientation, the color bar is adjusted such that approaching velocities are blue and receding velocities are red. In the face-on orientation, the post-kick stellar distribution exhibits distinct circular patterns in line-of-sight velocity due to the fact that the stars all gain eccentricities and tilt over their major axes. With the recoil kick pointed toward us, the outer edge of the ring is preferentially red-shifted and the inner regions are preferentially blue-shifted. In the edge-on orientation, we see that the dome-shaped, puffy disk still maintains the same angular momentum axis since stars all remain prograde post-kick: the right side is preferentially red-shifted and the left side is preferentially blue-shifted. We note that this is very similar to Figure 2 in \citet{Lippai2008} who show an aerial view of the density profile of a tightly-bound gaseous disk a week after responding to an out-of-plane recoil kick.

\section{Adding Initial Eccentricities}
\label{sec:ecc}

\subsection{Analytics: disk with initial eccentricities, arbitrary kick direction} 

We now expand our analyses to incorporate initial eccentricities. The initial position vector of a body on an eccentric orbit is given by

\[ \vec{r} = \frac{a(1-e^2)}{1+e \cos(\theta - \varpi)} \ (\cos \theta \ \hat{x} + \sin \theta \ \hat{y}) \ , \]

\noindent where $a$ is the semi-major axis, $e$ is the eccentricity, $\theta$ is the true anomaly, and $\varpi$ is the longitude of periapsis. The initial velocity vector is given by

\[ \begin{aligned}
\vec{v} = \frac{1}{\sqrt{a(1-e^2)}} & [-(\sin\theta + e\sin\varpi) \ \hat{x} \\ & + (\cos\theta + e\cos\varpi) \ \hat{y}] \ . \end{aligned} \]

\noindent We move into the recoiling black hole's reference frame by shifting the velocity vector as we have done in Equation \ref{eqn:post_kick_vel} before, and calculate the eccentricity vector. We average over mean anomaly, $\mathcal{M}$, and $\varpi$, so we obtain the analytic expectation starting with an axi-symmetric disk of stellar orbits with a certain given initial eccentricity. 

\subsubsection{Limit: $e = 0$, a circular orbit}

When we set $e = 0$, the initial position and velocity vectors above reduce to

\[ \vec{r} = a \cos \theta \ \hat{x} + a \sin \theta \ \hat{y} \ , \]

\[ \vec{v} = - \frac{1}{\sqrt{a}} \sin \theta \ \hat{x} + \frac{1}{\sqrt{a}} \cos \theta \ \hat{y} \ , \]

\noindent and since $a = r$ and $v_* = \sqrt{\frac{GM_\bullet}{r}} = \frac{1}{\sqrt{r}}$ for a circular orbit, we recover the analytics from Section \ref{sec:anal_out}.

\subsubsection{Limit: $e \rightarrow 1$, a high eccentricity orbit}
\label{sec:lim_high_ecc}

Stars on eccentric orbits spend a majority of their time near apoapsis. In the limit of $e \rightarrow 1$, we can capture the disk's behavior by replacing $v_*$ in Equation \ref{eqn:mean_ecc_vec} with $v_{\rm{apo}}$, the star's speed at apoapsis. This is given by $v_{\rm{apo}} = \sqrt{\frac{(1-e)G\Mbh}{(1+e)a}}$. Substituting this and setting $|\langle \vec{e} \rangle| \sim 1$ yields a characteristic semi-major axis,

\begin{equation}
    a_c = \frac{1-e}{1+e} \ r_c \ .
\label{eqn:characteristic_sma}
\end{equation}

\noindent This effectively reduces the semi-major axis at which the strongest apsidal alignment occurs. For an initial eccentricity $e = 0.99$, for instance, $v_{\rm{kick}} = 100$ km/s gives $a_c = 4 \times 10^{-3}$ pc as opposed to $r_c = 0.8$ pc for a circular disk, a factor of $\sim$200 difference.

\paragraph{In-plane kick}

Re-investigating kinematically distinct radial regions as in Section \ref{sec:circ_in_rad} but for an initially high eccentricity orbit, we see a significant change due to the fact that the initial apoapsis speed, $v_{\rm{apo}} = \sqrt{\frac{(1-e)G\Mbh}{(1+e)a}} \rightarrow 0$ is much slower than the circular velocity at that radius, $v_{\rm{circ}} = \sqrt{\frac{G\Mbh}{(1+e)a}} \rightarrow \sqrt{\frac{G\Mbh}{2a}}$. We find that:

\begin{itemize}
    \item A star with its apoapsis located along the $y$-axis will remain bound if $\vk < \sqrt{2} v_{\rm{circ}} \pm v_{\rm{apo}} \rightarrow \sqrt{2} v_{\rm{circ}}$ which translates to $a < \frac{G\Mbh}{\vk^2}$.
    \item A star with its apoapsis initially located toward the $-y$-direction will only remain prograde for kicks $\vk < v_{\rm{apo}} \rightarrow 0$, so roughly half of the stars will flip to retrograde at most radii.
    \item A star along the $y$-axis will have the same apoapsis pre- and post-kick if $\vk < v_{\rm{circ}} \pm v_{\rm{apo}} \rightarrow v_{\rm{circ}}$ or $a < \frac{G\Mbh}{2\vk^2}$.
    \item The same star will move onto a larger semi-major axis orbit where the pre-kick apoapsis becomes the post-kick periapsis if $v_{\rm{circ}} < \vk < \sqrt{2} v_{\rm{circ}}$ or at $\frac{G\Mbh}{2\vk^2} < a < \frac{G\Mbh}{\vk^2}$.
\end{itemize}

\noindent In other words, for initially high eccentricity orbits, the initial speed of the star is negligible, and their post-kick evolution is dictated by the recoil kick at most radii. Stars initially located toward the $-y$-direction will flip to retrograde orientation with eccentricity vectors pointing in the $+y$-direction at small semi-major axes and $-y$-direction at larger semi-major axes. Stars initially in the $+y$-direction will have eccentricity vectors pointing in the $-y$-direction at small semi-major axes and $+y$-direction at larger semi-major axes. Hence, for a given radial range, we expect a prograde eccentric nuclear disk and an anti-aligned retrograde eccentric disk creating a linear, bar-like structure. 

\paragraph{Out-of-plane kick}

It is fairly straightforward to apply the same analytic limit for purely out-of-plane kicks. As with the circular disk case, the recoil kick will tilt all of the orbits coherently over their major axes, but since $v_{\rm{circ}} \gg v_{\rm{apo}}$ for high eccentricity orbits, most stars will begin their new orbit at the apoapsis which coincides with the descending node. Thus, while we get $\omega$-clustering near $\pi$ for an initially circular disk, we expect clustering near 0 when starting with a high eccentricity disk. In addition, at larger semi-major axes, $\vk \gg v_{\rm{apo}}$, so the recoil kick should cause significant inclinations of $\sim 90^\circ$. 

\subsection{Numerical Results: parameter study of initial eccentricities}

\begin{figure}[t!]
\centering
\includegraphics[width=\linewidth]{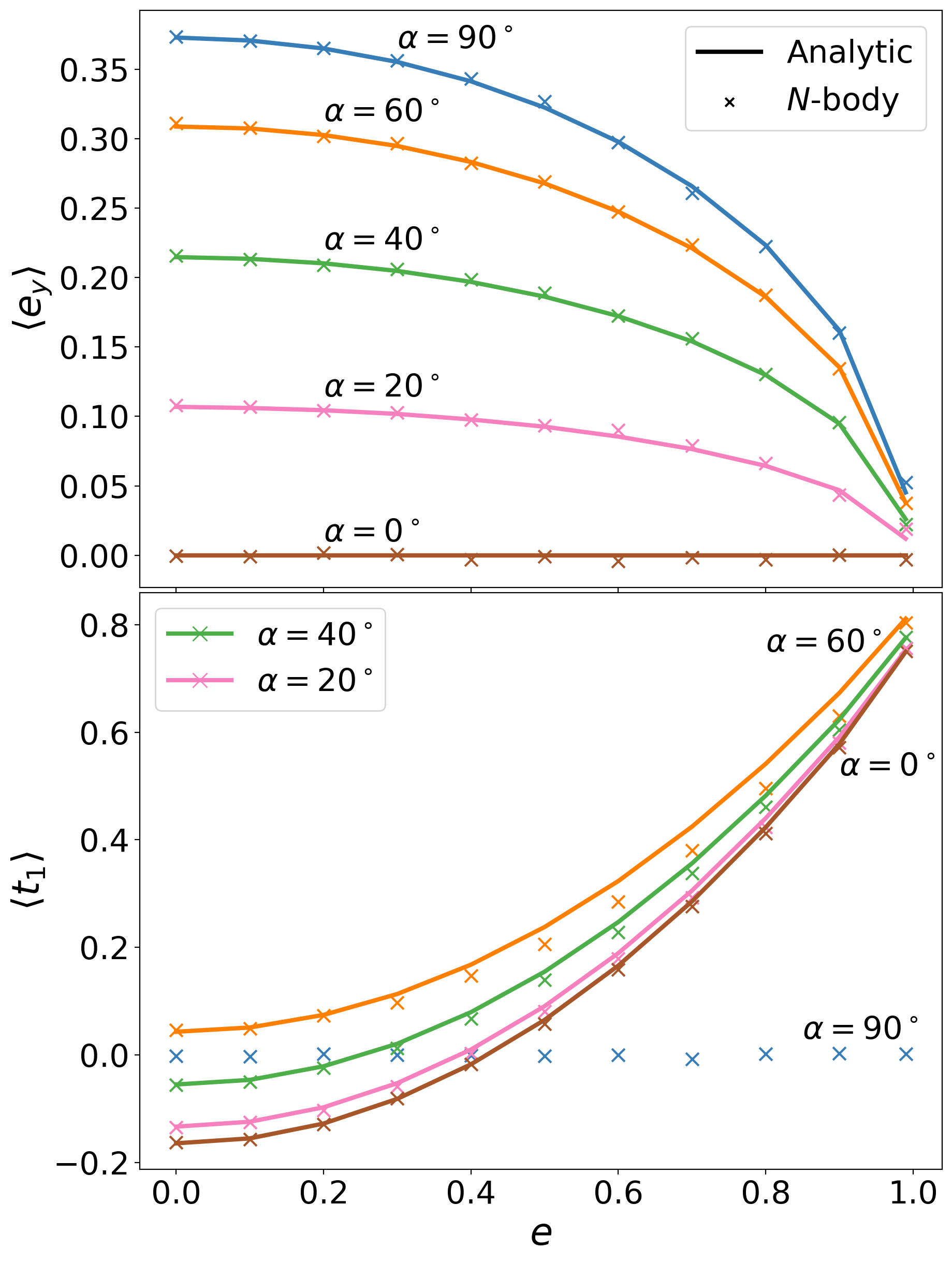}
\caption{\textbf{Parameter study of an initially thin disk with varying eccentricities and kick directions.} (\textit{Top:}) The mean $y$-component of the eccentricity vector, $\langle e_y \rangle$, and (\textit{bottom:}) the mean first component of the tilt vector, $\langle t_1 \rangle$, which measures roll, as a function of the initial eccentricity with different lines showing different out-of-plane kick components. We show $\langle e_y \rangle$ and $\langle t_1 \rangle$ because they dominate the mean eccentricity vector and the mean tilt vector, respectively. The other components, $\langle e_x \rangle$, $\langle e_z \rangle$, $\langle t_2 \rangle$, and $\langle t_3 \rangle$, are all $\sim$0. $\alpha = 0^\circ$ is a completely out-of-plane recoil kick whereas $\alpha = 90^\circ$ is a completely in-plane kick. The solid lines are analytic expectations and the ``x'' markers show the numerical results. We omit the analytic expectation for the purely in-plane tilt vector, since the longitude of ascending node is not well-defined for a two-dimensional orbit in the $x$-$y$ plane. Eccentricity vector alignment becomes weaker with increasing initial eccentricities and larger out-of-plane kick components. Tilt vector alignment happens in all cases except when the kick is purely in-plane.}
\label{fig:ecc_alpha_inc3}
\end{figure}

We study the instantaneous alignments of the eccentricity vector and the tilt vector as we change initial eccentricities and the out-of-plane kick component with a suite of numerical simulations. Once again, the set-up is the same as what is described in Section \ref{sec:sim_setup} except we incrementally change the initial eccentricity of orbits and the kick direction. For simulations starting with high eccentricities, we adjust the inner edge of the disk to be $a_c/2$ where $a_c$ is given by Equation \ref{eqn:characteristic_sma} to capture the behavior near the characteristic semi-major axis. For instance, for $e=0.99$, $a_c \approx 1/200 \ r_c$, so the inner edge of the disk is set to $a_c/2 \approx 0.0025$. In the top panel of Figure \ref{fig:ecc_alpha_inc3}, we plot the numerical results and corresponding analytic calculations of the mean $y$-component of the eccentricity vector, $\langle e_y \rangle$, as a function of the initial eccentricity of orbits for various kick angles. We plot $\langle e_y \rangle$ because it is the component that dominates the mean eccentricity vector: $\langle e_x \rangle$ and $\langle e_z \rangle$ are both $\sim$0 for all simulations. These results are averaged over the semi-major axis range, $a=0.005$--$50$, in our simulations with the given surface density profile. Numerical results are shown with the ``x'' markers whereas the analytic results are shown in solid lines. $\alpha = 90^\circ$ is an in-plane kick and $\alpha = 0^\circ$ is an out-of-plane kick. The circular, in-plane ($\alpha = 90^\circ$) case induces the most apsidal alignment, and the alignment becomes weaker as the out-of-plane kick component and initial eccentricities increase. Nonetheless, the preferential direction for apsidal alignment remains the same: if the in-plane recoil kick component is in the $+x$-direction, the eccentricity vectors will preferentially become aligned in the $+y$-direction, orthogonal to the in-plane kick. A statistically significant ($> 3 \sigma$) alignment is induced in most cases except when the kick is purely out-of-plane ($\alpha = 0^\circ$).

In the bottom panel of Figure \ref{fig:ecc_alpha_inc3}, we show the mean first component of the tilt vector, $\langle t_1 \rangle$, as a function of the initial eccentricity for various kick directions. $\langle t_1 \rangle$ is the component that dominates the mean tilt vector which means that all of the stellar orbits are coherently tilting over their major axes. $\langle t_2 \rangle$ and $\langle t_3 \rangle$ are $\sim$0 for all simulation runs. The solid lines show our analytic expectations and the ``x'' markers show our numerical results. We omit the analytic estimate for the purely in-plane kick case, since the longitude of ascending node is not well-defined for an orbit in the reference plane. As we have shown before, the tilt vectors are aligned for a circular, out-of-plane case, but $\langle t_1 \rangle < 0$ due to the fact that stars begin at periapsis. Pointing in the direction of the eccentricity vector, each orbit rolls to the left, so $t_1$ is preferentially negative. On the other hand, there is also a very strong $\vec{t}$-alignment induced in the limit of initially high eccentricity orbits provided that there is at least some out-of-plane kick component. This is due to the $\omega$-clustering at 0 predicted in Section \ref{sec:lim_high_ecc}, where orbits roll over their major axes but the stars begin their new orbits preferentially at apoapsis. Pointing in the direction of the eccentricity vector, each orbit rolls to the right, so $t_1$ is predominantly positive. It is important to note that while we only capture the average tilt alignment here, nearly every star begins at periapsis or apoapsis and rolls over its major axis following the kick. In general, we see two clusters at $\omega = 0$ and $\pi$ in competition, so $\langle t_x \rangle \sim 0$ simply means that these two clusters are similar in size. It does not mean that there is no tilt alignment. $\omega$ tends to cluster at $\pi$ for circular orbits and at 0 for eccentric orbits, and the turning point when $\langle t_x \rangle$ crosses 0 depends largely on $\alpha$. Statistically significant ($> 3 \sigma$) alignment emerges in most cases except when the kick is completely in-plane ($\alpha = 90^\circ$).

\subsection{Numerical Results: disk with initial $e=0.99$, in-plane kick}
\label{ss:highecc-inplane}

\subsubsection{Radial Regions and Apsidal Alignment}

\begin{figure*}[t!]
\centering
\includegraphics[width=0.85\linewidth]{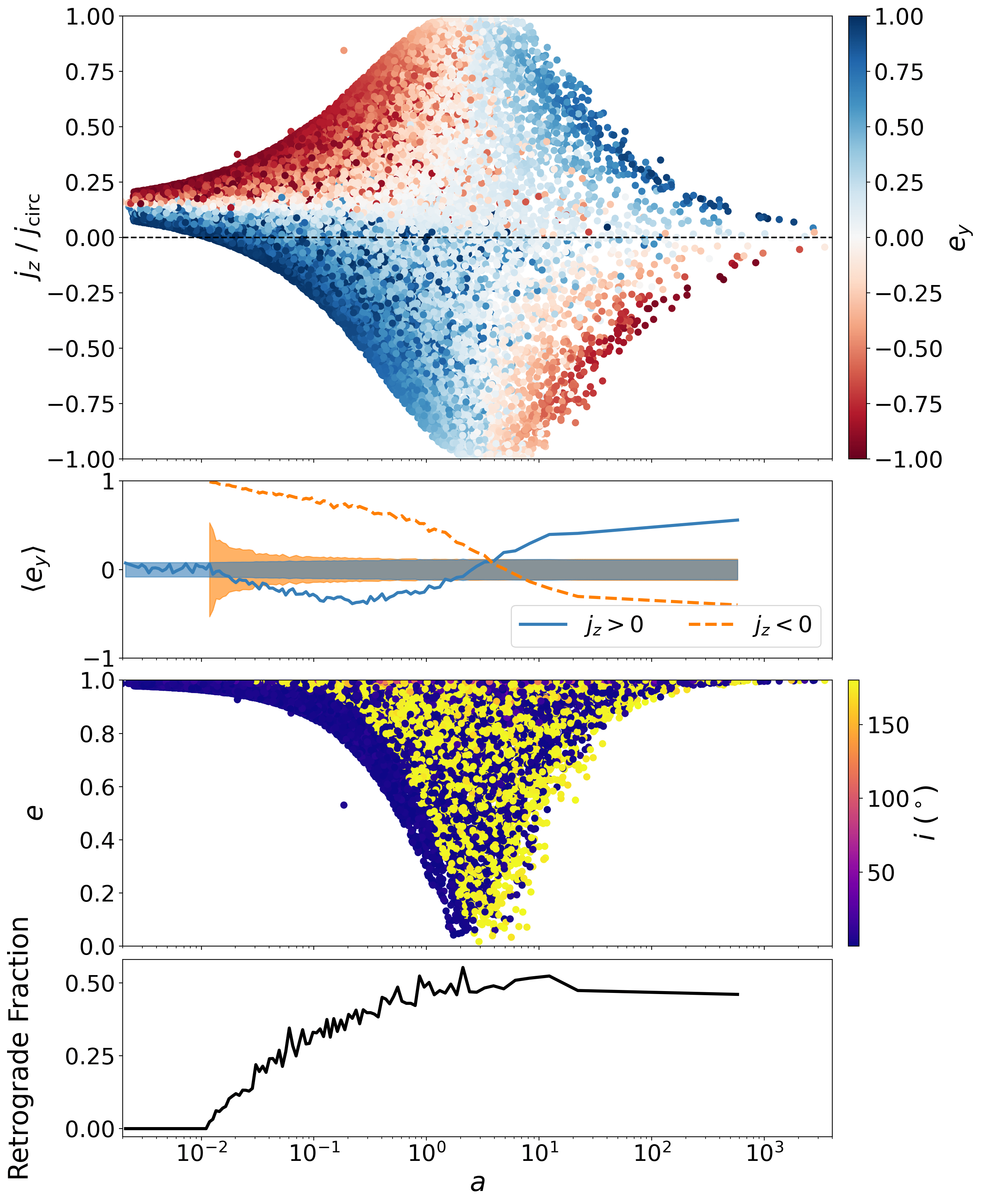}
\caption{\textbf{Disk with high initial eccentricities ($e=0.99$) with an in-plane recoil kick.} (\textit{Top:}) The post-kick distribution of the $z$-component of angular momentum (normalized by the circular angular momentum at a given radius) as a function of semi-major axis, with the color bar showing the $y$-component of the eccentricity vector. (\textit{Second from top:}) The $y$-component of the mean eccentricity vector, $\langle e_y \rangle$, binned by semi-major axis and separated into prograde ($j_z > 0$) and retrograde ($j_z < 0$) stars. The mean for prograde orbits is shown in the blue solid line whereas the mean for retrograde orbits is shown in the orange dashed line. The shaded noise floor shows the maximum expected deviation from 0 (at a $3 \sigma$ level) if the distribution were isotropic. (\textit{Second from bottom:}) The eccentricity profile as a function of semi-major axis with the inclination color-coded. (\textit{Bottom:}) The retrograde fraction of the disk at a given semi-major axis range. Two anti-aligned eccentric nuclear disks create a linear, bar-like structure with a $\gtrsim 40 \%$ retrograde fraction at most semi-major axis ranges. The eccentricity gradient, $de/da$, is negative inward of $a \sim 3$ and positive outward of $a \sim 5$.}
\label{fig:sma_ang_mom_mean_ecc_e=0.99}
\end{figure*}

We numerically study the case of a thin disk of high eccentricity orbits receiving an in-plane kick. This set-up is not motivated by an astrophysically realistic scenario. Rather we are building orbital complexity in a sequential manner.
In the top panel of Figure \ref{fig:sma_ang_mom_mean_ecc_e=0.99}, we show the post-kick distribution of the $z$-component of angular momentum (normalized by the circular angular momentum at a given radius), $j_z/j_{\rm{circ}}$, as a function of the post-kick semi-major axis for a simulation starting with $e=0.99$ orbits with the black hole recoiling in-plane. The color bar shows the $y$-component of the eccentricity vector. As predicted analytically, the prograde and retrograde sub-populations are nearly mirror images of each other except at very small semi-major axes ($a \lesssim 10^{-2}$) when $\vk \lesssim v_{\rm{apo}}$. At these very small semi-major axes, there exists two anti-aligned lopsided modes but both are prograde. When $10^{-2} \lesssim a \lesssim 4$, there is a prograde eccentric nuclear disk in the $-y$-direction and a retrograde eccentric disk in the $+y$-direction. At $a \gtrsim 4$, there is a prograde eccentric disk in the $+y$-direction and a retrograde eccentric disk in the $-y$-direction.

\begin{figure*}[t!]
\centering
\includegraphics[width=\linewidth]{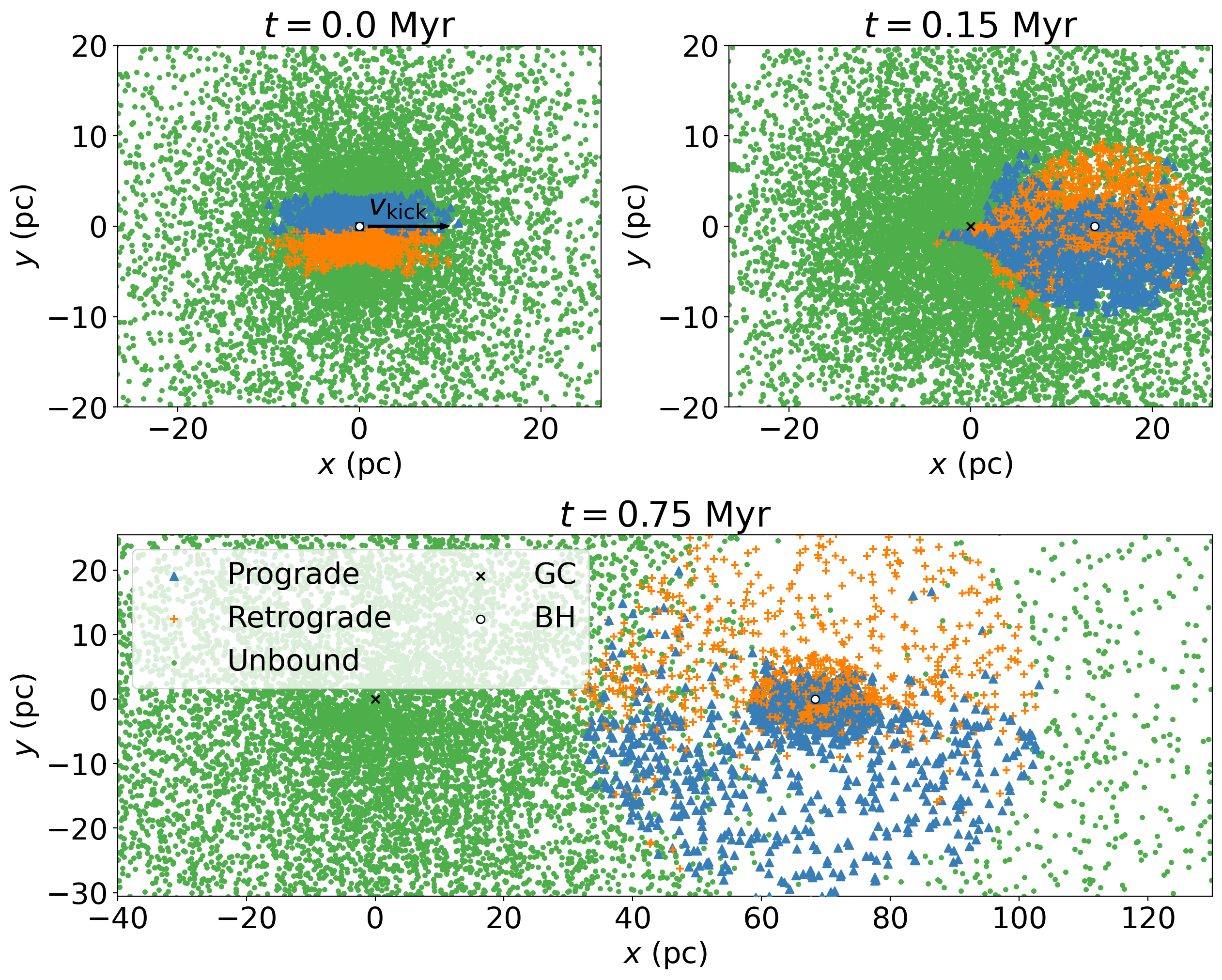}
\caption{\textbf{Disk with high initial eccentricities ($e=0.99$) with an in-plane recoil kick.} The face-on, post-kick distribution of stars at (\textit{top left:}) $t = 0$, (\textit{top right:}) $t = 0.15$ Myr, and (\textit{bottom:}) $t = 0.75$ Myr. Bound stars are separated into prograde (blue triangle markers) and retrograde (orange ``+'' markers). The unbound stars are marked with green dots. The ``x'' marker shows the center of the galaxy and the ``o'' marker shows the recoiling black hole. The arrow labeled $\vk$ indicates the recoil direction. Positions are computed in the galactic center frame of reference. Length scale and time units assume a black hole mass of $M_\bullet = 4 \times 10^6 \ M_{\odot}$ and a kick of $\vk = 100$ km/s. The prograde and retrograde populations are nearly mirror images of each other about the $x$-axis. Prograde stars are aligned in one direction and retrograde stars are anti-aligned with the prograde orbits.}
\label{fig:pos_stars_e=0.99_in}
\end{figure*}

\begin{figure*}[t!]
\centering
\includegraphics[width=\linewidth]{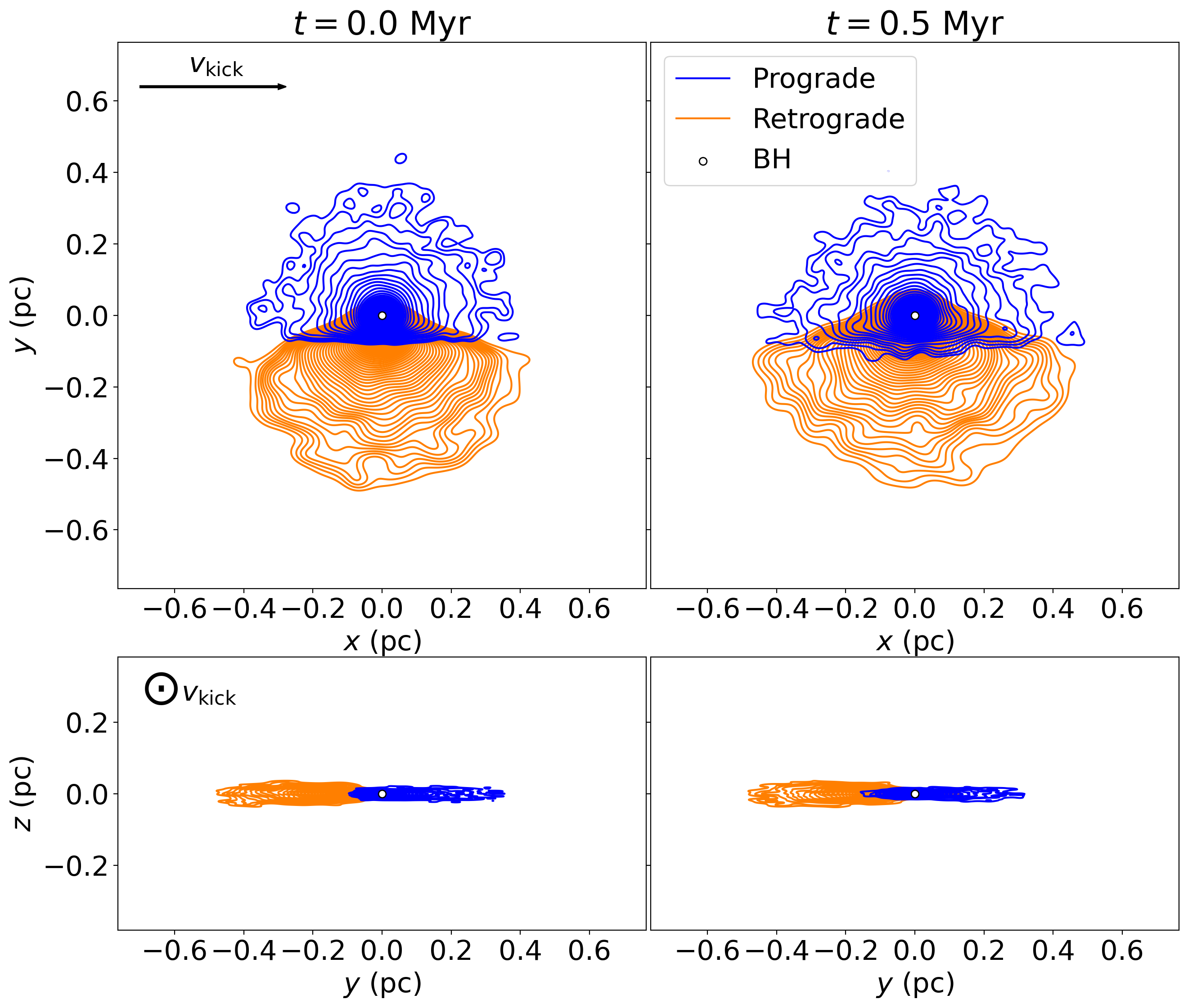}
\caption{\textbf{Disk with high initial eccentricities ($e=0.99$) with an in-plane recoil kick.} The post-kick surface density distribution of bound stars in the (\textit{top:}) face-on and (\textit{bottom:}) edge-on orientation at (\textit{left:}) $t = 0$ and (\textit{right:}) $t = 0.5$ Myr, where the stars have been separated into prograde (blue contours) and retrograde (orange contours). The ``o'' marker shows the recoiling black hole, and density contours are calculated in the frame co-moving with the black hole. In each orientation, the recoil kick direction is marked and labeled as $\vk$. Length scale and time units assume a black hole mass of $M_\bullet = 4 \times 10^6 \ M_{\odot}$ and a kick of $\vk = 100$ km/s. The density contours reveal a prograde eccentric disk and an anti-aligned retrograde disk forming a bar-like structure.}
\label{fig:sigma_zoom_e=0.99_in}
\end{figure*}

\begin{figure*}[t!]
\centering
\includegraphics[width=\linewidth]{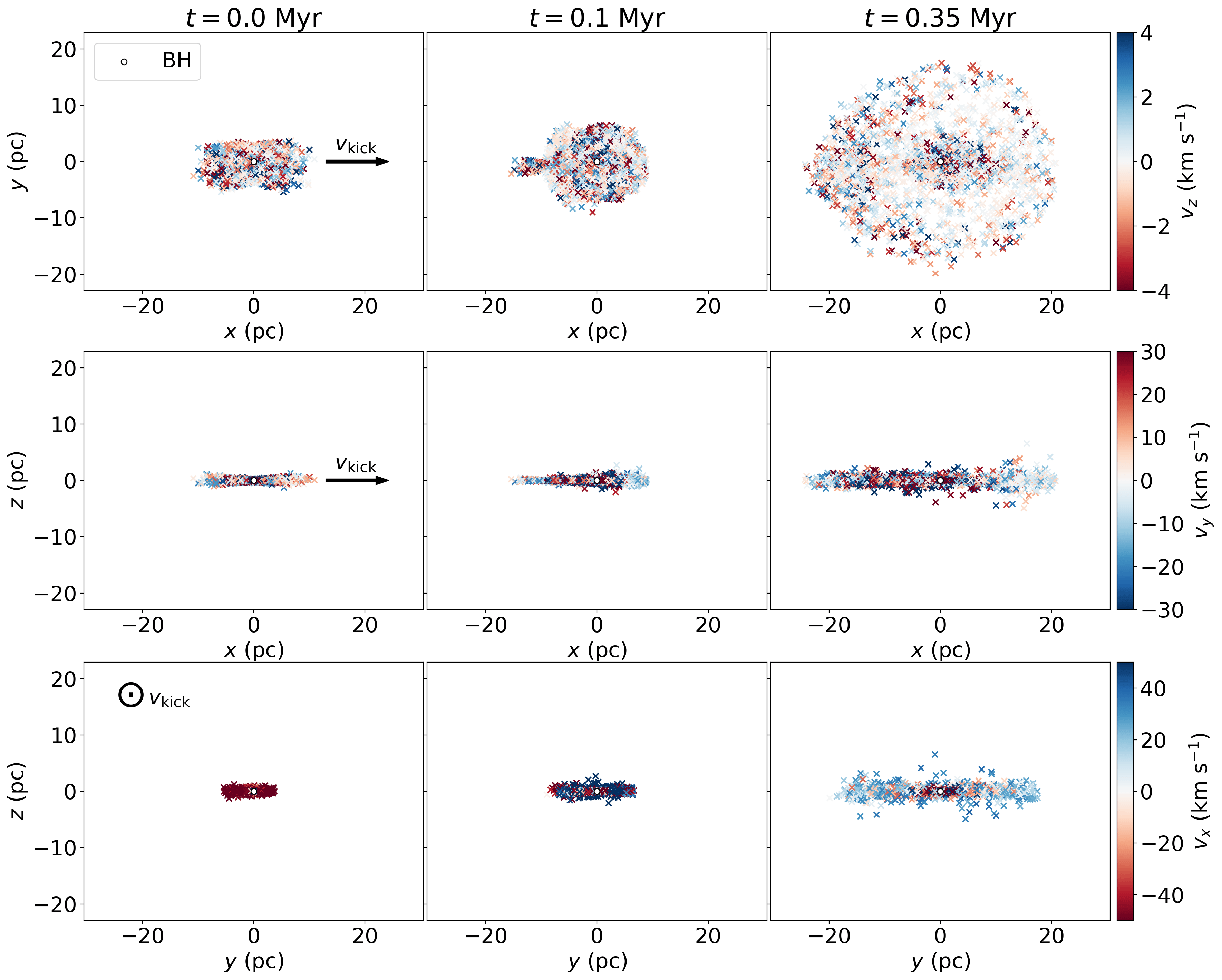}
\caption{\textbf{Disk with high initial eccentricities ($e=0.99$) with an in-plane recoil kick.} The line-of-sight velocity maps in the frame co-moving with the recoiling black hole (``o'' marker) in the (\textit{top:}) $x$-$y$, (\textit{center:}) $x$-$z$, and (\textit{bottom:}) $y$-$z$ plane at (\textit{left:}) $t=0$, (\textit{center:}) $t=0.1$ Myr, and (\textit{right:}) $t=0.35$ Myr. In each orientation, the color bar is adjusted such that approaching velocities are blue and receding velocities are red. The recoil kick direction is marked and labeled as $\vk$ for each row. Length scale and time units assume a black hole mass of $M_\bullet = 4 \times 10^6 \ M_{\odot}$ and a kick of $\vk = 100$ km/s. In the $x$-$z$ plane, there appears a cluster of stars with high line-of-sight velocity immediately behind the recoiling black hole. In the $y$-$z$ plane, different radial regions become coherently blue-shifted over time as both prograde and retrograde populations evolve along their orbits.}
\label{fig:velocity_maps_e=0.99_in}
\end{figure*}

This can also be seen by plotting the $y$-component of the mean eccentricity vector, $\langle e_y \rangle$, separated into prograde and retrograde orbits as we have done in the second from the top panel of Figure \ref{fig:sma_ang_mom_mean_ecc_e=0.99}. The mean $e_y$ among prograde stars is shown in the blue solid line whereas the mean for retrograde stars is shown in the orange dashed line. Statistically significant apsidal alignment is seen in both anti-aligned eccentric disks outward of $a \sim 10^{-2}$, and the flip in alignment occurs nearly simultaneously: around $a \approx 3$ for the prograde population and $a \approx 5$ for the retrograde population (with $r_c = 1$). We also plot the eccentricity profile as a function of semi-major axis with the inclination distribution color-coded in the second from the bottom panel of Figure \ref{fig:sma_ang_mom_mean_ecc_e=0.99}. The prograde and retrograde populations have very similar eccentricity distributions, hitting a minimum at around $a \approx 3$ and $a \approx 5$, respectively, and eccentricities increase both inward and outward. In the bottom panel of Figure \ref{fig:sma_ang_mom_mean_ecc_e=0.99}, we plot the retrograde fraction in the disk as a function of semi-major axis. Compared to the retrograde profile we saw in Figure \ref{fig:sma_ang_mom_mean_ecc_circ_in}, here the retrograde population emerges at much smaller semi-major axes ($a \sim 10^{-2}$) and is $\gtrsim 40 \%$ for $a > r_c$. This makes sense since $a_c \sim 10^{-2} \ r_c$ for $e=0.99$, so orbits are affected much closer in.

\newpage

\subsubsection{Post-Kick Density and Velocity Profile}

For a disk with initially $e=0.99$ orbits with an in-plane kick to the black hole, we plot the distribution of stars in the face-on orientation at three different times post-kick in Figure \ref{fig:pos_stars_e=0.99_in}. Length scale and time units assume a black hole mass of $M_\bullet = 4 \times 10^6 \ M_{\odot}$ and a kick of $\vk = 100$ km/s. Once again the unbound (green dots) and bound populations are separated and the bound population is further split into prograde (blue triangles) and retrograde (orange ``+'' markers) orbits, with the galactic center marked with the ``x'' symbol and the recoiling black hole shown by the ``o'' symbol. The positions are in the galactic center frame of reference. At this zoomed-out scale of $\sim 10$ pc, we mainly see stars on large semi-major axis orbits. At $t=0$, the prograde orbits are located in the $+y$-direction and the retrograde orbits in the $-y$-direction. The bound population is split $\sim$equally about the $x$-axis in this radial range. Over time, the prograde and retrograde populations switch places: at $t=0.15$ Myr and $t=0.75$ Myr, the prograde population is now located toward the $-y$-direction and the retrograde population toward the $+y$-direction. This is because at large semi-major axis, the stars begin their new orbits at periapsis, which is located in the $+y$-direction for prograde orbits and $-y$-direction for retrograde orbits. As the stars evolve along their Keplerian orbits, they begin to show clustering near their collective apoapses, which is in the opposite direction ($-y$-direction for prograde and $+y$ for retrograde orbits).

We investigate the density structure with a kernel density estimation in Figure \ref{fig:sigma_zoom_e=0.99_in}, where we plot surface density contours in both the face-on and edge-on orientations at two different times post-kick, zoomed-in to sub-parsec scales. We separate the population into prograde (blue) and retrograde (orange) orbits with the recoiling black hole shown with the ``o'' marker. Density contours here are in the black hole frame. We see that in this region of small semi-major axes, the prograde and retrograde populations are initially located in the $+y$ and $-y$-directions, respectively, and after $0.5$ Myr, the clustering is unchanged. This is because at small semi-major axes, stars begin their new orbit at apoapsis.

\begin{figure}[t!]
\centering
\includegraphics[width=\linewidth]{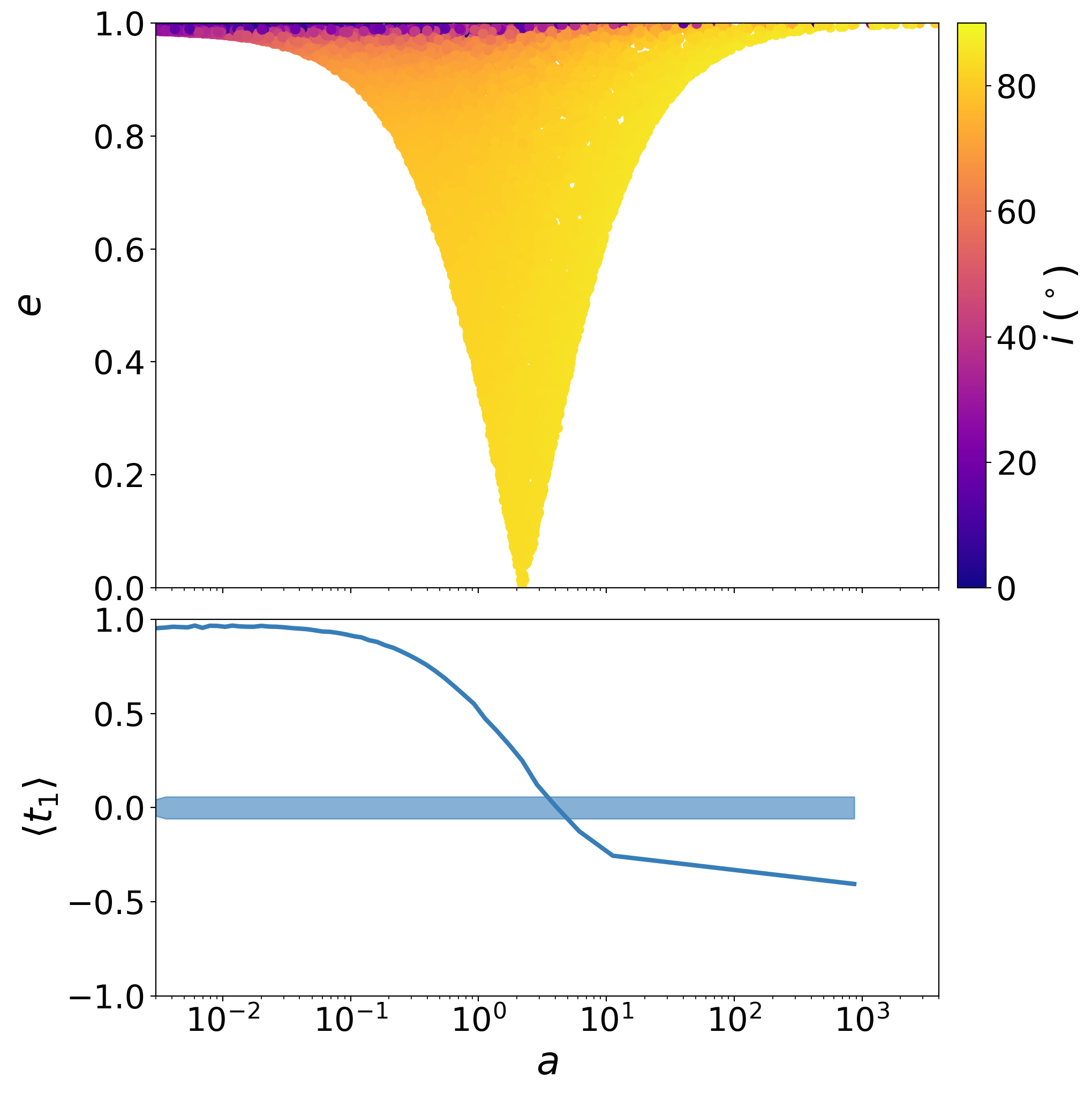}
\caption{\textbf{Disk with high initial eccentricities ($e=0.99$) with an out-of-plane recoil kick.} (\textit{Top:}) The post-kick eccentricity distribution as a function of semi-major axis with the inclination on the color bar. (\textit{Bottom:}) The post-kick mean first component of the tilt vector, $\langle t_1 \rangle$, which measures roll, as a function of semi-major axis. The shaded noise floor shows the maximum expected deviation from 0 (at a $3 \sigma$ level) if the distribution were isotropic. The eccentricity gradient, $de/da$, is negative inward of 
$a \sim 2$ and positive outward, with most stars taking on a high inclination $\sim 90^\circ$. The tilt alignment is statistically significant at most semi-major axis ranges.}
\label{fig:ecc_sma_mean_tilt_e=0.99}
\end{figure}

\begin{figure*}[t!]
\centering
\includegraphics[width=\linewidth]{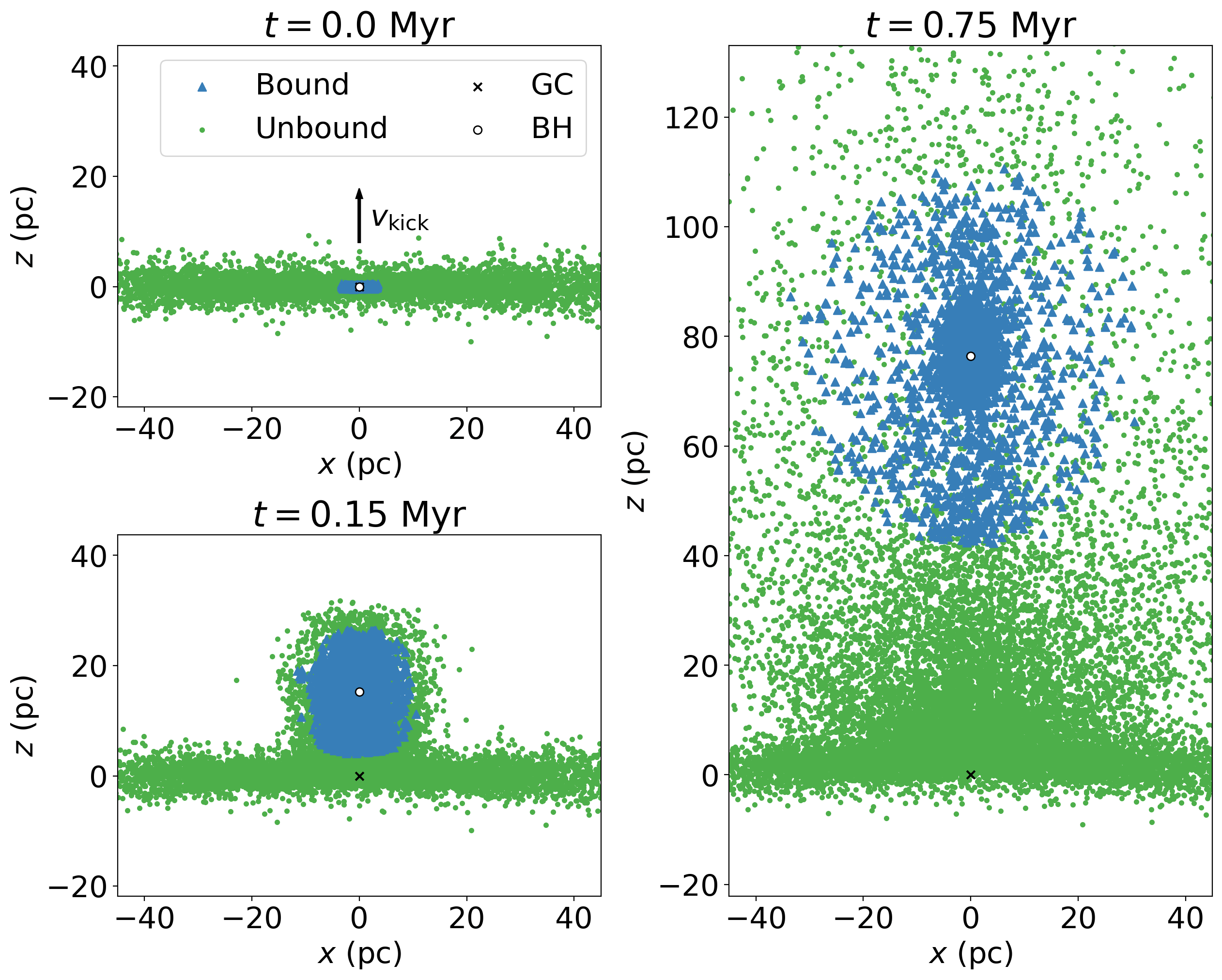}
\caption{\textbf{Disk with high initial eccentricities ($e=0.99$) with an out-of-plane recoil kick.} The edge-on, post-kick distribution of stars at (\textit{top left:}) $t = 0$, (\textit{bottom left:}) $t = 0.15$ Myr, and (\textit{right:}) $t = 0.75$ Myr. Bound and unbound stars are shown with blue triangle markers and green dots, respectively. The ``x'' marker shows the center of the galaxy and the ``o'' marker shows the recoiling black hole. The arrow labeled $\vk$ indicates the recoil kick direction. Positions are computed in the galactic center frame of reference. Length scale and time units assume a black hole mass of $M_\bullet = 4 \times 10^6 \ M_{\odot}$ and a kick of $\vk = 100$ km/s. Both bound and unbound stars form an elongated ball around the recoiling black hole.}
\label{fig:pos_stars_e=0.99_out_z}
\end{figure*}

\begin{figure*}[t!]
\centering
\includegraphics[width=\linewidth]{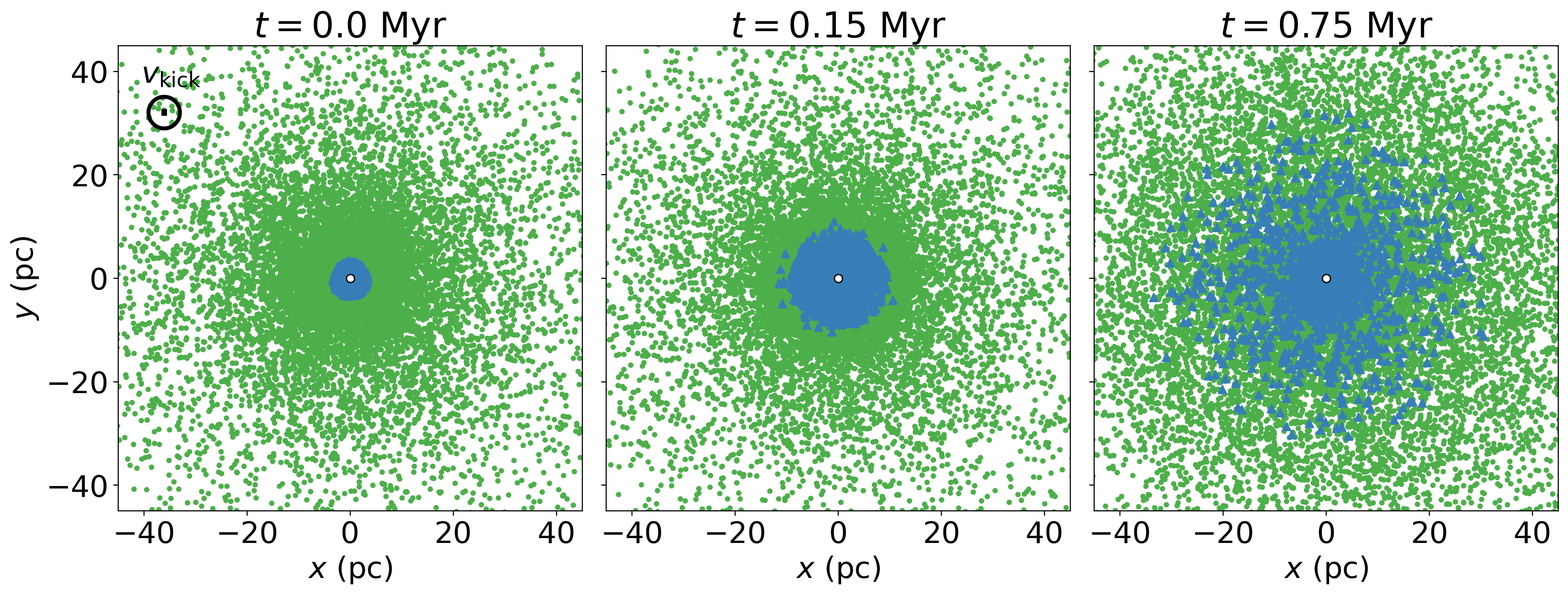}
\caption{\textbf{Disk with high initial eccentricities ($e=0.99$) with an out-of-plane recoil kick.} The face-on, post-kick distribution of stars at (\textit{left:}) $t = 0$, (\textit{center:}) $t = 0.15$ Myr, and (\textit{right:}) $t = 0.75$ Myr. Bound and unbound stars are shown with blue triangle markers and green dots, respectively. The ``x'' marker shows the center of the galaxy and the ``o'' marker shows the recoiling black hole. The recoil kick direction is out of the plane, marked and labeled as $\vk$. Length scale and time units assume a black hole mass of $M_\bullet = 4 \times 10^6 \ M_{\odot}$ and a kick of $\vk = 100$ km/s. Both bound and unbound populations seem to radially expand over time.}
\label{fig:pos_stars_e=0.99_out_y}
\end{figure*}

\begin{figure*}[t!]
\centering
\includegraphics[width=\linewidth]{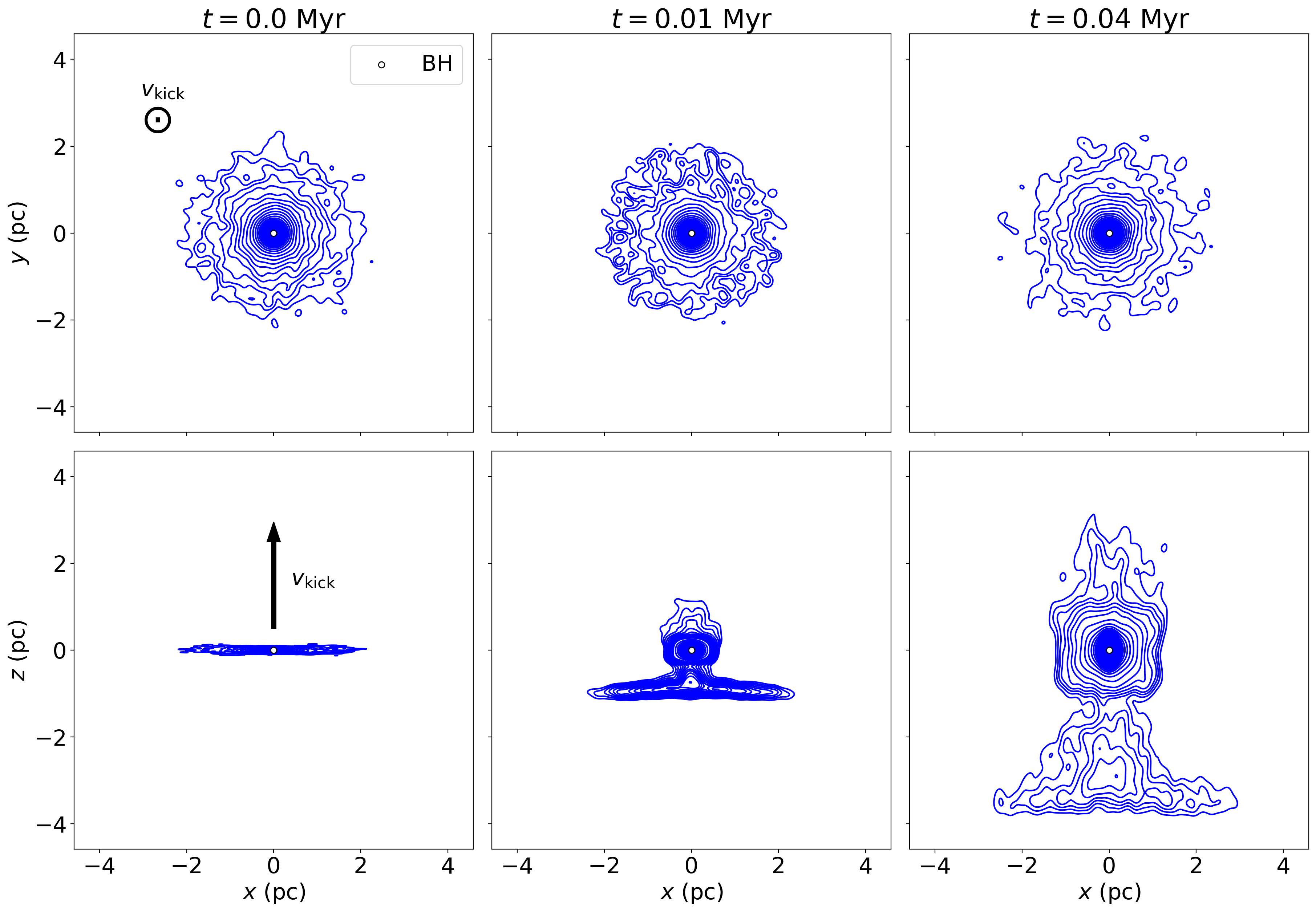}
\caption{\textbf{Disk with high initial eccentricities ($e=0.99$) with an out-of-plane recoil kick.} The post-kick surface density distribution of bound stars in the (\textit{top:}) face-on and (\textit{bottom:}) edge-on orientations at (\textit{left:}) $t = 0$, (\textit{center:}) $t = 0.01$ Myr, and (\textit{right:}) $t = 0.04$ Myr. The ``o'' marker shows the recoiling black hole, and the contours are computed in the frame co-moving with the black hole. In each orientation, the recoil kick direction is marked and labeled as $\vk$. Length scale and time units assume a black hole mass of $M_\bullet = 4 \times 10^6 \ M_{\odot}$ and a kick of $\vk = 100$ km/s. The edge-on orientation reveals an over-density behind the recoiling black hole.}
\label{fig:sigma_zoom_e=0.99_out}
\end{figure*}

\begin{figure*}[t!]
\centering
\includegraphics[width=\linewidth]{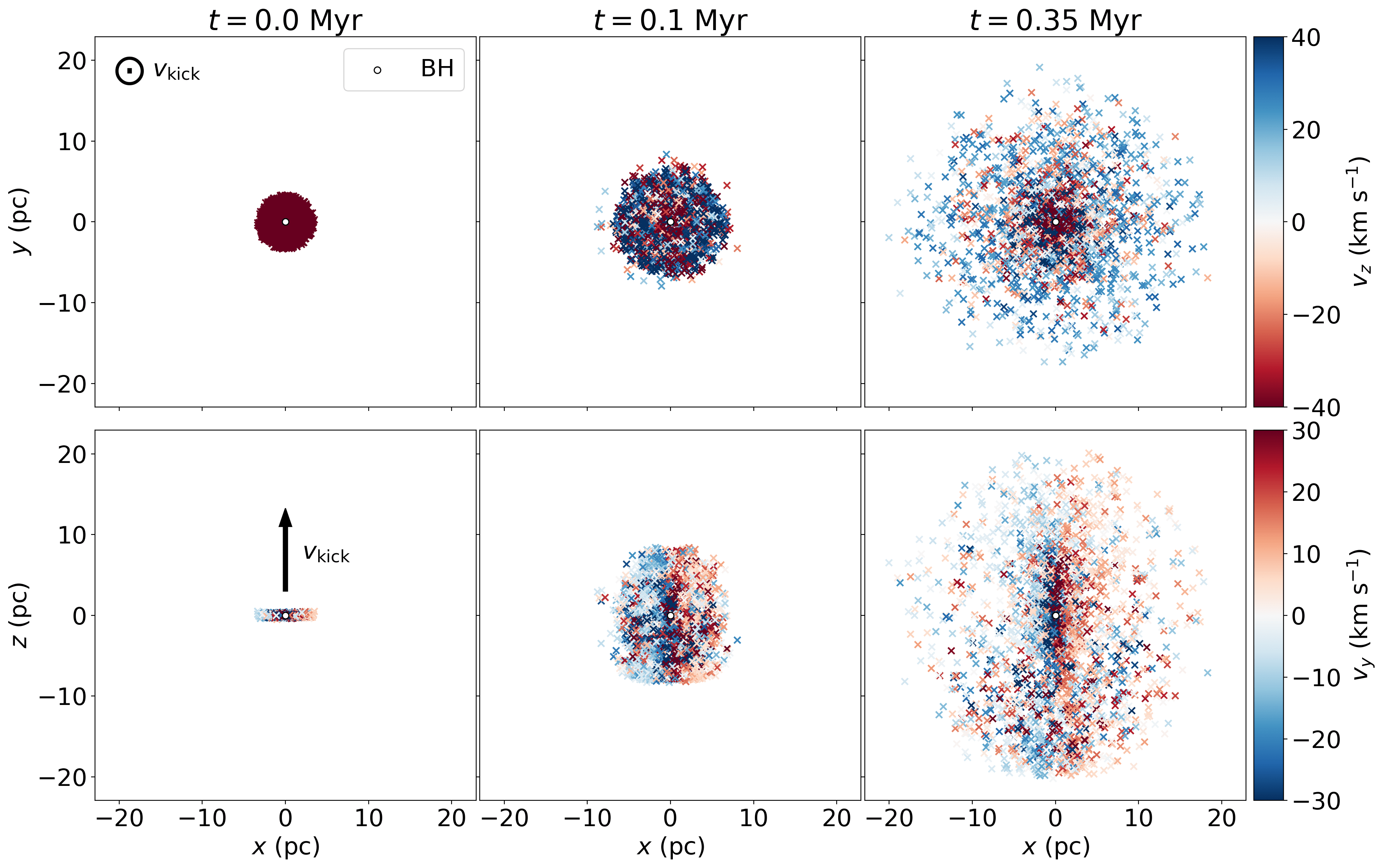}
\caption{\textbf{Disk with high initial eccentricities ($e=0.99$) with an out-of-plane recoil kick.} The line-of-sight velocity maps in the frame co-moving with the recoiling black hole in the (\textit{top:}) face-on and (\textit{bottom:}) edge-on orientations at (\textit{left:}) $t=0$, (\textit{center:}) $t=0.1$ Myr, and (\textit{right:}) $t=0.35$ Myr. In each orientation, the color bar is adjusted such that approaching velocities are blue and receding velocities are red. The recoil kick direction is marked and labeled as $\vk$ for each row. Length scale and time units assume a black hole mass of $M_\bullet = 4 \times 10^6 \ M_{\odot}$ and a kick of $\vk = 100$ km/s. The face-on orientation shows circular patterns where the edges of the distribution becomes more blue-shifted over time. The edge-on orientation demonstrates that the elongated ball-structure maintains coherent rotation as it evolves.}
\label{fig:velocity_maps_e=0.99_out}
\end{figure*}

Finally, in Figure \ref{fig:velocity_maps_e=0.99_in}, we plot the line-of-sight velocity maps in the face-on ($x$-$y$) and both edge-on ($x$-$z$ and $y$-$z$) orientations at three different times post-kick. Velocities are in the frame co-moving with the recoiling black hole, and the color bar in each orientation is adjusted so that approaching velocities are blue and receding velocities are red. We do not see much velocity structure in the face-on orientation. In the $x$-$z$ plane, a cluster of stars with high line-of-sight velocity seems to form right behind the recoiling black hole after some time ($t=0.35$ Myr). This is due to high eccentricity, large semi-major axis stars that initially begin at periapsis. Since their eccentricity vectors are aligned along the $y$-axis and they have high eccentricity orbits, shortly after the kick they have velocities almost purely in the line-of-sight with a magnitude close to its high periapsis speed. This creates the collection of high line-of-sight velocity stars behind the recoiling black hole. In the $y$-$z$ plane, we clearly see the symmetry about the $x$-axis caused by the $\sim$equal split between prograde and retrograde stars. At $t=0$, all stars are preferentially red-shifted since stars toward the $+y$-direction are on prograde orbits and stars toward the $-y$-direction are on retrograde orbits. At $t=0.1$ Myr, the two populations begin to switch places and we see several blue-shifted stars. Finally, at $t=0.35$ Myr, the edges are now preferentially blue-shifted since the retrograde population has migrated toward the $+y$-direction and the prograde population has moved toward the $-y$-direction.

\subsection{Numerical Results: disk with initial $e=0.99$, out-of-plane kick}

\subsubsection{Distribution of Orbital Elements and Tilt Vector Alignment}

In the top panel of Figure \ref{fig:ecc_sma_mean_tilt_e=0.99}, we plot the post-kick eccentricity distribution as a function of the semi-major axis with the inclination color-coded. Circular orbits are found in a narrow region around $a \approx 2$ (in units where $r_c = 1$), and eccentricities grow both inward and outward. Since $v_{\rm{circ}} \gg v_{\rm{apo}}$, the post-kick inclinations of these orbits are generally very high, $i \sim 90^\circ$: one only finds low inclination orbits at smaller semi-major axes. In the bottom panel of Figure \ref{fig:ecc_sma_mean_tilt_e=0.99}, we plot the first component of the mean tilt vector as a function of semi-major axis. The high, positive $\langle t_1 \rangle$ at small semi-major axes is due to $\omega$-clustering at 0 since the stars begin at their apoapsis which also corresponds to their descending node (so periapsis coincides with the ascending node). On the other hand, the negative $\langle t_1 \rangle$ at large semi-major axes is due to $\omega$-clustering at $\pi$, since the descending node now corresponds to the periapsis.

\subsubsection{Post-Kick Density and Velocity Profile}

We show the edge-on distribution of stars in Figure \ref{fig:pos_stars_e=0.99_out_z} at three different times post-kick. Stars are split into bound (blue triangles) and unbound (green dots) orbits, the galactic center is indicated by a ``x'' marker, and the black hole with a ``o'' marker. The positions are computed in the galactic center frame of reference. At $t=0$, the bound stars are confined to a compact disk surrounding the recoiling black hole. As the system evolves, both bound and unbound stars seem to form an elongated ball surrounding the black hole that expands in size over time. This elongated appearance is due to the high eccentricity, high inclination orbits with large semi-major axes we see in Figure \ref{fig:ecc_sma_mean_tilt_e=0.99}. In Figure \ref{fig:pos_stars_e=0.99_out_y}, we show the same distribution of stars in the face-on orientation. The structure is axi-symmetric at all times, as expected, and we see the ball of bound and unbound stars expand in radius over time. In Figure \ref{fig:sigma_zoom_e=0.99_out}, we show surface density contours in the face-on and edge-on orientations zoomed-in to $\sim$parsec-scales at three different times following the kick, using a kernel density estimation. The ``o'' marker shows the recoiling black hole, and these density contours are computed in the frame co-moving with the black hole. In the face-on orientation, the structure always remains axi-symmetric in appearance. In the edge-on orientation, a puffy disk of tightly bound stars surround the recoiling black hole as it lifts above the disk plane, and the more loosely bound stars seem to cluster behind the black hole.

Finally, we plot the line-of-sight velocity profile in both face-on and edge-on orientations in Figure \ref{fig:velocity_maps_e=0.99_out}. Velocities are computed in the frame co-moving with the black hole, and in each orientation, the color bar is adjusted such that approaching velocities are blue and receding velocities are red. In the face-on orientation, the stars are initially all red-shifted since they all start moving away from us on their new orbits; every star begins their new orbit at its descending node. At $t=0.1$ Myr, there appears to be a circular pattern in the line-of-sight velocity once again caused by the coherent roll of the orbits over their major axes, but this pattern is less pronounced than the circular case (Figure \ref{fig:velocity_maps_circ_out}) and seems to disperse at earlier times ($t = 0.35$ Myr). In the edge-on orientation, one can see that as the black hole begins to recoil and the disk puffs up into an elongated ball, the rotation profile of the initial disk is maintained. While the inclination distribution of the cluster significantly changes, the right side is always red-shifted and the left side blue-shifted.

\section{Adding Initial Inclinations}
\label{sec:inc}

\subsection{Numerical Results: parameter study of initial
inclinations}

While we do not perform further calculations here, we can make predictions based on our analyses thus far. As we increase the inclinations within the disk, stellar orbital angular momentum vectors will have a wider range of angles with respect to the recoil kick. This will reduce the overall alignment of the eccentricity vector. However, we expect to see non-zero tilt vector alignment even in a puffy disk. As shown in Figure~\ref{fig:ecc_alpha_inc3}, if there is some out-of-plane component to the kick, the post-kick orbits have significant $\omega$-clustering near 0 or $\pi$. 

We numerically study the instantaneous alignment of the eccentricity vector and the tilt vector as we change the initial inclination distribution and the initial eccentricity distribution. Once again, we only slightly alter the set-up described in Section \ref{sec:sim_setup}. We vary the scale parameter, $\sigma_i$, that dictates the inclination range and gradually puff-up the disk while simultaneously changing initial eccentricities incrementally. We additionally simulate an isotropic distribution with a $N(e)=2e \ de$ eccentricity profile relevant for Keplerian orbits with isotropic velocities \citep{Bin87} as the most general case. In addition, we vary the initial density profile. We first use the same number density as before which translates to a steep volume density profile of $\rho \sim a^{-2}$, but we also run simulations with the Bahcall-Wolf cusp density profile ($\rho \sim a^{-7/4}$) \citep{Bah76} following the simulations presented in \citet{Merritt2009}.

\begin{figure}[t!]
\centering
\includegraphics[width=\linewidth]{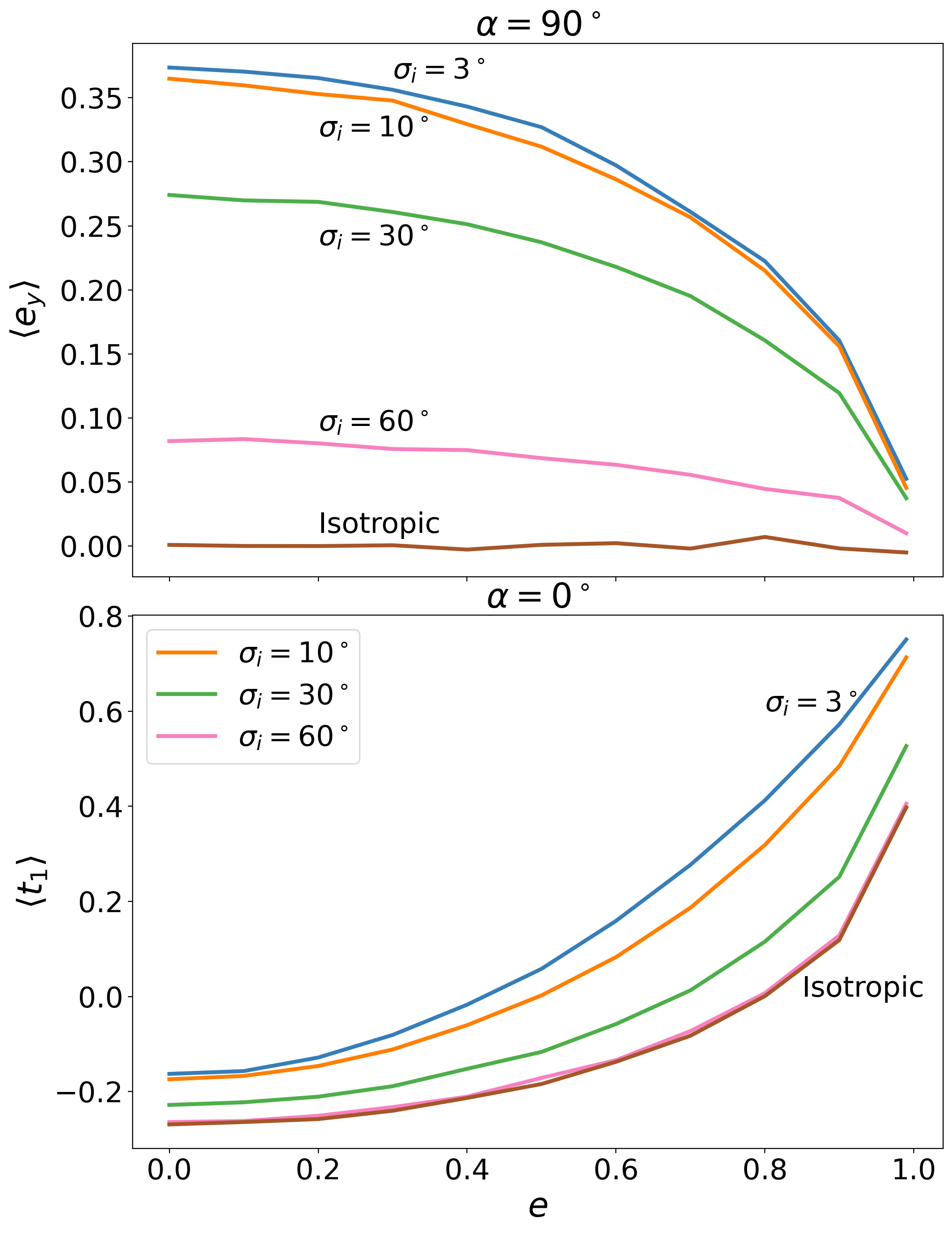}
\caption{\textbf{Parameter study with disks of varying initial inclination distributions.} (\textit{Top}:) The average $y$-component of the eccentricity vector, $\langle e_y \rangle$, for an in-plane kick and (\textit{bottom}:) the average first component of the tilt vector, $\langle t_1 \rangle$, for an out-of-plane kick, as a function of the initial eccentricity with different lines showing different inclination distributions taken from a suite array of numerical simulations. Apsidal alignment becomes weaker with increased disk thickness, but tilt alignment is expected regardless of the inclination distribution.}
\label{fig:ecc_inc_alpha}
\end{figure}

In the top panel of Figure \ref{fig:ecc_inc_alpha}, we show $\langle {e_y} \rangle$ as a function of eccentricity for a variety of inclination distributions for an in-plane kick. As expected, the apsidal alignment becomes weaker as we increase $\sigma_i$, the scale parameter of the Rayleigh distribution. $\langle e_y \rangle$ seems to show relatively weak apsidal alignment in puffy disks (e.g. $\sigma_i = 60^\circ$) and no alignment in the isotropic case, but as we discuss in the next section, the mean eccentricity vector fails to capture symmetric anisotropies, so the isotropic case warrants further investigation. In the bottom panel of Figure \ref{fig:ecc_inc_alpha}, we show $\langle t_1 \rangle$ as a function of eccentricity for a variety of inclination distributions for an out-of-plane kick. As expected, $\omega$-clustering is seen at various eccentricities either at 0 or $\pi$ with the turning point having a dependence on the inclination distribution. This makes sense since changing $\sigma_i$ changes the distribution of kick angles with respect to the orbital angular momenta of individual stars. Notably, $\langle t_1 \rangle$ predicts tilt alignment even in the isotropic case, but we emphasize that this measure is dependent on our choice of reference plane: specifically, tilt alignment is expected in the plane perpendicular to the recoil kick.

\begin{figure*}[t!]
\centering
\includegraphics[width=\linewidth]{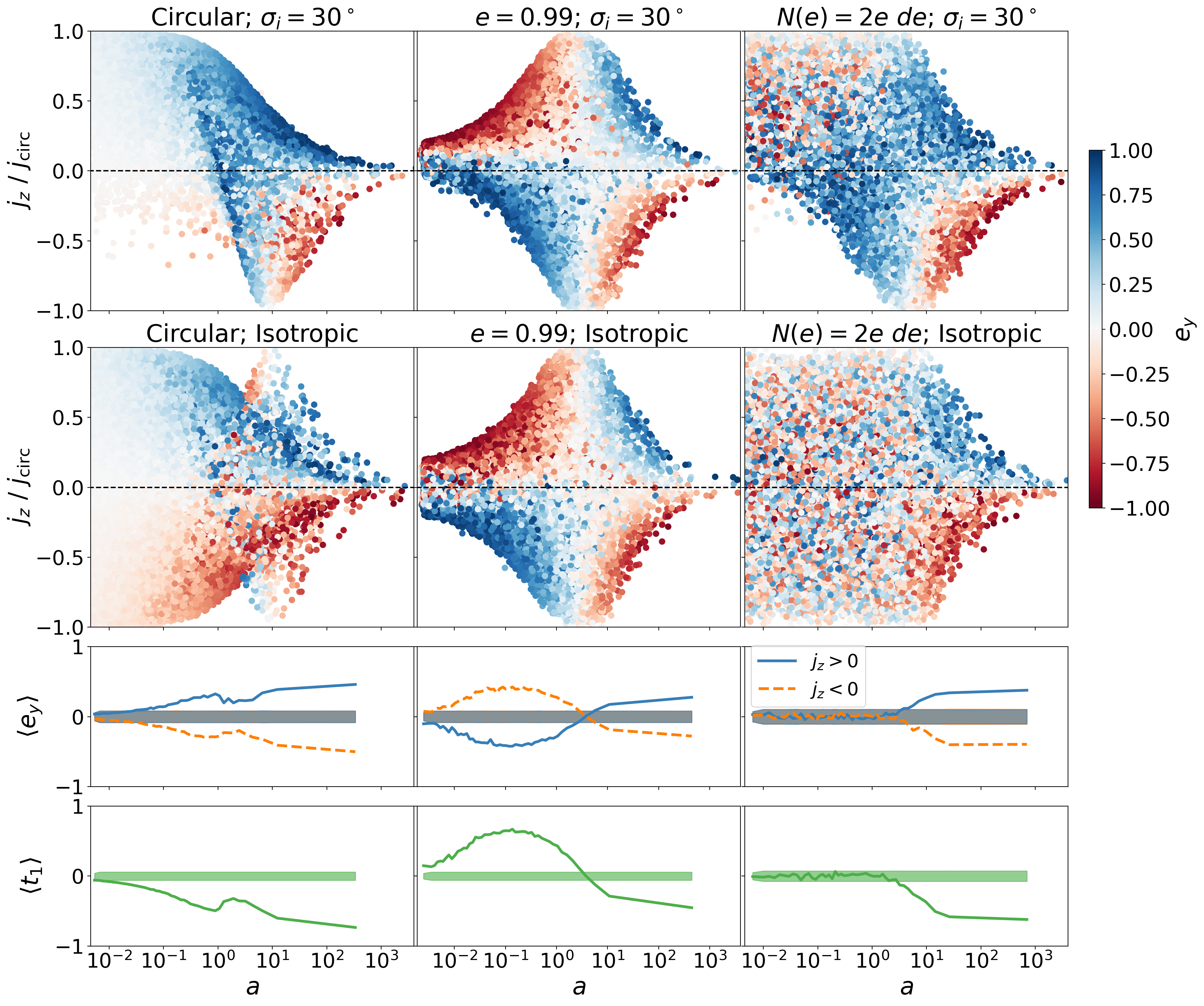}
\caption{\textbf{Parameter study with stellar distributions of varying initial inclinations.} (\textit{Top two rows:}) The post-kick distribution of the $z$-component of angular momentum (normalized by the circular angular momentum at a given radius) as a function of semi-major axis, with the color bar showing the $y$-component of the eccentricity vector for six different simulation runs. Rows correspond to (\textit{top:}) inclination Rayleigh-distributed with $\sigma_i = 30^\circ$ and (\textit{second from top:}) isotropically-distributed orbits. Columns correspond to (\textit{left:}) circular, (\textit{center:}) $e=0.99$, and (\textit{right:}) eccentricity profile $N(e) = 2e \ de$ with a Bahcall-Wolf cusp density profile ($\rho \sim a^{-1.75}$) \citep{Bah76}. The first two columns use the same number density as before which translates to a volume density $\rho \sim a^{-2}$. (\textit{Second from bottom:}) The $y$-component of the mean eccentricity vector separated into prograde and retrograde orbits (about the $z$-axis) as a function of semi-major axis corresponding to the isotropic simulations in the panels above. (\textit{Bottom:}) The mean first component of the tilt vector, $\langle t_1 \rangle$, binned by semi-major axis where the plane of reference is the plane perpendicular to the kick ($y$-$z$) and the reference direction is $+y$. In this rotated coordinated system, $t_1$ still measures roll over the major axis but now with respect to the new plane of reference ($y$-$z$). For the bottom two rows, the shaded noise floor shows the maximum expected deviation from 0 (at a $3 \sigma$ level) for an isotropic distribution. All of these simulations develop post-kick anisotropies. The isotropic simulations form a torus-like structure of eccentric orbits that preferentially align in the plane perpendicular to the recoil kick.}
\label{fig:sma_ang_mom_inc}
\end{figure*}

We compare the post-kick distribution of the $z$-component of angular momentum (normalized by the circular angular momentum at a given radius) as a function of semi-major axis for six different simulations in Figure \ref{fig:sma_ang_mom_inc}. The top row shows simulations where the inclination is initially Rayleigh-distributed with $\sigma_i = 30^\circ$, and the second from the top row shows simulations with initially isotropically-distributed stellar orbits. Columns correspond to different eccentricity distributions: the left column is initially circular orbits, the center column is initially $e=0.99$, and the right column is an eccentricity profile $N(e)=2e \ de$. The first two use the same number density profile as before which is a volume density of $\rho \sim a^{-2}$. For the right-most column, we change the density profile to that of a Bahcall-Wolf cusp ($\rho \sim a^{-7/4}$) \citep{Bah76}. There is no net angular momentum axis for an isotropic configuration, but we pick the $z$-axis around which we determine whether the orbit is prograde or retrograde, and the kick in this case is orthogonal to that axis ($+x$). This is a convenient way to visualize how anisotropies manifest in different ways as we alter the initial eccentricity and inclination distributions. Focusing on the top row, which is a puffy disk with $\sigma_i = 30^\circ$, the initially circular case is fairly similar to the top panel of Figure \ref{fig:sma_ang_mom_mean_ecc_circ_in}. Inward of $a \sim 10$, the majority of stars have a significant, positive $e_y$ component so we have a single eccentric nuclear disk. At $a \gtrsim 10$, there is a retrograde population with apsidal alignment in the opposite direction. The $e=0.99$ case is similar to the top panel of Figure \ref{fig:sma_ang_mom_mean_ecc_e=0.99}. Inward of $a \sim 4$, we have a prograde eccentric nuclear disk in the $-y$-direction and a retrograde eccentric disk in the $+y$-direction forming a bar-like structure. At $a \gtrsim 4$, they flip so we have a prograde eccentric disk in the $+y$-direction and a retrograde eccentric disk in the $-y$-direction. The $N(e) = 2e \ de$ simulation (right panel) contains eccentricities from the most circular to the most eccentric, so it is effectively an intermediate between the left two panels. Closer in ($a \lesssim 5$), the retrograde population shows alignment in the $+y$-direction since both circular and eccentric cases show coherent alignment in this quadrant. The prograde orbits, however, have seemingly randomized since initially circular orbits align in the opposite direction ($+y$) as initially eccentric orbits ($-y$). At $a \gtrsim 5$, both circular and eccentric cases exhibit similar structure so we obtain the bar-like structure again.

For the isotropic cases (second from top row), the prograde and retrograde sub-populations must be  mirror images of each other due to the axi-symmetry about the kick vector. For a given eccentricity distribution, one can see how the isotropic cases are obtained by combining the disk (top row) cases with their mirror images. For instance for the circular cases (right column), the ``V'' shape of the retrograde population around $a \sim 10$ in the disk case is still very clearly seen in the isotropic case, with the prograde orbits now showing the same mirror-image structure: an inverse ``V'' shape around $a \sim 10$. In a similar manner, the $e=0.99$ isotropic case is a mirror-image version of the disk case. Finally, the $N(e) = 2e \ de$ case can be understood as an intermediate between the circular and $e=0.99$ isotropic cases. Inward of $a \sim 10$, the alignment is opposite for the initially circular and eccentric simulations, so the post-kick eccentricity vectors are seemingly random. At $a \gtrsim 10$, however, the alignment in the circular and eccentric cases are the same, so we get $+y$-alignment for prograde orbits and $-y$-alignment for retrograde orbits. It is clear from these plots that the recoil kick induces distinct orbital structures even when starting from an isotropic configuration. One point of clarification is that the initially isotropic simulations exhibit these anisotropies in an axi-symmetric fashion, so statistically-significant apsidal alignment presents itself as a torus-like structure of eccentric orbits rather than a bar-like structure.

\subsubsection{Limitation of the Mean Eccentricity Vector}
\label{sec:drawback}

While the mean eccentricity vector is a convenient measure of lopsidedness (i.e., if the entire disk is eccentric and apse-aligned), it fails to capture \textit{symmetric} anisotropies. As a quick proof of concept let's consider the bar-like structure of two anti-aligned eccentric nuclear disks. The mean eccentricity vector of two anti-aligned eccentric disks of approximately the same strength in apsidal alignment is $\sim$0 even though the stellar distribution is clearly anisotropic. This poses a problem when starting with an isotropic distribution of orbits. By design, the post-kick configuration will be axi-symmetric about the kick vector, so the mean eccentricity vector will always average out to $\sim$0.

To capture symmetric anisotropies, we use two measures. The first measure
separates orbits into prograde and retrograde populations about an arbitrary axis perpendicular to the kick ($+z$). Then, the $y$-component of the mean eccentricity vector binned by semi-major axis can detect axi-symmetric anisotropies at different radii. The second measure looks at the tilt vector alignment with respect to the plane perpendicular to the recoil kick ($y$-$z$). We choose $+y$ to be our reference direction, so one can obtain the tilt vector components from Equation \ref{eqn:tilt_vec} in this new plane of reference by simply rotating the axes once over: replacing $e_x$ with $e_y$, $e_y$ with $e_z$, and $e_z$ with $e_x$.

\subsection{Spherically Isotropic Distribution}

We show our two measures of anisotropy in the bottom two rows of Figure \ref{fig:sma_ang_mom_inc} with these panels corresponding to the isotropic simulations in the second from top row. The circular case is in the left panel, the $e = 0.99$ case is in the center panel, and the $N(e) = 2e \ de$ case is in the right panel. For both measures, the shaded noise floor shows the maximum expected deviation from 0 (at a $3 \sigma$ level) for an isotropic distribution. The second from the bottom row shows the mean $y$-component of the eccentricity vector separated into prograde and retrograde orbits about the $z$-axis and binned by semi-major axis. When the eccentricity vector is separated in this way, we notice that statistically-significant apsidal alignment is induced along the $y$-axis. These panels are also consistent with the distribution of angular momenta in the second from top row. For the circular case, the prograde orbits are preferentially aligned in the $+y$-direction for most stars and the retrograde orbits are preferentially aligned in the $-y$-direction. Correspondingly, $\langle e_y \rangle$ is positive for prograde orbits and negative for retrograde orbits. For the $e = 0.99$ case, the flip in alignment between prograde and retrograde stars at $a \sim 5$ is captured by $\langle e_y \rangle$ as well. Finally, for the $N(e) = 2e \ de$ case, we see alignment emerging at large semi-major axes, once again consistent with the color-coded angular momenta in the panel above. Keeping in mind that there is always axi-symmetry along the kick vector ($+x$), the same eccentricity pattern exists along the $z$-axis (or any axis perpendicular to the kick vector), so we can understand these anisotropy patterns as a \textit{torus} of eccentric orbits that is perpendicular to the kick vector. We note, however, that this is not a typical torus in which individual orbits have clustered angular momentum vectors. The eccentricity vectors lie in the plane perpendicular to the kick, but the angular momentum vectors are not clustered. 

In the bottom row of Figure \ref{fig:sma_ang_mom_inc}, we show $\langle t_1 \rangle$, the tilt vector component that measures roll, with respect to the plane perpendicular to the recoil ($y$-$z$) binned by semi-major axis. We immediately see that this measure detects the same anisotropic structure as above. This makes sense based on how the tilt vector is computed. If we, for example, take the circular, isotropic case (the bottom three panels of the left-most column), the prograde orbits (with respect to the $z$-axis) have eccentricity vectors coherently aligned in $+y$-direction. This means that the ascending node with respect to the $y$-$z$ plane is preferentially in the $-y$-direction, and since the eccentricity vector ($+y$) and the ascending node ($-y$) are misaligned, $\langle t_1 \rangle < 0$ in this case. Another way to understand this is that $\Omega$ (with respect to the $y$-$z$ plane) is distributed axi-symmetrically following the kick, but since the tilt vector rotates every eccentricity vector by $- \Omega$, the axi-symmetry is erased and the torus-like anisotropies are uncovered.

\subsubsection{Numerical Results: spherically isotropic, circular}

\begin{figure*}[t]
\centering
\includegraphics[width=0.95\linewidth]{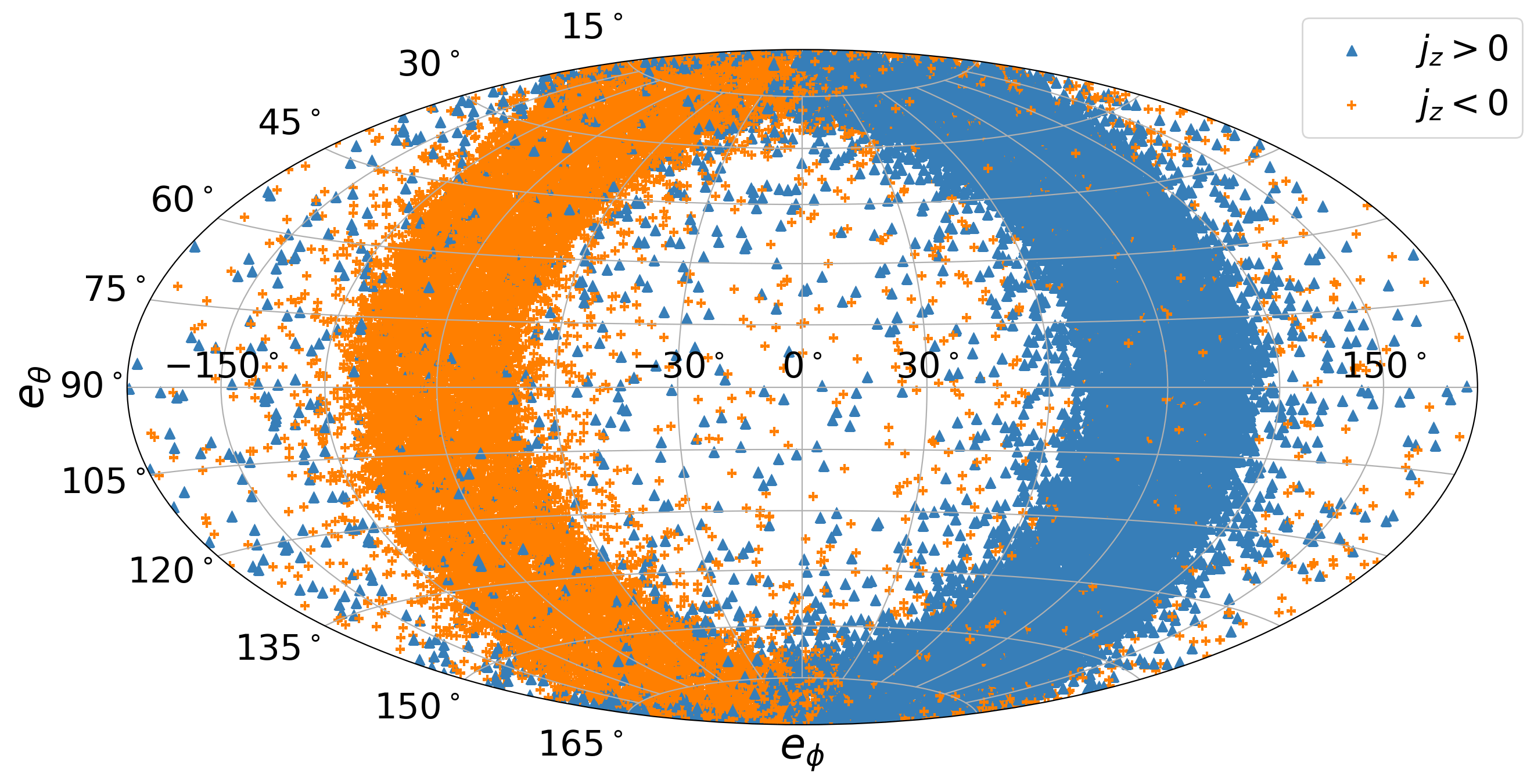}
\caption{\textbf{Initially circular orbits in a spherically isotropic distribution; recoil kick along the $x$-axis.} A two-dimensional projection of the post-kick eccentricity vectors in spherical coordinates where $\theta$ is the polar angle and $\phi$ is the azimuthal angle. With respect to the $z$-axis, prograde stars are marked with blue triangles and retrograde stars are marked with orange ``+'' markers. Eccentricity vectors preferentially lie in the plane perpendicular to the recoil kick ($y$-$z$), and this torus-like structure has a clear division when separated into prograde and retrograde stars about a certain axis within that plane ($z$-axis). Prograde stars have preferentially positive $e_y$ and retrograde stars have preferentially negative $e_y$.}
\label{fig:e_phi_theta_dist_iso}
\end{figure*}

\begin{figure*}[t!]
\centering
\includegraphics[width=0.95\linewidth]{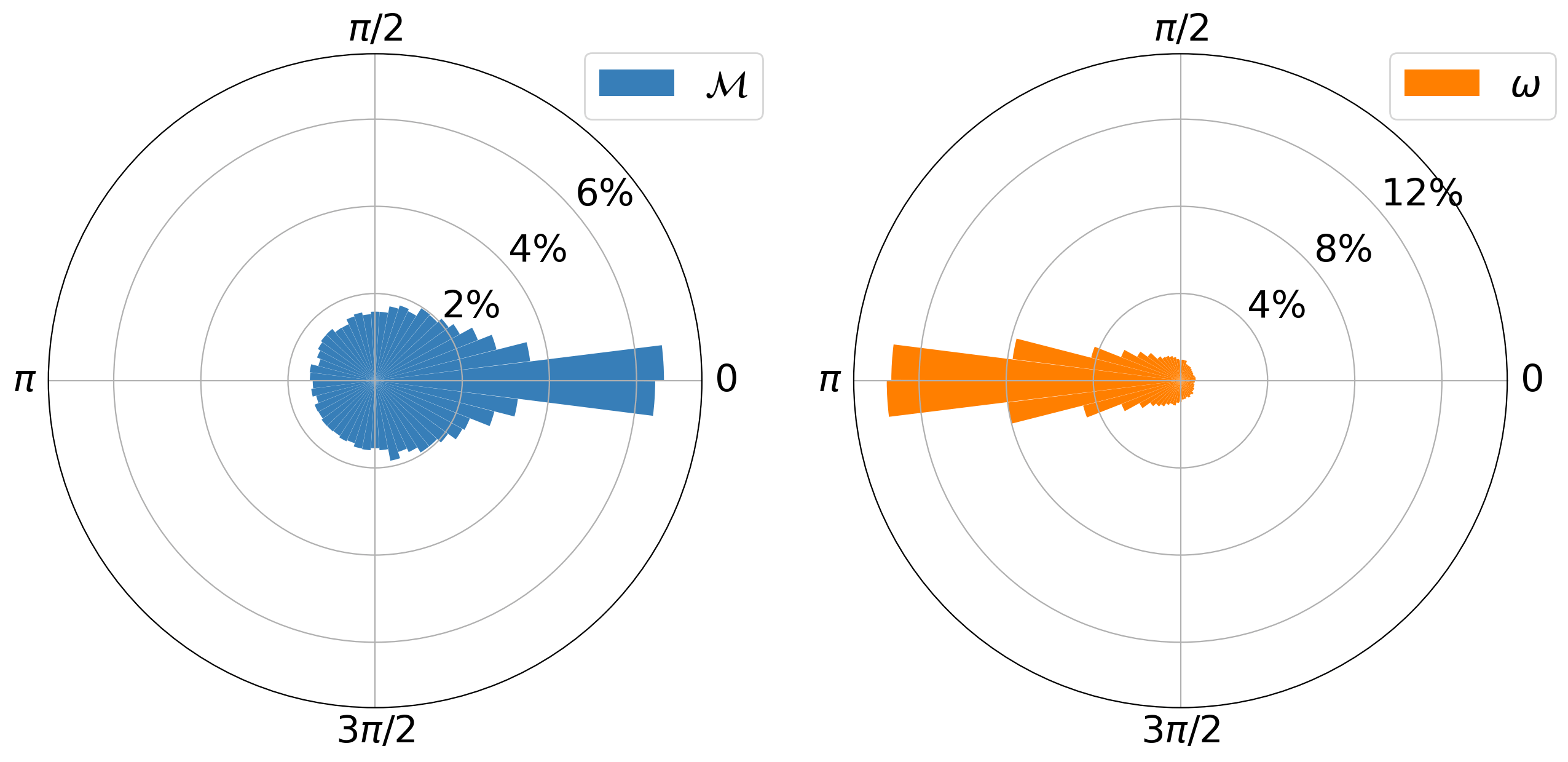}
\caption{\textbf{Spherically isotropic distribution with a $N(e) = 2e \ de$ eccentricity profile and a Bahcall-Wolf cusp density profile.} A histogram of the post-kick (\textit{left:}) mean anomaly, $\mathcal{M}$, and (\textit{right:}) argument of periapsis, $\omega$. Mean anomaly is clustered at 0 and $\omega$ is clustered at $\pi$, even for an initially isotropic distribution.}
\label{fig:omega_anom_hist_2ede_iso}
\end{figure*}

To illustrate the torus-like anisotropy, we show a projection of the post-kick eccentricity vectors in spherical coordinates for an initially isotropic distribution of circular orbits in Figure \ref{fig:e_phi_theta_dist_iso}, where we have separated prograde (blue triangle) and retrograde (orange ``+'' marker) orbits with respect to the $z$-axis. Just as we saw in Figure \ref{fig:sma_ang_mom_inc} in the bottom three panels of the left-most column, the eccentricity vectors show coherent alignment preferentially perpendicular to the kick, forming a torus of eccentric orbits. The torus has a clear division between prograde and retrograde stars: prograde stars have eccentricity vectors that preferentially point in the $+y$-direction and retrograde stars have eccentricity vectors preferentially pointing in the $-y$-direction, consistent with Figure \ref{fig:sma_ang_mom_inc}. This intuitively makes sense according to our in-plane and out-of-plane kick simulation results for initially circular orbits. We have seen that a circular orbit in the $x$-$y$ plane preferentially develops eccentricities along the $y$-direction from Section \ref{sec:circ_in}. Similarly a circular orbit in the $x$-$z$ plane should develop eccentricities along the $z$-direction. Finally, for an out-of-plane kick with an orbit initially in the $y$-$z$ plane, eccentricities develop along their initial position vector which is also contained within the $y$-$z$ plane (Section \ref{sec:circ_out}). It makes sense, then, that an isotropic distribution of circular orbits will predominantly develop eccentricities in the plane perpendicular to the kick vector.

\subsubsection{Numerical Results: spherically isotropic, eccentric}

As the most general case, we employ a $N(e) = 2e \ de$ eccentricity distribution and run a recoil kick simulation starting with an isotropic star cluster. As shown in the bottom three panels of the right-most column in Figure \ref{fig:sma_ang_mom_inc}, the post-kick distribution seems to have no discernible structure before $a=1$--$10$, and beyond $a \gtrsim 10$, a clear eccentric torus forms. As shown in the bottom row of this figure, tilt vector alignment (with respect to the plane perpendicular to the kick) is still strong in the isotropic cases. This is because tilt vector alignment occurs even for a relatively small out-of-plane kick component (see Figure \ref{fig:ecc_inc_alpha}). In Figure \ref{fig:omega_anom_hist_2ede_iso}, we show the post-kick histogram of mean anomalies, $\mathcal{M}$, in the left panel and arguments of periapsis, $\omega$, on the right panel. While the distribution is much more spread-out compared to what we saw for the circular, out-of-plane case in Figure \ref{fig:omega_anom_hist_circ_out}, both distributions are clearly anisotropic: mean anomalies skew toward 0, which means stars are preferentially starting their new orbits at periapsis, and the distribution of $\omega$ is skewed toward $\pi$, which means that the stars periapses preferentially correspond to their descending node (i.e., moving toward the $-x$-direction at periapsis). 

\paragraph{Post-Kick Density and Velocity Profile}

\begin{figure*}[t!]
\centering
\includegraphics[width=\linewidth]{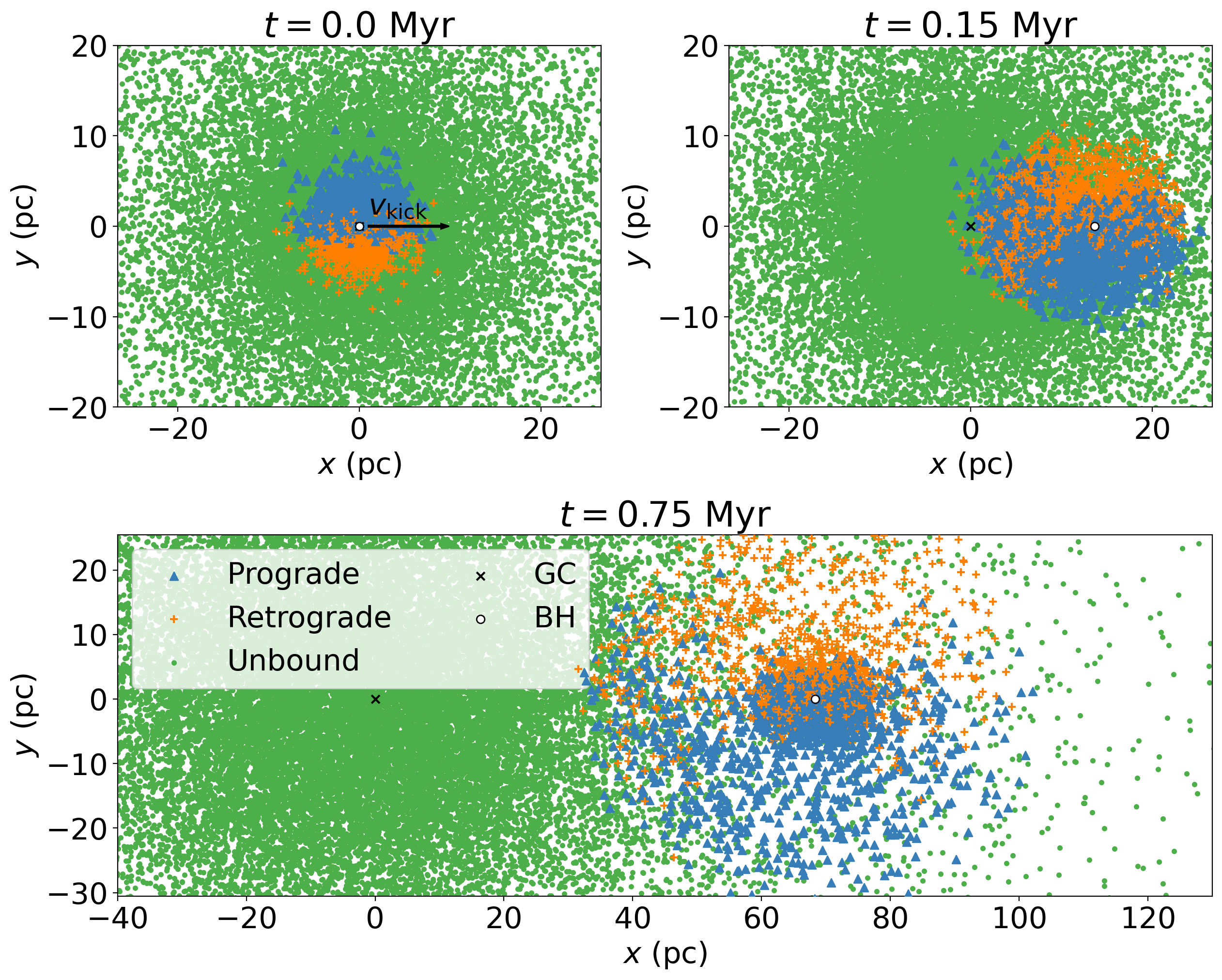}
\caption{\textbf{Spherically isotropic distribution with a $N(e) = 2e \ de$ eccentricity profile and a Bahcall-Wolf cusp density profile.} The post-kick distribution of stars in the $x$-$y$ plane at (\textit{top left:}) $t = 0$, (\textit{top right:}) $t = 0.15$ Myr, and (\textit{bottom:}) $t = 0.75$ Myr. Bound stars are separated into prograde (blue triangles) and retrograde (orange ``+'' markers) about the $z$-axis. The unbound stars are marked with green dots. The ``x'' marker shows the center of the galaxy and the ``o'' marker shows the recoiling black hole. The recoil kick direction is marked by the arrow labeled $\vk$. All positions are computed in the galactic center frame of reference. Length scale and time units assume a black hole mass of $M_\bullet = 4 \times 10^6 \ M_{\odot}$ and a kick of $\vk = 100$ km/s. Even for an isotropic distribution, the prograde and retrograde stars seem to have a clear division at large radii.}
\label{fig:pos_stars_2ede_iso}
\end{figure*}

For the most general case of an initially isotropic star cluster with a $N(e)=2e \ de$ eccentricity profile, we take a look at the density and velocity profiles to explore any potentially observable structures. In Figure \ref{fig:pos_stars_2ede_iso}, we show the distribution of stars separated into bound and unbound (green dots) stars, with the bound stars further separated into prograde (blue triangles) and retrograde (orange ``+'' markers) about the $z$-axis, at three different times post-kick. The ``x'' marker shows the galactic center and the ``o'' marker shows the recoiling black hole. All positions are computed in the galactic center frame of reference. The evolution is quite similar to what was shown in Figure \ref{fig:pos_stars_e=0.99_in}. At $t=0$ the stars on prograde orbits start toward the $+y$-direction and the stars on retrograde orbits start in the $-y$-direction. As the stars move along their Keplerian orbits, the two populations switch places such that at $t=0.15$ Myr and $t=0.75$ Myr, the prograde population is now in the $-y$-direction and the retrograde orbits are in the $+y$-direction. Of course, one key difference between the two simulations is that this evolution only occurs in the $x$-$y$ plane for the disk case whereas the initially isotropic case exhibits this structure axi-symmetrically about the $x$-axis, so the same structure should be observed in the $x$-$z$ plane, for instance.

\begin{figure*}[t!]
\centering
\includegraphics[width=\linewidth]{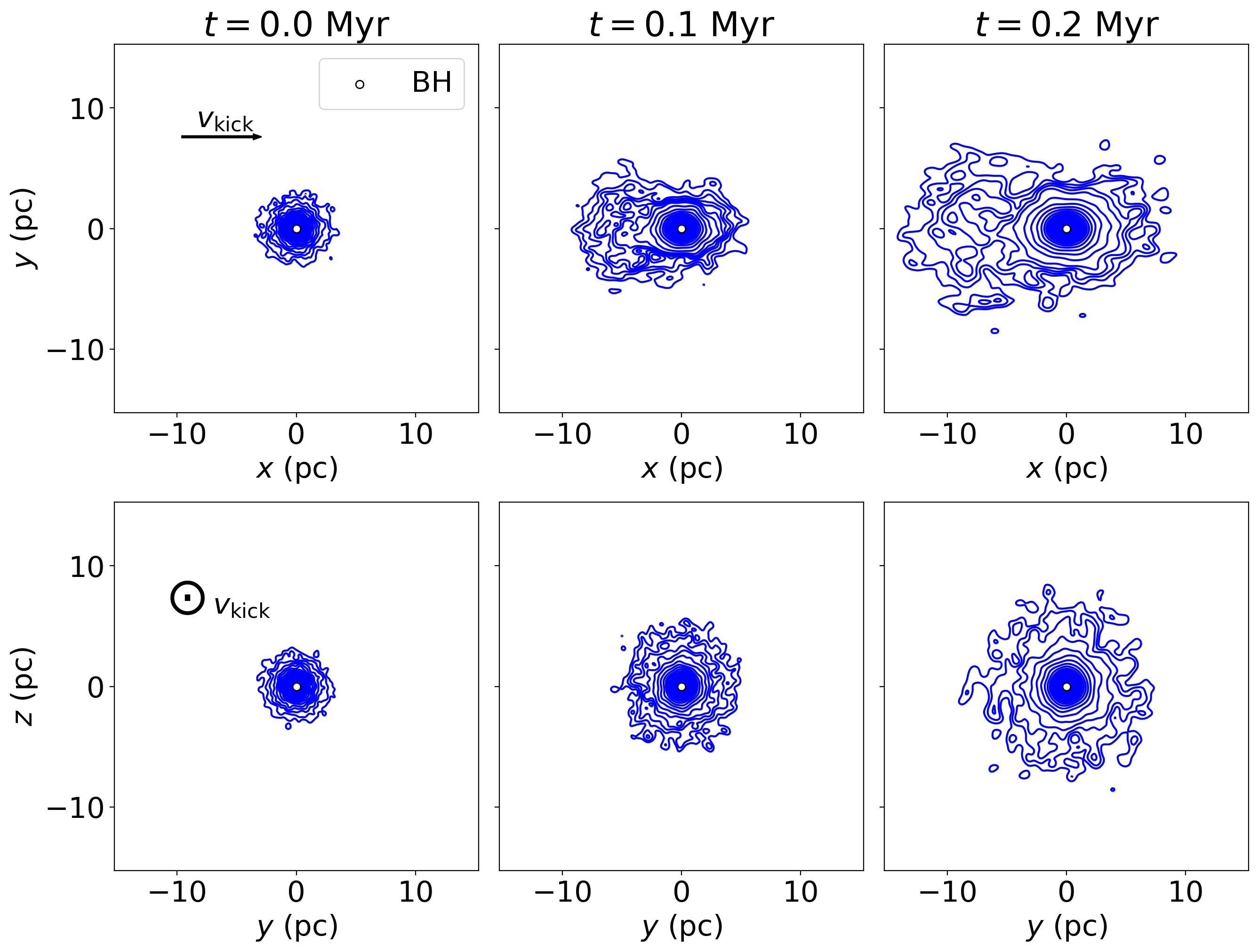}
\caption{\textbf{Spherically isotropic distribution with a $N(e) = 2e \ de$ eccentricity profile and a Bahcall-Wolf cusp density profile.} The post-kick surface density distribution of bound stars in the (\textit{top:}) $x$-$y$ plane and (\textit{bottom:}) $y$-$z$ plane at (\textit{left:}) $t = 0$, (\textit{center:}) $t=0.1$ Myr, and (\textit{right:}) $t = 0.2$ Myr. The ``o'' marker shows the recoiling black hole, and the density contours are computed in the frame co-moving with the black hole. In each orientation, the recoil kick direction is marked and labeled as $\vk$. Length scale and time units assume a black hole mass of $M_\bullet = 4 \times 10^6 \ M_{\odot}$ and a kick of $\vk = 100$ km/s. The $x$-$y$ plane density contours reveal the ``fan''-shape behind the recoiling black hole found in \citet{Merritt2009}.}
\label{fig:sigma_2ede_iso}
\end{figure*}

\begin{figure*}[t!]
\centering
\includegraphics[width=\linewidth]{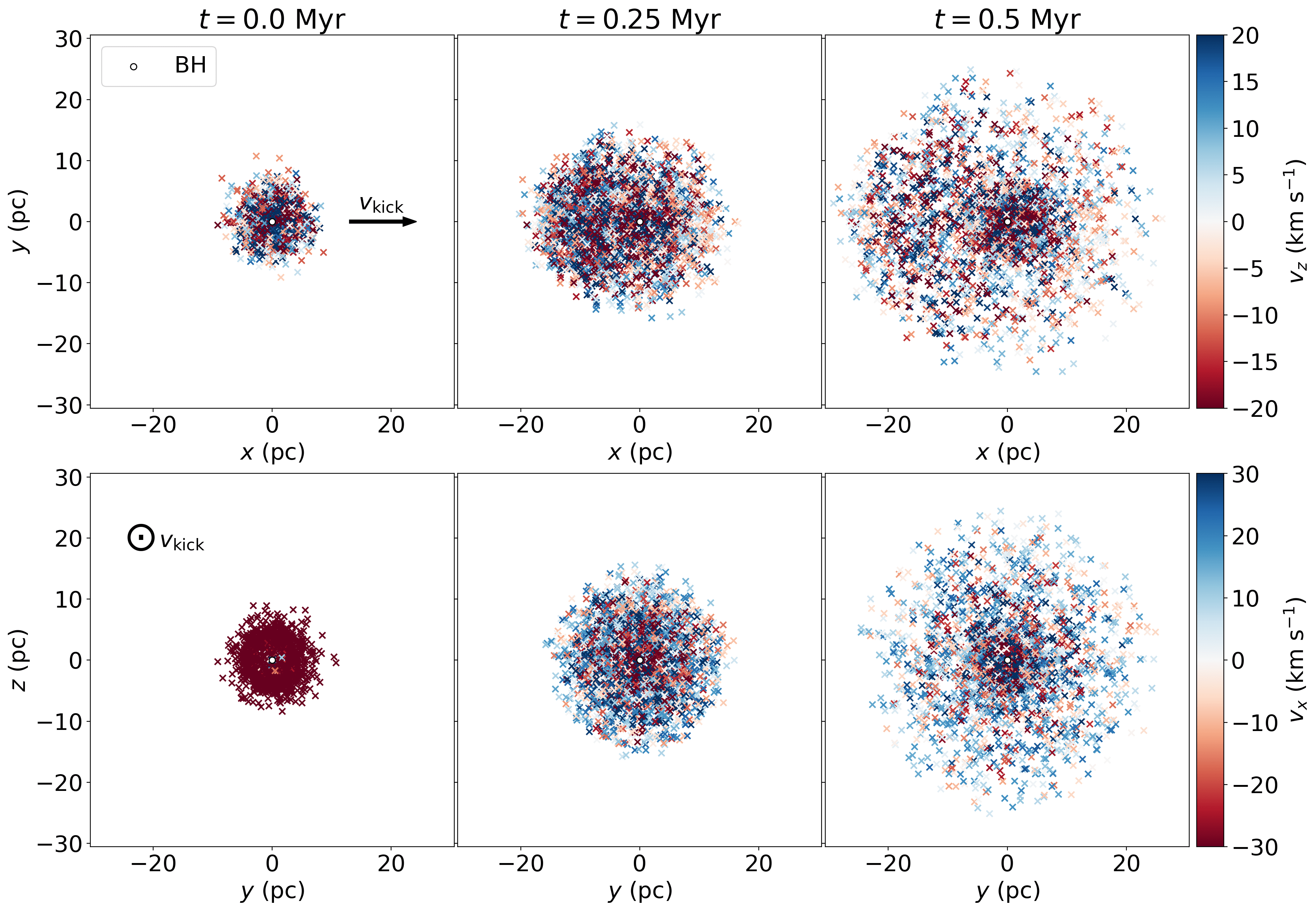}
\caption{\textbf{Spherically isotropic distribution with a $N(e) = 2e \ de$ eccentricity profile and a Bahcall-Wolf cusp density profile.} The line-of-sight velocities in the frame co-moving with the recoiling black hole in the (\textit{top:}) $x$-$y$ plane and (\textit{bottom:}) $y$-$z$ plane at (\textit{left:}) $t=0$, (\textit{center:}) $t=0.25$ Myr, and (\textit{right:}) $t=0.5$ Myr. In each orientation, the color bar is adjusted such that approaching velocities are blue and receding velocities are red. The recoil kick direction is marked and labeled as $\vk$ in each row. Length scale and time units assume a black hole mass of $M_\bullet = 4 \times 10^6 \ M_{\odot}$ and a kick of $\vk = 100$ km/s. In the $x$-$y$ plane, there appears a cluster of stars with high line-of-sight velocities right behind the recoiling black hole. In the $y$-$z$ plane, the edges become more blue-shifted over time.}
\label{fig:velocity_maps_2ede_iso}
\end{figure*}

In Figure \ref{fig:sigma_2ede_iso}, we show the density profiles in two orientations ($x$-$y$ and $y$-$z$) using a kernel density estimation at three different times following the recoil. The ``o'' marker shows the recoiling black hole, and these contours are in the frame of reference co-moving with the black hole. In the top row, we see that an initially isotropic distribution forms a ``fan'' shape behind the black hole as seen at $t=0.1$ Myr and $t=0.2$ Myr, which is consistent with density plots presented in \citet{Merritt2009}. The density profile projected onto the $y$-$z$ plane is always axi-symmetric, as expected, and expands in size over time. Finally, we show the line-of-sight velocity distribution in the $x$-$y$ and $y$-$z$ planes at three different times post-kick in Figure \ref{fig:velocity_maps_2ede_iso}. Velocities are calculated in the frame co-moving with the recoiling black hole, and the color bar is adjusted such that approaching velocities are in blue and receding velocities are in red. In the $x$-$y$ plane, the fan-shape behind the black hole seems to exhibit higher line-of-sight velocities than the rest of the stellar distribution. This is caused by extremely eccentric and large semi-major axis orbits that begin their new orbits at their periapsis, similar to what we found in the center row of Figure \ref{fig:velocity_maps_e=0.99_in}. They develop eccentricities perpendicular to the kick vector and tilt over their major axes such that the periapsis is their descending node ($-x$-direction). Then, shortly after the kick, these stars form a cluster right behind the black hole in this fan-shape where they exhibit line-of-sight velocities close to their high periapsis speed. The $y$-$z$ plane shows unique line-of-sight velocity structures as well. Initially, stars start at their descending node, so every star is red-shifted. Over time, stars will switch to being blue-shifted at particular radii depending on their post-kick orbital period, so that the edges of the cluster are preferentially blue-shifted ($t=0.25$ Myr). This pattern gets weaker over time as the cluster expands in size and the anomalies are randomized ($t=0.5$ Myr).

\section{Discussion} 
\label{sec:conclusion}

In \citet{Akiba2021}, we explored the post-kick orbits of stars that remain bound to a recoiling black hole. For a narrow range of initial conditions, we showed that a lopsided stellar disk emerges post-kick. In this paper, we expand to more general initial stellar distributions and kicks. Our findings can be summarized as follows: 

\begin{enumerate}

    \item Circular disks which experience an in-plane kick of the black hole map to a lopsided, eccentric disk with a rich, complex orbital structure. Maximum orbital alignment occurs at a characteristic post-kick radius 
    \begin{equation}
        r_c \equiv \frac{4G\Mbh}{9 \vk^2} \ ,
    \end{equation}
    and significant sections of the post-kick disk are counter-rotating with respect to the disk's initial angular momentum. The fraction of retrograde orbits within the bound population tends toward 50\% for large post-kick semi-major axes.
    The azimuthally-averaged post-kick density profile remains similar to its pre-kick profile out to $\sim 4 \ r_c$, then drops sharply. 
    
    \item Circular disks which experience an out-of-plane kick map to a distinctive dome-shape. Post-kick, all stars start at the periapsis of their new elliptical orbits which results in a clustering of mean anomalies at 0. The angular momentum vectors of the orbits collectively roll over their major axes in the direction of their initial velocity vectors, which results in clustering of arguments of periapsis at $\pi$. Eccentricities and inclinations increase with semi-major axis from the black hole reaching a maximum of 1 and $45^{\circ}$ for bound orbits, respectively. Stars within a narrow radial range will have identical values for 5 out of their 6 Kepler elements ($a, e, i, \omega, \mathcal{M}$). Only longitude of ascending node, $\Omega$, will differ. 

    \item  The critical post-kick semi-major axis in which apsidal alignment is strongest is much smaller for high eccentricity initial conditions,
    \begin{equation}
    a_c = \frac{1-e}{1+e} \ r_c \ .
    \end{equation}
    
    In disks of high eccentricity orbits which experience an in-plane kick, roughly half of the stars will map to retrograde orientations post-kick, except at the smallest semi-major axes. At these very small distances from the black hole, there exist two anti-aligned lopsided prograde modes. At larger semi-major axes, there is a prograde eccentric nuclear disk apsidally-aligned in one direction and a retrograde eccentric disk aligned in the  other. This forms a bar-like structure. The orientations reverse around $a \sim 4$ (Figure~\ref{fig:sma_ang_mom_mean_ecc_e=0.99}).

    \item In disks of high eccentricity orbits which experience an out-of-plane kick, all orbits tilt significantly. Stellar orbits at low semi-major axes cluster with $\omega = 0$ while those at large semi-major axes cluster with $\omega = \pi$ and gain inclinations of up to 90$^\circ$. 
     
    \item  Even for the most general case of stellar orbits being distributed isotropically about the black hole before recoil, the post-kick distribution has complex kinematic structure. In particular, irrespective of the initial stellar orbital eccentricities, the recoil kick produces a torus of eccentric orbits perpendicular to the kick vector at large semi-major axes (bottom three rows of Figure~\ref{fig:sma_ang_mom_inc} and Figure~\ref{fig:omega_anom_hist_2ede_iso}). Furthermore, mean anomalies and arguments of periapsis are non-uniformly distributed.

\end{enumerate}

\citet{Mastrobuono-Battisti2023} recently showed that mergers of supermassive black holes embedded in nuclear star clusters can leave an imprint on the shape, density profile, rotation, and velocity structure of the post-merger nuclear star cluster. While the density profile is sensitive to the initial merger conditions, the post-merger nuclear star cluster is always found to be flattened and rotating (expected due to the conserved orbital angular momentum of its progenitors). The post-merger cluster is also tangentially anisotropic at its center. These are the initial conditions --- flattened, rotating, with central tangential anisotropy --- that we might expect for a cluster that then responds to a recoil kick when the supermassive black holes merge at the center. The simulation in this paper that best matches this initial condition is a low-eccentricity disk with $\sigma_i = 30^\circ$ which corresponds to an oblate spheroid with an axial ratio between $0.6$--$0.7$. As can be seen in Figure~\ref{fig:ecc_inc_alpha} and the top panel of Figure~\ref{fig:sma_ang_mom_inc}, these initial conditions result in statistically significant apsidal and/or tilt alignment of post-kick orbits.  In particular, at large post-kick semi-major axes we would expect anti-aligned prograde and retrograde disks of high eccentricity orbits. 

Disks of stars responding to in-plane components of recoil kicks end up with a significant fraction of their outer orbits with retrograde orientation (see bottom panels of Figures~\ref{fig:sma_ang_mom_mean_ecc_circ_in} and \ref{fig:sma_ang_mom_mean_ecc_e=0.99}). This type of  counter-rotation in an edge-on stellar ring surrounding a central black hole has already been observed in an early-type galaxy NGC 3706 \citep{Gultekin2011}.

In this paper, we have presented the instantaneous orbital distributions of stars following a recoil kick of the central black hole. We leave as future work the effects of mass loss due to gravitational wave emission \citep[e.g.][]{Penoyre2018} and the long-term evolution of these systems. We expect many configurations to evolve dramatically with time once self-gravity between the stars is included. For example, the anti-aligned eccentric nuclear disks we find in Section~\ref{ss:highecc-inplane} should gravitationally torque each other to extreme eccentricities. Furthermore, counter-rotating disks are  unstable to buckling out of the plane \citep{toomre64,merritt&sellwood94}. 
For the case in which stars are initially on circular orbits, those that map to retrograde orientations originate further from the black hole. If the stellar population is initially mass segregated with heavier stars and compact objects located closer to the black hole, we would expect the post-kick prograde population to be composed of heavier bodies and the retrograde population to be lighter. 

\section{Acknowledgements} \label{sec:acknowledgements}

AM gratefully acknowledges support from the David and Lucile Packard Foundation. 
This work utilized resources from the University of Colorado Boulder Research Computing Group, which is supported by the National Science Foundation (awards ACI-1532235 and ACI-1532236), the University of Colorado Boulder, and Colorado State University. 

\software{\texttt{REBOUND} \citep{Rein2012}}

\bibliographystyle{aasjournal}
\bibliography{ms}

\appendix

\section{Tilt Vector and $\protect\ia$, $\protect\ib$, and $\protect\ie$ Relations}
\label{sec:appA}

Two angles are required to orient the plane of an orbit in space. In \citet{Madigan2016} we introduced two inclination angles, $i_a$ and $i_b$.
These angles represent rotations of an orbit with angular momentum vector ($\hat{j}$) about its semi-major axis ($\hat{a}$) and its semi-minor axis (or semi-latus rectum) ($\hat{b} \equiv \hat{j} \times \hat{a}$), respectively. An additional angle is needed to specify the direction of the orbit within the orbital plane. We use $i_e$ which is the rotation of an orbit about the angular momentum vector. The angles $i_a$, $i_b$, and $i_e$ are equivalent to that of an aircraft’s roll, pitch and yaw.

Here we relate these angles to the tilt vector as introduced in Section~\ref{ss:circ-out}. The tilt vector results from the rotation of the eccentricity vector by an angle $-\Omega$, so taking the $x$-$y$ plane to be our reference plane and $+x$ to be our reference direction,

\begin{equation*}
\begin{aligned}
& t_1 = \cos(\Omega) e_x + \sin(\Omega) e_y \ , \\
& t_2 = -\sin(\Omega) e_x + \cos(\Omega) e_y \ , \\
& t_3 = e_z \ .
\end{aligned}
\end{equation*}

\noindent This allows us to view the eccentricity vector from the ascending node of the orbit rather than from a common reference direction. The ratio between $t_1$ and $t_3$ indicates the degree to which the orbit is tilting over their major axis versus minor axis. 

\subsection{$i_a$ is related to $t_1$ and they measure roll}

$i_a$ is defined as

\[ i_a = \arctan \left( \frac{\hat{b}_z}{\sqrt{1 - \hat{b}_z^2}} \right) \ . \]

\noindent We note that the longitude of ascending node, $\Omega$, is always $90^\circ$ counter-clockwise in the $x$-$y$ plane from $\vec{j}_\perp$, the projection of $\vec{j}$ onto the $x$-$y$ plane. Since the tilt vector rotates the orbit such that the ascending node coincides with the positive $x$-axis, this means that the angular momentum vector becomes rotated such that it resides in the $y$-$z$ plane. The angular momentum vector post-rotation is

\[ \vec{j}' = \left\langle 0, -\sqrt{j_x^2 + j_y^2}, j_z \right\rangle \ . \]

\noindent The definition of the tilt vector is the same rotation applied to the eccentricity vector, so $\vec{e}' = \vec{t}$. Then, we can calculate $\hat{b}_z$ as

\[ \begin{aligned}
\hat{b}_z &= \hat{j}'_x \hat{t}_2 - \hat{j}'_y \hat{t}_1 \\
        &= - \frac{j_\perp}{j} \frac{t_1}{e} \ ,
\end{aligned} \]

\noindent so $i_a$ can be written in terms the first component of the tilt vector, $t_1$.

\subsection{$i_b$ is related to $t_3$ and they measure pitch}

$i_b$ is given by

\[ i_b = \arctan \left( -\frac{\hat{a}_z}{\sqrt{1 - \hat{a}_z^2}} \right) \ , \]

\noindent and since $\hat{a} = \hat{e}$, the unit eccentricity vector, $i_b$ can be equivalently written as

\[ i_b = \arctan \left( -\frac{\hat{e}_z}{\sqrt{1 - \hat{e}_z^2}} \right) \ . \]

\noindent Furthermore, $\hat{e}_z = \hat{t}_3$, so

\[ i_b = \arctan \left( -\frac{\hat{t}_3}{\sqrt{1 - \hat{t}_3^2}} \right) \ . \]

\subsection{$i_e$ is related to the ratio $t_2/t_1$ and they measure yaw}

$i_e$ is defined as

\[ i_e = \arctan \left( \frac{\hat{e}_y}{\hat{e}_x} \right) \ , \]

\noindent and this measures yaw with respect to the $x$-axis. A similar definition of $i_e$ using the tilt vector would be

\[ i_e = \arctan \left( \frac{\hat{t}_2}{\hat{t}_1} \right) \ , \]

\noindent which measures yaw with respect to the longitude of ascending node. This is effectively a measure of yaw that is independent of our choice of reference direction.

\end{document}